%% file: main.tex
\documentclass[11pt,letterpaper]{article}

\usepackage[margin=1in]{geometry}
\usepackage{setspace}
\usepackage{microtype}

\usepackage{amsmath}
\usepackage{amssymb}
\usepackage{amsthm}
\usepackage{mathtools}

\numberwithin{equation}{section}

\theoremstyle{definition}

\theoremstyle{remark}

\usepackage{graphicx}
\usepackage{booktabs}
\usepackage{threeparttable}
\usepackage{tabularx}
\usepackage{caption}
\usepackage{subcaption}

\graphicspath{{figures/}}

\usepackage[round,authoryear]{natbib}

\usepackage[hidelinks]{hyperref}
\usepackage[nameinlink,capitalise]{cleveref}

\title{
Dependency Network-Based Portfolio Design:\\
A Theoretical and Empirical Assessment of Pairwise GFEVD
}

\author{
Zihan Lin$^{1,*}$ \quad
Haojie Liu$^{1}$ \quad
Randall R. Rojas$^{2}$ \\[0.8em]
\small $^{1}$Department of Economics,
University of California, Riverside\\
\small Riverside, CA, USA\\[0.3em]
\small $^{2}$Department of Economics,
University of California, Los Angeles\\
\small Los Angeles, CA, USA\\[0.5em]
\small $^{*}$Corresponding author:
\texttt{zlin169@ucr.edu}\\
\small \texttt{hliu332@ucr.edu}\\
\small \texttt{rrojas@econ.ucla.edu}
}

\date{\today}

\begin{document}

\maketitle


\input{sections/00_abstract}


\input{sections/01_introduction}


\input{sections/02_literature}


\input{sections/03_theory}


\input{sections/04_data}


\input{sections/05_methodology}


\input{sections/06_results}


\input{sections/07_robustness}


\input{sections/08_discussion_and_conclustion}


\bibliographystyle{plainnat}
\bibliography{bib}

\section*{Statements and Declarations}

\subsection*{Funding}
The authors received no specific funding for this research.

\subsection*{Competing Interests}
The authors declare no competing interests.

\subsection*{Data Availability}
The historical adjusted closing prices used in this study
were obtained from Yahoo Finance
(\url{https://finance.yahoo.com/}).
The three-month Treasury bill rate was obtained from
the Federal Reserve Economic Data (FRED) database
(\url{https://fred.stlouisfed.org/series/DTB3}).
Details of data construction, sample selection, and
computational procedures are provided in Appendix C.
Access to and redistribution of the underlying third-party
data are subject to the respective data providers'
terms of use.

\subsection*{Code Availability}
The code used in this study is available from the
corresponding author upon reasonable request.


\appendix

\input{appendix/A_proofs}

\input{appendix/B_additional_results}

\input{appendix/C_data_reproducibility}

\end{document}

%% file: sections/00_abstract.tex
\begin{abstract}
This study examines whether pairwise Generalized Forecast Error
Variance Decomposition (GFEVD) provides additional information
for equity portfolio selection beyond contemporaneous Pearson
correlation. We establish the conditions under which pairwise
GFEVD and absolute innovation correlation produce equivalent
Minimum Spanning Trees (MSTs), and derive sufficient conditions
for network and stock selection stability under perturbations
in edge costs. We further develop a rolling-window estimation
error bound that characterizes the trade-off between sampling
error and time variation in dependence relationships.
Using 451 U.S. stocks and 293 weekly formation dates, we
compare GFEVD networks with absolute and signed Pearson
networks under common portfolio construction and trading rules.
Although the estimated MSTs change across all 876 adjacent-week
comparisons, the selected five-stock sets remain unchanged
in 194 cases. Under the primary specification, GFEVD achieves
higher final wealth than absolute Pearson but lower final
wealth than signed Pearson under both weighting methods.
The intersection--union test fails to establish a consistent
investment advantage ($p=0.6649$). The empirical results are
sensitive to the estimation window, portfolio size, and
network construction rule, and remain subject to the ex-post
stock universe and retrospective evaluation design.
Our findings distinguish the additional dependence information
captured by GFEVD from its transmission into network structure,
stock selection, and realized portfolio performance.
\end{abstract}

\noindent\textbf{Keywords:} Generalized Forecast Error Variance
Decomposition; Minimum Spanning Tree; Network Stability;
Portfolio Selection; Time-Varying Dependence.

\noindent\textbf{JEL Classification:} C32, C58, G11.

%% file: sections/01_introduction.tex
\section{Introduction}
\label{sec:introduction}

Understanding the relationships among stocks is an important part of
portfolio construction. Investors need to consider not only the behavior
of individual assets but also how their returns move together. In the
traditional mean--variance framework, this dependence enters through the
covariance matrix and affects the risk of holding several assets in the
same portfolio \citep{Markowitz1952}. Correlation provides a familiar way
to summarize these relationships. However, when we consider hundreds of
stocks, it becomes difficult to understand the structure of the market
from a large correlation matrix alone. Moreover, contemporaneous
correlation does not directly describe whether the past movements of one
stock contain information about the future movements of another.
Common return factors can also affect many stock pairs simultaneously,
which complicates the economic interpretation of individual links
\citep{FamaFrench1993}.

One way to address the first problem is to represent the market as a
network. Each stock can be treated as a node, while the relationships
between stocks determine the costs or strengths of the edges. Instead of
examining all possible connections, we can use a minimum spanning tree
(MST) to identify a smaller set of links that connects the entire stock
universe. \citet{Mantegna1999} introduced this approach to study the
hierarchical structure of financial markets, and \citet{Onnela2003}
showed that correlation-based asset trees can change with market
conditions. Related research has examined the topology of empirical
trees and alternative ways to filter dense dependence matrices
\citep{Bonanno2003,Tumminello2005}. An MST is therefore a useful way
to summarize a complicated dependency structure. Still, the tree can
only reflect the information contained in the measure used to construct
its edges. If the input is based solely on contemporaneous correlation,
the resulting network does not explicitly incorporate lagged
relationships.

Time-series models provide another perspective. Models of dynamic
conditional correlation allow contemporaneous comovement to change
over time \citep{Engle2002}, while a vector autoregression (VAR)
allows the return of one stock to depend on its own past returns
and on the past returns of another stock. Forecast error variance
decomposition (FEVD) then describes how innovations in these series
are associated with forecast uncertainty over a chosen horizon.
The generalized forecast error variance decomposition (GFEVD)
approach related to \citet{PesaranShin1998} avoids the need to
choose a particular ordering of variables for recursive
orthogonalization. \citet{DieboldYilmaz2012} and
\citet{DieboldYilmaz2014} further developed the use of generalized
variance decompositions to study directional connectedness in
financial systems. These developments suggest a way to construct
a stock network that incorporates estimated dynamic dependence
in addition to contemporaneous innovation covariance.

Nevertheless, adding a time-series model does not necessarily
improve the network or the portfolio constructed from it.
In some settings, a GFEVD-based measure can contain essentially
the same ranking information as a measure based on innovation
correlations. In other settings, lagged interactions can change
the relative importance of stock pairs. Even when the estimated
edge costs differ, the MST may remain unchanged. Furthermore,
a change in the MST does not always change the stocks selected
from it, and a change in the selected stocks does not necessarily
improve portfolio performance. These distinctions are important
because the ultimate purpose of our network is not only to describe
relationships, but also to support an investment decision. The
relevant question is therefore not simply whether GFEVD produces
different numbers from correlation, but under what conditions
these differences affect the network and the selected stocks,
and whether they provide additional economic value.
Existing work on variance-decomposition normalization
provides additional reasons for caution in interpreting these
measures \citep{ChanLau2017,LanneNyberg2016}.

To study this question, we construct a financial network from
pairwise generalized forecast error variance decompositions.
For each pair of stocks, we estimate a bivariate VAR(1) using
a rolling window of past returns and calculate GFEVD over a
fixed forecast horizon. We convert the two directional measures
into a single undirected edge cost, apply the MST algorithm,
and rank stocks according to their degree within the resulting
tree. A portfolio is then formed from the highest-degree stocks
under a predetermined selection and allocation procedure.
This approach follows the broad motivation of network-based
portfolio research, but it is important to distinguish a
network ranking from a portfolio optimization result.
Previous studies, including \citet{PeraltaZareei2016} and
\citet{VyrostLyocsaBaumohl2019}, have examined the use of
centrality information for asset allocation. In our setting,
a high degree indicates that a stock has many connections
in the filtered tree; it does not establish that the stock
has a higher expected return, a lower risk contribution,
or a causal influence on other assets. Whether this selection
rule is useful must be examined empirically. In particular,
\citet{Pozzi2013} reported diversification benefits from
choosing more peripheral rather than central stocks in
filtered financial networks.

Our first objective is to clarify what GFEVD can add to a
correlation-based financial network. We derive conditions
under which the normalized pairwise GFEVD measure is
determined entirely by the contemporaneous correlation
of VAR innovations. This provides a useful benchmark:
in such a setting, the additional modeling complexity
does not identify a distinct form of cross-variable dynamic
dependence. We then consider cases in which lagged interactions
generate positive GFEVD dependence even when the
innovations are uncorrelated. Together, these results
explain both the potential value and the limits of using
a dynamic dependence measure. They do not imply that
GFEVD must outperform a correlation network for every
stock universe or market period.

Our second objective is to understand how changes in
estimated dependence affect the network and the selected
stocks. The edge costs are estimated from finite rolling
windows, so different samples or market conditions may
change their values and rankings. We use the replacement
margins of minimum spanning trees to describe when a
perturbation in estimated costs is sufficiently small
to leave the network unchanged. We then relate the number
of edge replacements to the degree ranking used to select
the portfolio. In particular, the degree difference
between the last selected stock and the first excluded
stock provides a way to examine how sensitive stock
selection is to changes in the tree. This allows us to
distinguish the stability of the entire network from
the stability of the selected stocks. The resulting
conditions are sufficient rather than necessary, so a
failure to satisfy them does not necessarily imply
that the network or the selected stocks will change.

We examine these theoretical conditions empirically by
reconstructing the complete set of estimated candidate
edge costs, including those that are not selected by the
MST algorithm. This allows us to calculate replacement
margins and compare them with the changes in edge costs
observed between consecutive estimation dates. We also
examine how the replacement of tree edges is associated
with changes in the degree ranking and the selected
stock sets. These calculations describe the sensitivity
of the estimated networks to changes in their costs.
They do not establish that the corresponding population
networks have been recovered, since the estimation
errors relative to the underlying population costs
are unknown. We also discuss the trade-off involved in
choosing the estimation-window length. The broader
econometric literature shows that structural breaks
can change the appropriate estimation-window length
\citep{PesaranTimmermann2007}.

Our third objective is empirical. We evaluate the pairwise
GFEVD--MST selection rule using a time-ordered portfolio
simulation and compare it with two correlation-based
MST benchmarks under identical portfolio selection,
weighting, and trading rules. The analysis uses a common
universe of 451 stocks and 293 completed-week estimation
dates. Each portfolio is rebalanced at the next trading
session close following the weekly signal. Between
rebalancing dates, the holdings remain unchanged in
terms of adjusted-price units, while their portfolio
weights are allowed to change with asset returns.
Transaction costs are calculated from the gross value
of the assets traded at each rebalancing date.

We also examine how the results respond to different
rolling-window lengths and numbers of selected stocks.
The baseline specification is fixed in advance of this
sensitivity analysis; alternative parameter choices are
reported as sensitivity results rather than treated as
independently validated improvements. Our evaluation
therefore considers not only investment performance,
but also the extent to which it depends on modeling
assumptions and portfolio design choices. This separation
is important because repeated selection among many
historical strategies can create data-snooping problems
\citep{White2000,Hansen2005}, while transaction costs
can change the economic attractiveness of active
equity strategies \citep{KorajczykSadka2004}.

Our empirical findings show that pairwise GFEVD and
correlation can produce different stock selections
even when their networks share a large proportion
of their edges. Under the primary specification of
five selected stocks and a 250-day estimation window,
the GFEVD and absolute Pearson networks have an average
edge Jaccard index of approximately 0.813 and an
average five-stock selection overlap of approximately
0.799. However, the two methods select exactly the
same five stocks on only 70 of the 293 estimation
dates. This suggests that a high degree of similarity
between two networks does not necessarily imply that
the stocks selected from them are the same.

The stability analysis provides further evidence of
the distinction between network stability and stock
selection stability. Across the three network methods,
the sufficient condition for preserving the MST is
not satisfied in any of the 876 comparisons between
consecutive weekly networks. Although the estimated
tree changes in every comparison, the selected
five-stock set remains unchanged in 194 cases.
Moreover, ties at the degree cutoff occur on more
than 60 percent of the estimation dates under each
network method. These results show that changes in
the network do not always pass through to the final
stock selection, while the separation between selected
and excluded stocks is often limited. They also
illustrate why the theoretical sufficient conditions
can be conservative when applied to observed changes
in financial networks.

The differences in network structure and stock selection
do not lead to a consistent investment advantage for
pairwise GFEVD. Under the primary specification,
GFEVD achieves higher reserved-period final wealth
than the absolute Pearson benchmark under both
weighting methods, but lower final wealth than the
signed Pearson benchmark. The predetermined
four-comparison criterion for incremental network
value is therefore not satisfied. The paired bootstrap
analysis also does not provide statistically significant
evidence of a consistent advantage, with an
intersection--union $p$-value of 0.6649. This conclusion
remains unchanged when alternative bootstrap block
lengths are used. Since the statistical analysis is
conducted on a historical period whose portfolio
outcomes have already been examined, these results
are interpreted as an assessment of statistical
uncertainty rather than as independent out-of-sample
confirmation.

The robustness analyses further show that portfolio
performance depends on the estimation window and
portfolio size. Under equal weighting, several
alternative parameter combinations produce favorable
final-wealth differences for GFEVD, although these
results are not maintained across all specifications.
Additional comparisons using a one-step GFEVD horizon
and alternative symmetrization rules examine the
contribution of multi-step propagation and the
treatment of directional dependence. The resulting
differences in portfolio performance are not uniformly
favorable across weighting methods. Comparisons with
a broad equal-weight portfolio and a market benchmark
based on vendor-adjusted SPY prices also show that
the network-based portfolios do not necessarily
improve upon simpler investment procedures.
Taken together, these findings highlight the distinction
between identifying additional dependence information
and obtaining reliable investment value from that
information.

The interpretation of the empirical results is subject
to several limitations. In particular, the common
stock universe is reconstructed using eligibility
information from the full sample rather than obtained
from a point-in-time record of historical membership.
The portfolio simulations are also based on
vendor-adjusted closing prices rather than a verified
transaction record. Consequently, the empirical
comparisons describe the relative behavior of the
specified selection rules within a fixed historical
stock universe. They do not establish how these rules
would have performed if the available investment
universe had been determined separately at each
historical date.

The contribution of this paper is an integrated theoretical
and empirical assessment of pairwise GFEVD for network-based
portfolio selection. First, we characterize the relationship
between GFEVD and innovation correlation and examine how
lagged dependence can affect network construction.
Second, we connect rolling-window estimation uncertainty
with MST replacement margins and degree-based selection
stability. This provides sufficient conditions under which
changes in estimated dependence leave the network or the
selected stock set unchanged. Third, we evaluate the economic
value of these differences under common portfolio selection,
weighting, and trading rules. Rather than introducing a new
portfolio optimization method, our analysis distinguishes
the additional information contained in a dependence
measure from its transmission into network structure,
stock selection, and realized investment performance.

The remainder of the paper is organized as follows.
Section~\ref{sec:literature} reviews the related
literature on financial networks, variance decompositions,
and portfolio selection. Section~\ref{sec:theory}
develops the theoretical framework.
Section~\ref{sec:data} describes the data, and
Section~\ref{sec:methodology} presents the empirical
methodology. Section~\ref{sec:results} reports the main
empirical results, while Section~\ref{sec:robustness}
examines their sensitivity to alternative specifications
and implementation choices.
Section~\ref{sec:discussion_conclusion} discusses
the implications and limitations of the findings
and concludes the paper.

%% file: sections/02_literature.tex
\section{Related Literature}
\label{sec:literature}

Our study connects two questions in financial research. The first is
how to describe the relationships among a large number of stocks.
The second is whether the structure of these relationships provides
useful information for portfolio selection. Financial networks offer
a convenient way to connect the two questions, but the answer depends
on both the measure used to construct the network and the rule used
to select assets from it. We organize the relevant literature around
these two decisions, the stability of the resulting networks and
stock selections, and the evaluation of their portfolio performance.

\subsection{Correlation-Based Financial Networks}
\label{subsec:literature_correlation}

Correlation is one of the most common measures of dependence between stock
returns. In a traditional portfolio problem, correlations enter the
covariance matrix and therefore affect the risk of combining different
assets \citep{Markowitz1952}. The same information can also be represented
as a network, where stocks are nodes and the strength of their relationships
determines the connections between them. This representation is especially
useful when the number of stocks is large and it becomes difficult to
interpret a complete correlation matrix directly.

\citet{Mantegna1999} introduced a hierarchical representation of financial
markets based on correlations and the minimum spanning tree (MST). Rather
than displaying every possible relationship, the MST retains a set of
$N-1$ edges that connects all $N$ stocks without cycles. In this way, a
large dependency matrix can be reduced to a simpler structure that is easier
to examine. \citet{Onnela2003} further studied the changing topology of
correlation-based asset trees and showed that the network structure can vary
with market conditions. These studies established the MST as a useful tool
for summarizing financial dependence, although the tree itself is a
filtered representation of the underlying data rather than a complete
description of the market.

Subsequent research has examined both the information captured by an MST
and the information lost during network filtering. \citet{Bonanno2003}
compared the topology of empirical stock-market trees with those obtained
from simulated market models, while \citet{Tumminello2005} proposed a
different filtering method that retains a richer set of representative
links. These results are relevant because a minimum spanning tree is only
one possible summary of the original dependence matrix. Changes in the
network measure and changes in the filtering algorithm should therefore
be treated as separate modeling decisions. We use the MST to obtain a
sparse and reproducible selection structure, not because it is known to
preserve every economically relevant relationship.

The main limitation for our purpose is that contemporaneous correlation
does not directly describe how information contained in one return series
may be associated with another series at later dates. Two stocks may have
little contemporaneous correlation and still exhibit substantial lagged
dependence. Conversely, their correlation may be strong even when a model
of their dynamic relationship adds little new information. This distinction
motivates the use of a time-series-based network measure. However, even
when such a measure identifies different dependence relationships, these
differences may not affect the edges retained by the MST. Whether dynamic
information changes the filtered network is therefore a separate question
from whether it changes the original dependence matrix.

Correlation-based methods are not necessarily static. For example,
\citet{Engle2002} developed dynamic conditional correlation models that
allow the covariance structure to vary over time. Our distinction is
therefore between measures of contemporaneous comovement and measures
that explicitly use the lagged cross-response structure of a fitted VAR,
not between models that are always static and models that are always
dynamic. In addition, pairwise dependence does not identify the
underlying economic source of comovement. Common return factors are
well established in asset pricing \citep{FamaFrench1993}, and factor-based
covariance estimation provides a way to separate pervasive components
from more specific residual relationships \citep{FanLiaoMincheva2013}.
A GFEVD edge may consequently reflect common exposures rather than a
direct economic connection between two firms. This is an important
limitation for the interpretation of our estimated network.

\subsection{Variance Decompositions and Financial Connectedness}
\label{subsec:literature_connectedness}

Vector autoregressions (VARs) provide a framework for describing how a
collection of variables depends on its own past values and on the past
values of other variables. Forecast error variance decomposition (FEVD)
uses the fitted model to characterize the sources of forecast uncertainty
over a specified horizon. The generalized approach associated with
\citet{PesaranShin1998} avoids the dependence on variable ordering that
arises from a particular recursive identification. Building on this
approach, \citet{DieboldYilmaz2009} studied return and volatility
spillovers across global equity markets. \citet{DieboldYilmaz2012}
subsequently developed generalized measures of directional spillovers,
while \citet{DieboldYilmaz2014} explicitly connected variance
decompositions to weighted and directed financial networks. Related
econometric approaches have also represented interdependence through
principal components, Granger-causality relations, and tail-risk
connections \citep{BillioEtAl2012,HautschEtAl2015}. The broader message
of this literature is that a financial network is defined partly by
what kind of dependence it is designed to measure.

This literature provides an important foundation for our study. A GFEVD
measure incorporates both the innovation covariance and the way shocks
propagate through the fitted VAR over the forecast horizon. Therefore, it
can describe some forms of dynamic dependence that are not directly
represented by a contemporaneous correlation coefficient. However, the
presence of a dynamic model does not necessarily mean that its variance
decomposition contains different ranking information from an
innovation-correlation measure. Understanding when these two measures
provide similar or different information is one of the questions examined
in our theoretical analysis.

Generalized variance shares also require careful interpretation. In
particular, their normalization can affect the resulting connectedness
measures, and the network should not be interpreted as an identified
causal transmission system. The limitations of conventional
variance-decomposition networks and alternative constructions are
discussed by \citet{ChanLau2017}. \citet{LanneNyberg2016} proposed a
generalized decomposition with a different normalization property,
illustrating that the way forecast uncertainty is attributed across
variables is not a purely mechanical choice. These alternatives do not
establish that one normalization is universally best for portfolio
selection.

The size of the financial system creates an additional estimation
problem. Fully multivariate time-series networks can become difficult
to estimate when the number of assets is large relative to the available
sample. \citet{BarigozziBrownlees2019}, for example, developed a
regularized approach to network estimation in high-dimensional time
series. \citet{BrunnerHipp2023} examined the estimation of
large-dimensional connectedness tables based on forecast error
variance decompositions. Our pairwise construction offers a
different trade-off.
 We fit
a separate bivariate VAR to each stock pair, which avoids estimating
one large VAR but does not condition each estimated pair on all other
stocks. This makes the estimation problem manageable for a broad equity
universe, while leaving the possibility that common factors or omitted
variables affect the resulting pairwise measures.

Our implementation therefore differs from a conventional system-wide
connectedness model. Although the pairwise GFEVD measures are assembled
into a common dependence matrix, this matrix is not the FEVD of a
single high-dimensional VAR. We then convert the two directional
measures for each pair into one undirected cost before constructing
the MST. As a result, the final tree summarizes a particular aspect
of forecast-error dependence; it does not preserve the complete
direction of the original pairwise relationships. These distinctions
matter when interpreting both the network and the stocks selected
from it.

\subsection{Network-Based Portfolio Selection and Stability}
\label{subsec:literature_portfolio}

The use of network information in portfolio construction has also received
considerable attention. \citet{PeraltaZareei2016} studied the relationship
between financial network centrality and portfolio selection, providing a
connection between network characteristics and conventional portfolio
allocation. \citet{VyrostLyocsaBaumohl2019} examined several network
representations and centrality measures in out-of-sample asset allocation.
Together, these studies show that financial networks can serve as inputs to
investment decisions. They also make clear that the way assets are ranked
and weighted is part of the strategy, rather than an automatic consequence
of network construction.

There is also evidence that network position can have different
implications depending on the investment objective. \citet{Pozzi2013}
found that selecting stocks in more peripheral parts of filtered
financial networks could improve diversification relative to selecting
central stocks. More recent work has explicitly distinguished
idiosyncratic and network risk when constructing portfolios
\citep{HardleEtAl2026}. A central position may indicate broad
connectedness, but it may also imply greater exposure to common market
movements. In particular, a stock with a high degree in an MST has
more connections in the filtered tree, but this property does not by
itself imply a higher expected return or a lower contribution to
portfolio risk. The economic meaning of centrality depends on both
the network definition and the investment objective
\citep{PeraltaZareei2016}.

In our study, selecting the highest-degree stocks is a predetermined
and transparent selection rule. We use this rule to examine whether
differences in the dependence measure lead to differences in portfolio
selection. We do not treat it as a solution to a mean--variance
optimization problem or claim that greater centrality must lead to
better investment outcomes. The distinction is important because the
performance of a network-based portfolio depends not only on the
information contained in the network, but also on how that information
is used to select and allocate assets.

Conventional portfolio research provides another reason to examine the
allocation decision separately from the stock-selection decision.
\citet{JagannathanMa2003} showed that portfolio constraints can reduce
risk when covariance estimates are noisy, while
\citet{DeMiguelGarlappiUppal2009} documented how difficult it can be for
estimated optimal portfolios to improve reliably on simple allocation
rules. This literature supports the use of clearly defined weighting
procedures and simple portfolio benchmarks. It also suggests that the
economic value of a more complicated selection rule should be examined
relative to simpler alternatives rather than inferred from its
construction alone.

Another issue is the sensitivity of network-based decisions to changes
in estimated dependence. Since the edge costs are estimated from finite
samples, different observations may change their relative values.
However, not every change in edge costs affects the MST. The
mathematical sensitivity of minimum spanning trees has a long history
\citep{Tarjan1982,DixonRauchTarjan1992}.
 In particular, the replacement of a tree edge
depends on the cost of an alternative edge relative to the costs of
the edges along its corresponding tree path. This provides a way to
describe how large a perturbation in edge costs can be before the
stability of the original tree is no longer guaranteed.

The stability of node centrality under network perturbations
has also been studied in the network literature.
\citet{SegarraRibeiro2016} examined the stability and continuity
of centrality measures in weighted graphs, while
\citet{ShenKolaczyk2026} studied the recovery of degree-based
top-$k$ node sets under network noise. Our analysis considers
a related but different problem, in which perturbations in
estimated edge costs affect an MST and subsequently the
stocks selected from its degree ranking.

For portfolio selection, the stability of the MST is only part of the
problem. When stocks are selected according to their degree within
the tree, a change in the edge set matters only to the extent that
it affects the relevant degree ranking. Several edges may be replaced
without changing the selected stocks. Conversely, even a small
number of edge replacements may affect stock selection when the
degree values near the selection cutoff are close or tied. We use
these observations to connect established MST sensitivity arguments
with the degree-based selection rule considered in this study.

There is also a distinction between the sensitivity of an estimated
network and the recovery of an underlying population network.
Replacement margins describe the stability of a tree relative to
perturbations in a given cost matrix. Whether an estimated tree
recovers the population tree additionally depends on the estimation
errors in its edge costs. For financial data, rolling-window
estimation creates a familiar difficulty: longer windows may reduce
estimation noise but can also include observations generated under
earlier dependence structures. This is related to the bias--variance
problem in locally stationary time-series estimation
\citep{DahlhausGiraitis1998}. Recent work has also developed
estimation and inference methods for time-varying VAR models
\citep{GaoPengYan2024}. Our focus is on how rolling-window
estimation errors affect the GFEVD edge costs and the
resulting network-based selection rule.

The choice of a rolling-window length has also been studied directly
in econometric forecasting. \citet{PesaranTimmermann2007} examined how
the trade-off between estimation precision and structural breaks affects
window selection. For our application, the same tension operates through
the estimated network: an old observation may improve parameter
precision while making the estimated dependence structure less
representative of the current market. Neither a long window nor a short
window can be justified solely by the number of observations it contains.
These considerations suggest that an empirical evaluation should
examine not only portfolio returns, but also the sensitivity of the
estimated networks and the persistence of the selected stocks.

\subsection{Portfolio Evaluation and Research Design}
\label{subsec:literature_evaluation}

An additional concern is how a network strategy is evaluated after it has
been designed. Comparing many selection rules and then reporting only
the best historical return can make the apparent advantage sensitive
to the search procedure. \citet{White2000} developed a framework for
data-snooping adjustments, and \citet{Hansen2005} examined tests of
superior predictive ability in the presence of multiple alternatives.
Related research in finance emphasizes the danger of repeated testing
and the likelihood of discovering apparently attractive results by
chance \citep{HarveyLiuZhu2016,BaileyEtAl2017}. We draw a practical
distinction between the performance of a fixed baseline and the results
of sensitivity exercises. Examining several window lengths or portfolio
sizes helps reveal dependence on assumptions, but choosing the best
observed combination after examining the same evaluation period does not
create a new, untouched out-of-sample test.

The measurement of performance matters as well. A high historical
return does not necessarily imply a superior risk-adjusted strategy,
and the sampling properties of performance ratios may be nontrivial
when returns are serially dependent \citep{Lo2002}. A strategy that
reconstitutes its holdings frequently can also incur substantial
trading costs. \citet{KorajczykSadka2004} demonstrated the relevance of
such costs for evaluating active equity strategies; the network-based
asset allocation evidence of \citet{VyrostLyocsaBaumohl2019} similarly
highlights the role of turnover.

In a network-based portfolio, changes in the selected stock set are
one source of turnover, but they are not the only source. Even when
the same stocks are selected at two consecutive rebalancing dates,
changes in asset prices or target weights may require additional
trades. Conversely, between scheduled rebalancing dates, portfolio
weights may change without any trading taking place. It is therefore
important to distinguish changes in selected stocks from changes in
portfolio weights and from the actual value of the assets traded.
Our empirical assessment uses the same portfolio accounting and
transaction-cost assumptions across the compared network methods.

The interpretation of a historical portfolio evaluation also depends
on how the investment universe is defined. A common stock universe
allows different selection rules to be compared using the same set
of assets. However, when this universe is constructed using
information from the full sample, the comparison does not necessarily
represent the investment opportunities that were available at each
historical date. These considerations limit the conclusions that
can be drawn from the observed portfolio returns, even when the
estimation and trading procedures follow a chronological order.

We build on these literatures in three related ways. First, we characterize
conditions under which pairwise GFEVD conveys the same network-ranking
information as an innovation-correlation measure, and conditions under
which lagged dependence can produce a different result. Second, we use
MST sensitivity and degree-ranking arguments to distinguish changes in
estimated edge costs from changes in the selected stock set. These
arguments rely on established variance-decomposition and graph-theoretic
properties; our purpose is to make their implications explicit for the
selection rule considered here. By examining replacement margins,
edge replacements, and degree cutoff gaps in the empirical networks,
we also investigate the extent to which the theoretical stability
conditions describe the observed network and selection changes.

Third, we evaluate the resulting strategy under a fixed empirical
specification, comparing it with relevant alternatives and examining
its sensitivity to portfolio cardinality, estimation-window length,
and other modeling choices. We distinguish the structural differences
between the estimated networks from the economic performance of the
portfolios constructed from them. We therefore treat the value of
additional dynamic information as a theoretical and empirical question,
not as a guaranteed improvement over correlation-based selection.

Section~\ref{sec:theory} develops the corresponding theoretical framework.

%% file: sections/03_theory.tex
\section{Theoretical Framework}
\label{sec:theory}

To examine whether pairwise GFEVD provides additional information for stock selection, we first study how dependence between stocks is measured and how the resulting network is constructed. Unlike a correlation-based approach, GFEVD allows lagged relationships to enter the measurement of forecast-error dependence. However, the presence of additional dependence information does not necessarily imply that the resulting network or portfolio will be different. We therefore examine the relationship between GFEVD and correlation, the conditions under which changes in edge costs affect the Minimum Spanning Tree (MST), and the role of estimation windows and portfolio size in stock selection. The objective is to establish the theoretical properties of the network construction procedure rather than to assume that a particular network must generate higher investment returns.

\subsection{Pairwise VAR and Generalized Forecast Error Variance Decomposition}
\label{subsec:pairwise_var_gfevd}

We use a pairwise Vector Autoregression (VAR) model to describe the relationship between the daily returns of two stocks. Unlike Pearson correlation, which measures contemporaneous linear dependence, a VAR also incorporates past returns and can therefore describe lagged predictive relationships. Let $\mathcal U=\{1,\ldots,N\}$ denote the fixed stock universe and let $r_{i,s}$ represent the daily return of stock $i$ at time $s$. The universe is held constant across network formation dates, although its composition is determined from the available historical sample rather than from a point-in-time constituent list. At each formation date $t$, we estimate the network using only observations available through that date. For a rolling window containing $W$ completed daily returns, the estimation sample is defined as
\begin{equation}
\mathcal T_t(W)=\{t-W+1,\ldots,t\}.
\label{eq:rolling_window}
\end{equation}
For each distinct stock pair $(i,j)$, we define $\mathbf x_s^{(ij)}=(r_{i,s},r_{j,s})^\top$ and estimate the following bivariate VAR(1):
\begin{equation}
\mathbf x_s^{(ij)}
=\boldsymbol\mu_{ij,t}
+\mathbf A_{ij,t}\mathbf x_{s-1}^{(ij)}
+\mathbf u_s^{(ij)},
\qquad
\mathbf A_{ij,t}
=
\begin{pmatrix}
a_{ii,t}&a_{ij,t}\\
a_{ji,t}&a_{jj,t}
\end{pmatrix}.
\label{eq:pairwise_var}
\end{equation}
Here, $\boldsymbol\mu_{ij,t}$ contains the intercepts, while the diagonal elements of $\mathbf A_{ij,t}$ describe the dependence of each stock on its own previous return. The off-diagonal elements represent lagged cross-stock predictive relationships. Since the model uses one lag, a window containing $W$ return observations provides $W-1$ usable response--lag pairs.

We assume that the innovation vector is a martingale difference with a positive-definite covariance matrix:
\begin{equation}
\mathbb E[\mathbf u_s^{(ij)}\mid\mathcal F_{s-1}]
=\mathbf 0,
\qquad
\mathbb E[
\mathbf u_s^{(ij)}
\mathbf u_s^{(ij)\top}
]
=
\boldsymbol\Sigma_{ij,t}.
\label{eq:var_innovations}
\end{equation}
The covariance matrix captures contemporaneous dependence between innovations, while the coefficient matrix describes how previous returns enter the current return equations. Both components are relevant for measuring forecast-error dependence. For the population analysis, we consider the parameters at a given formation date to be locally fixed and assume that the VAR is stable, so the spectral radius of $\mathbf A_{ij,t}$ is smaller than one. The associated fixed-parameter process then has the moving-average representation
\begin{equation}
\mathbf x_s^{(ij)}-\overline{\mathbf x}_{ij,t}
=
\sum_{h=0}^{\infty}
\boldsymbol\Phi_{ij,t}(h)
\mathbf u_{s-h}^{(ij)},
\qquad
\boldsymbol\Phi_{ij,t}(0)=\mathbf I_2,
\quad
\boldsymbol\Phi_{ij,t}(h)=\mathbf A_{ij,t}^{h}.
\label{eq:var_ma_representation}
\end{equation}
The matrices $\boldsymbol\Phi_{ij,t}(h)$ describe how innovations are propagated over subsequent periods. Therefore, the variance of a forecast error depends not only on the contemporaneous innovation covariance but also on the dynamic responses represented by these matrices.

To measure this dependence, we use the Generalized Forecast Error Variance Decomposition (GFEVD) developed by \citet{PesaranShin1998} and applied in the connectedness literature by \citet{DieboldYilmaz2012}. For variables $a,b\in\{1,2\}$ and forecast horizon $H$, the generalized variance contribution is defined as
\begin{equation}
\theta^{(ij)}_{a\leftarrow b}(H;t,W)
=
\frac{
\sigma_{bb,t}^{-1}
\displaystyle\sum_{h=0}^{H-1}
\left(
\mathbf e_a^\top
\boldsymbol\Phi_{ij,t}(h)
\boldsymbol\Sigma_{ij,t}
\mathbf e_b
\right)^2
}{
\displaystyle\sum_{h=0}^{H-1}
\mathbf e_a^\top
\boldsymbol\Phi_{ij,t}(h)
\boldsymbol\Sigma_{ij,t}
\boldsymbol\Phi_{ij,t}(h)^\top
\mathbf e_a
}.
\label{eq:generalized_fevd}
\end{equation}
Here, $\mathbf e_1=(1,0)^\top$ and $\mathbf e_2=(0,1)^\top$ are selection vectors, while $\sigma_{bb,t}$ denotes the corresponding diagonal element of the innovation covariance matrix. The numerator measures the generalized contribution associated with innovations in variable $b$, and the denominator represents the forecast-error variance of variable $a$. Unlike an orthogonalized decomposition, GFEVD does not require a particular ordering of the variables. However, because the innovations may be correlated, the two generalized contributions within a bivariate system do not necessarily sum to one. Following the row-normalization approach, we calculate
\begin{equation}
\widetilde\theta^{(ij)}_{a\leftarrow b}(H;t,W)
=
\frac{
\theta^{(ij)}_{a\leftarrow b}(H;t,W)
}{
\theta^{(ij)}_{a\leftarrow1}(H;t,W)
+
\theta^{(ij)}_{a\leftarrow2}(H;t,W)
}.
\label{eq:normalized_gfevd}
\end{equation}
For a stock pair ordered as $(i,j)$, we denote the normalized directional relationships by $\omega_{j\to i,t}=\widetilde\theta^{(ij)}_{1\leftarrow2}$ and $\omega_{i\to j,t}=\widetilde\theta^{(ij)}_{2\leftarrow1}$. Since the MST requires an undirected cost matrix, we combine the two directions using
\begin{equation}
d_{ij,t}
=
\min\{
1-\omega_{j\to i,t},
1-\omega_{i\to j,t}
\}
=
1-\max\{
\omega_{j\to i,t},
\omega_{i\to j,t}
\}.
\label{eq:symmetric_network_cost}
\end{equation}
Under this definition, a smaller cost indicates stronger forecast-error dependence in at least one direction. The minimum-cost rule emphasizes the stronger directional relationship rather than requiring the dependence to be equally strong in both directions. This is a modeling choice that allows us to construct an undirected network from directional GFEVD shares. The resulting cost matrix defines the dependency network $\mathcal G_t(W,H)=(\mathcal U,\mathcal E,\mathbf D_t)$, with $d_{ii,t}=0$. Since each pair is estimated separately, the assembled matrix is not the variance decomposition of a single $N$-dimensional VAR, and its entries should not be interpreted as system-wide variance shares. Moreover, the estimated relationships describe statistical forecast-error dependence rather than identified causal effects.

\subsection{GFEVD--Correlation Equivalence}
\label{subsec:gfevd_correlation}

Although GFEVD incorporates lagged relationships, it is useful to determine when its network contains information beyond what can already be obtained from correlation. We begin with a setting in which the two stocks do not have cross-lag relationships. In this case, the VAR coefficient matrix is diagonal, so each stock depends on its own previous return but not directly on the previous return of the other stock. This provides a benchmark for understanding which part of the GFEVD network comes from contemporaneous innovation dependence and which part can arise from dynamic interactions.

\noindent\textbf{Proposition 1 (Correlation equivalence).}
Consider a stable bivariate VAR(1) with a diagonal coefficient matrix $\mathbf A=\operatorname{diag}(a_i,a_j)$ and a positive-definite innovation covariance matrix $\boldsymbol\Sigma$. Let
\begin{equation}
\rho_{\varepsilon,ij}
=
\frac{\sigma_{12}}
{\sqrt{\sigma_{11}\sigma_{22}}}
\label{eq:innovation_correlation}
\end{equation}
denote the contemporaneous correlation between the two innovations. Then, for every forecast horizon $H\geq1$,
\begin{equation}
\widetilde\theta_{1\leftarrow2}(H)
=
\widetilde\theta_{2\leftarrow1}(H)
=
\frac{\rho_{\varepsilon,ij}^{2}}
{1+\rho_{\varepsilon,ij}^{2}}.
\label{eq:gfevd_correlation_equivalence}
\end{equation}

To understand this result, note that a diagonal VAR produces diagonal impulse responses, so $\boldsymbol\Phi(h)=\operatorname{diag}(a_i^h,a_j^h)$. When these matrices are substituted into Equation~\eqref{eq:generalized_fevd}, the terms involving the own autoregressive coefficients cancel. The unnormalized contributions therefore become
\begin{equation}
\theta_{1\leftarrow2}(H)
=
\frac{\sigma_{12}^{2}}
{\sigma_{11}\sigma_{22}}
=
\rho_{\varepsilon,ij}^{2},
\qquad
\theta_{1\leftarrow1}(H)=1.
\label{eq:unnormalized_equivalence}
\end{equation}
Applying row normalization gives the expression in Proposition 1, and the reverse direction follows in the same way. A complete proof is provided in Appendix~\ref{app:theory_proofs}. The important implication is that, without cross-lag relationships, the GFEVD share depends only on the correlation between innovations. It does not depend on the forecast horizon or the individual own-lag coefficients. Using the symmetrization rule from Section~\ref{subsec:pairwise_var_gfevd}, the corresponding undirected edge cost becomes
\begin{equation}
d_{ij}
=
\frac{1}{1+\rho_{\varepsilon,ij}^{2}}.
\label{eq:correlation_edge_cost}
\end{equation}
This expression establishes a direct relationship between GFEVD and the ranking of edges used to construct the minimum spanning tree.

\noindent\textbf{Corollary 1 (Spanning-tree equivalence).}
Suppose that the conditions of Proposition 1 hold for every stock pair in the network. Then, a minimum spanning tree constructed from the GFEVD costs in Equation~\eqref{eq:correlation_edge_cost} is equivalent to a maximum spanning tree constructed from absolute pairwise innovation correlations.

The corollary follows because $d_{ij}$ is a strictly decreasing function of $|\rho_{\varepsilon,ij}|$. Ranking stock pairs from the smallest GFEVD cost to the largest is therefore equivalent to ranking them from the largest absolute innovation correlation to the smallest. Both methods produce the same set of optimal spanning trees. When the optimal tree is unique, the resulting MST is identical under the two approaches. If multiple edges have equal weights, the particular tree returned by an algorithm may depend on its tie-breaking rule. This result shows that GFEVD does not necessarily introduce a different network structure simply because it is calculated from a VAR model.

However, the equivalence result concerns innovation correlations rather than the Pearson correlations of observed stock returns. For a diagonal VAR with own-lag coefficients $a_i$ and $a_j$, the population correlation of observed returns is
\begin{equation}
\rho_{r,ij}
=
\rho_{\varepsilon,ij}
\frac{
\sqrt{(1-a_i^2)(1-a_j^2)}
}{
1-a_i a_j
}.
\label{eq:return_innovation_correlation}
\end{equation}
The two correlations coincide when $a_i=a_j$, but they need not be equal otherwise. When the own-lag coefficients differ across stock pairs, the adjustment factor may also change the ordering of their correlations. Therefore, the absence of cross-lag relationships alone is not sufficient to establish exact equivalence between GFEVD and a Pearson network constructed from observed returns. A related result holds for the one-step forecast horizon. Since $\boldsymbol\Phi(0)=\mathbf I_2$, the normalized one-step GFEVD depends only on the innovation covariance matrix even when the VAR coefficient matrix is not diagonal. Lagged propagation can enter the decomposition only when the forecast horizon extends beyond one period. This does not imply that estimated one-step GFEVD is independent of the fitted VAR, since the innovation covariance is estimated from that model. Together, these results identify the circumstances in which GFEVD and correlation provide equivalent edge rankings and clarify why their networks may differ when lagged relationships are present.

\subsection{Dynamic Spillovers and Network Differences}
\label{subsec:dynamic_spillovers}

We next consider how lagged relationships affect the GFEVD network. Even when the innovations of two stocks are uncorrelated at the same time, the previous return of one stock may contain information about the future return of another. A contemporaneous correlation does not directly measure this type of relationship, while a multi-step forecast-error decomposition can incorporate its effect. The forecast horizon is therefore important because it determines how many periods of dynamic propagation enter the dependence measure.

\noindent\textbf{Proposition 2 (Dependence with uncorrelated innovations).}
Consider a stable bivariate VAR(1) with $\boldsymbol\Sigma=\operatorname{diag}(\sigma_{11},\sigma_{22})$, where both variances are positive. Define $\phi_{ab}(h)=\mathbf e_a^\top\mathbf A^h\mathbf e_b$. Then, for any forecast horizon $H\geq1$,
\begin{equation}
\widetilde\theta_{1\leftarrow2}(H)
=
\frac{
\sigma_{22}
\displaystyle\sum_{h=0}^{H-1}
\phi_{12}(h)^2
}{
\sigma_{11}
\displaystyle\sum_{h=0}^{H-1}
\phi_{11}(h)^2
+
\sigma_{22}
\displaystyle\sum_{h=0}^{H-1}
\phi_{12}(h)^2
}.
\label{eq:dynamic_gfevd}
\end{equation}
It follows that $\widetilde\theta_{1\leftarrow2}(H)>0$ if and only if $\phi_{12}(h)\neq0$ for at least one $h\in\{1,\ldots,H-1\}$. Since the innovation covariance matrix is diagonal, the raw generalized contributions already sum to one, and row normalization does not change them. The result follows by substituting the diagonal covariance matrix into Equation~\eqref{eq:generalized_fevd}, with the complete derivation provided in Appendix~\ref{app:theory_proofs}.

To illustrate the result, consider a simple system in which the previous return of stock 2 affects stock 1, but not the reverse:
\begin{equation}
\mathbf A
=
\begin{pmatrix}
0&\beta\\
0&0
\end{pmatrix},
\qquad
\boldsymbol\Sigma=\mathbf I_2,
\qquad
\beta\neq0.
\label{eq:one_way_dynamic_example}
\end{equation}
The corresponding return equations are $r_{1,t}=\beta r_{2,t-1}+\varepsilon_{1,t}$ and $r_{2,t}=\varepsilon_{2,t}$. Although the innovations are uncorrelated and the contemporaneous correlation between the observed returns is zero, stock 2 still has a lagged predictive relationship with stock 1. Specifically,
\begin{equation}
\operatorname{Cov}(r_{1,t},r_{2,t})=0,
\qquad
\operatorname{Cov}(r_{1,t},r_{2,t-1})=\beta.
\label{eq:example_lagged_covariance}
\end{equation}
Since $\mathbf A^2=\mathbf0$, the directional GFEVD shares are
\begin{equation}
\widetilde\theta_{1\leftarrow2}(1)=0,
\qquad
\widetilde\theta_{1\leftarrow2}(H)
=
\frac{\beta^2}{1+\beta^2}>0
\quad(H\geq2),
\qquad
\widetilde\theta_{2\leftarrow1}(H)=0.
\label{eq:example_dynamic_gfevd}
\end{equation}
The corresponding undirected cost is therefore
\begin{equation}
d_{12}(H)
=
\begin{cases}
1, & H=1,\\[4pt]
\dfrac{1}{1+\beta^2}<1, & H\geq2.
\end{cases}
\label{eq:example_dynamic_edge_cost}
\end{equation}
A network constructed from contemporaneous absolute Pearson correlations would assign this stock pair a distance of $\sqrt{2}$, since their contemporaneous correlation is zero. This distance remains unchanged regardless of the forecast horizon. In contrast, the GFEVD cost decreases from one at the one-step horizon to $1/(1+\beta^2)$ when the horizon includes the lagged response. Since the two methods use different cost transformations, their numerical values are not directly comparable. Nevertheless, the example demonstrates that GFEVD can reflect a lagged predictive relationship that is not captured by contemporaneous Pearson correlation.

When innovations are correlated, contemporaneous covariance and lagged propagation can interact. To examine this relationship, consider the numerator of the generalized contribution from variable 2 to variable 1 in Equation~\eqref{eq:generalized_fevd}. Expanding the squared term gives
\begin{equation}
\begin{aligned}
N_{1\leftarrow2}(H)
&=
\sigma_{22}^{-1}
\sum_{h=0}^{H-1}
\bigl(
\phi_{11}(h)\sigma_{12}
+
\phi_{12}(h)\sigma_{22}
\bigr)^2\\
&=
\frac{\sigma_{12}^2}{\sigma_{22}}
\sum_{h=0}^{H-1}
\phi_{11}(h)^2
+
2\sigma_{12}
\sum_{h=0}^{H-1}
\phi_{11}(h)\phi_{12}(h)
+
\sigma_{22}
\sum_{h=0}^{H-1}
\phi_{12}(h)^2.
\end{aligned}
\label{eq:dependence_expansion}
\end{equation}
The first term captures contemporaneous innovation covariance,
the last term represents dynamic cross-variable propagation,
and the middle term captures their interaction. Both the
interaction term and the forecast-error variance depend on
the VAR coefficients. Therefore, the normalized GFEVD share
is not determined by the magnitude of a cross-lag coefficient
alone. These results explain why GFEVD may produce different
edge costs from contemporaneous correlation. However, a
difference in edge costs does not necessarily imply a
different MST or selected portfolio, since the MST depends
on the ordering of edges and the portfolio depends on the
degree ranking of the resulting tree.

\subsection{MST Stability and Centrality Ranking}
\label{subsec:mst_stability}

After obtaining the symmetric dependency costs, we construct a Minimum Spanning Tree to represent the network using a smaller number of edges. Let $\mathfrak T(\mathcal U)$ denote the set of all spanning trees on the fixed stock universe. The MST is defined by
\begin{equation}
T_t
\in
\arg\min_{T\in\mathfrak T(\mathcal U)}
\sum_{(i,j)\in E(T)}d_{ij,t}.
\label{eq:mst_optimization}
\end{equation}
The resulting tree contains exactly $N-1$ edges and connects every stock without forming cycles. Because the MST is determined by the relative ordering of edge costs, two different cost matrices may still produce the same network structure. This distinction is important when comparing GFEVD and correlation-based networks, since differences in dependence measures do not necessarily lead to differences in the selected tree. Following the classical sensitivity analysis of spanning trees \citep{Tarjan1982}, we examine how large a change in edge costs must be before the optimal tree can change.

Suppose that the cost matrix $\mathbf D$ has a unique MST $T$. For a non-tree edge $e=(i,j)$, let $P_T(e)$ denote the unique path connecting its endpoints within the tree. We define its replacement gap and the minimum replacement margin as
\begin{equation}
\gamma_e(T,\mathbf D)
=
d_e-\max_{f\in P_T(e)}d_f,
\qquad
\gamma(T,\mathbf D)
=
\min_{e\notin E(T)}
\gamma_e(T,\mathbf D).
\label{eq:global_mst_margin}
\end{equation}
For a unique MST with at least three vertices, the replacement gaps are strictly positive. A smaller margin indicates that at least one edge outside the tree has a cost close to that of an edge it could replace. Therefore, the margin provides a way to describe the sensitivity of an MST to estimation errors in its edge costs.

\noindent\textbf{Proposition 3 (MST stability).}
Let $T$ be the unique MST of $\mathbf D$ on a network with at least three vertices, and let the perturbed cost matrix satisfy $\widehat{\mathbf D}=\mathbf D+\mathbf E$, where $\mathbf E$ is symmetric and $\|\mathbf E\|_{\max}=\max_{i<j}|E_{ij}|\leq\varepsilon$. If
\begin{equation}
2\varepsilon<\gamma(T,\mathbf D),
\label{eq:mst_stability_condition}
\end{equation}
then $T$ remains the unique MST under $\widehat{\mathbf D}$.

The intuition is that a non-tree edge cannot enter the MST unless it becomes sufficiently inexpensive relative to an edge on the path connecting its endpoints. Under the stated error bound, for every $e\notin E(T)$ and $f\in P_T(e)$,
\begin{equation}
\widehat d_e-\widehat d_f
\geq
d_e-d_f-2\varepsilon
\geq
\gamma(T,\mathbf D)-2\varepsilon
>0.
\label{eq:mst_stability_proof}
\end{equation}
Thus, the ordering required by the cycle property remains unchanged, and the original tree continues to be the unique MST. A complete proof is provided in Appendix~\ref{app:theory_proofs}. This result can also be applied to GFEVD estimation because the symmetric cost is obtained from the maximum of two directional dependence shares. Let $\widehat\omega$ and $\omega$ denote estimated and reference GFEVD weights. Since the maximum operator is Lipschitz continuous,
\begin{equation}
\begin{aligned}
|\widehat d_{ij}-d_{ij}|
\leq
\max\bigl\{
&|\widehat\omega_{j\to i}
-\omega_{j\to i}|,\\
&|\widehat\omega_{i\to j}
-\omega_{i\to j}|
\bigr\}.
\end{aligned}
\label{eq:gfevd_cost_error_bound}
\end{equation}
A uniform bound on the errors in directional GFEVD shares therefore provides a bound on the resulting edge costs. When this bound is sufficiently small relative to the replacement margin, the MST remains unchanged. However, the condition is only a sufficient guarantee and does not establish that the estimation errors are small in any particular empirical window.

We next examine how changes in the MST affect stock selection. After constructing the tree, we rank stocks according to degree centrality, which counts the number of edges directly connected to each stock:
\begin{equation}
g_i(T)
=
\sum_{j\neq i}
\mathbf1\{(i,j)\in E(T)\}.
\label{eq:mst_degree}
\end{equation}
The portfolio contains the $k$ stocks with the highest degree values, with a predetermined rule used to resolve ties. Let $g_{(1)}(T)\geq\cdots\geq g_{(N)}(T)$ denote the ordered degrees. We define the degree-selection gap at the portfolio cutoff as
\begin{equation}
\Delta_k(T)
=
g_{(k)}(T)-g_{(k+1)}(T),
\qquad 1\leq k<N.
\label{eq:degree_selection_gap}
\end{equation}
This gap measures the separation between the final selected stock and the first excluded stock. When $\Delta_k(T)=0$, the two stocks have the same degree and the final selection depends on the tie-breaking rule. A positive gap indicates clearer separation, although the stability of the selected set also depends on how much the MST changes.

\noindent\textbf{Proposition 4 (Top-$k$ selection stability).}
Let $T$ and $T'$ be spanning trees on the same stock universe, and write
\begin{equation}
s(T,T')
=
|E(T)\setminus E(T')|
=
|E(T')\setminus E(T)|.
\label{eq:tree_edge_replacements}
\end{equation}
For every stock $i$, $|g_i(T')-g_i(T)|\leq s(T,T')$. Consequently, if
\begin{equation}
\Delta_k(T)>2s(T,T'),
\label{eq:topk_stability_condition}
\end{equation}
then the two trees produce the same top-$k$ selected stock set:
\begin{equation}
S_k(T')=S_k(T).
\label{eq:topk_set_invariance}
\end{equation}

To understand this condition, note that removing an edge can reduce the degree of a connected stock by at most one, while adding an edge can increase its degree by at most one. When $s$ edges are replaced, the degree of any stock can change by at most $s$. Therefore, the degree difference between an initially selected stock and an unselected stock can decrease by at most $2s$. If the original selection gap is larger than this value, the ranking across the cutoff cannot be reversed. The formal proof is given in Appendix~\ref{app:theory_proofs}. Together, Propositions 3 and 4 distinguish the stability of the network from the stability of the selected portfolio. An MST may remain unchanged despite differences in estimated edge costs, or its edges may change without affecting the top-$k$ stock set. These conditions are sufficient rather than necessary, and their failure does not imply that the network or the portfolio must change. They also do not establish that a more stable network or a larger degree-selection gap necessarily produces higher investment returns.

We can also relate the magnitude of a cost perturbation directly to
the number of edge replacements. For an MST $T$ of $\mathbf D$ and
a perturbed cost matrix $\mathbf D'$ satisfying
$\|\mathbf D'-\mathbf D\|_{\max}\leq\delta$, define
\[
q(T,\mathbf D;\delta)
=
\#\{e\notin E(T):\gamma_e(T,\mathbf D)\leq2\delta\}.
\]
If $\gamma_e(T,\mathbf D)>2\delta$, then under $\mathbf D'$
the non-tree edge $e$ remains strictly more expensive than every
edge on $P_T(e)$. By the cycle property, it cannot belong to any
MST of $\mathbf D'$. Consequently, for any MST $T'$ of
$\mathbf D'$,
\[
s(T,T')\leq q(T,\mathbf D;\delta).
\]
Combining this bound with Proposition 4 gives the sufficient
condition $\Delta_k(T)>2q(T,\mathbf D;\delta)$ for the selected
stock set to remain unchanged. If $q(T,\mathbf D;\delta)=0$,
the tree itself remains unchanged, and so does the selected stock
set under the same tie-breaking rule, even when $\Delta_k(T)=0$.
The bound may be conservative, since not every non-tree edge
counted by $q(T,\mathbf D;\delta)$ will enter the perturbed tree.

These conditions can also be applied to two estimated cost matrices
from consecutive formation dates. For example, if $\widehat{\mathbf
D}_t$ and $\widehat{\mathbf D}_{t+1}$ are the corresponding sample
cost matrices, we can use their observed difference
$\|\widehat{\mathbf D}_{t+1}-\widehat{\mathbf D}_t\|_{\max}$ as
the perturbation magnitude. The resulting replacement margins
and stability conditions describe changes in the estimated
networks, rather than estimation errors relative to an unknown
population cost matrix. We make use of this distinction in the
empirical analysis.

\subsection{Estimation Window Length and Portfolio Selection}
\label{subsec:time_varying_recovery}
\label{subsec:parameter_foundations}

The estimation window length $W$ and the number of selected stocks $k$ are two important parameters in our portfolio construction procedure. The estimation window determines how much historical information is used to measure stock dependence, while $k$ determines how many stocks are included in the portfolio after ranking their degree centrality. These parameters influence different stages of the methodology and therefore affect the portfolio in different ways. A longer estimation window generally provides more observations for fitting the pairwise VAR models, which may reduce sampling uncertainty. However, financial relationships can change over time, and older observations may no longer represent the current dependency structure. This creates a trade-off between estimation uncertainty and parameter variation, which is closely related to the analysis of locally stationary time series and estimation under structural changes \citep{DahlhausGiraitis1998,PesaranTimmermann2007}.

To examine this trade-off, we extend the pairwise VAR model by allowing its parameters to vary over time. For each stock pair $p=(i,j)$, consider
\begin{equation}
\mathbf x_s^{(p)}
=
\mathbf B_{p,s}
\mathbf z_{s-1}^{(p)}
+
\mathbf u_s^{(p)},
\label{eq:tv_pairwise_var}
\end{equation}
where
\begin{equation}
\mathbf z_{s-1}^{(p)}
=
\begin{pmatrix}
1\\
r_{i,s-1}\\
r_{j,s-1}
\end{pmatrix},
\qquad
\mathbf B_{p,s}
=
\begin{pmatrix}
\boldsymbol\mu_{p,s}&\mathbf A_{p,s}
\end{pmatrix}.
\label{eq:tv_pairwise_parameters}
\end{equation}
The matrix $\mathbf B_{p,s}$ contains both the intercepts and the autoregressive coefficients, while the innovation covariance matrix $\boldsymbol\Sigma_{p,s}$ is also allowed to change over time. At a network formation date $t$, we define the population GFEVD cost matrix $\mathbf D_t$ using the current parameters $(\mathbf A_{p,t},\boldsymbol\Sigma_{p,t})$ for each stock pair. These parameters are treated as locally fixed when calculating the forecast-error decomposition over horizon $H$. The resulting cost matrix serves as the population benchmark against which the rolling-window estimate is compared.

We assume that the population parameters
$(\mathbf A_{p,s},\boldsymbol\Sigma_{p,s})$
belong to a common compact subset $\mathcal K$
of the stable and positive-definite parameter region,
uniformly across stock pairs and dates.
The parameter bounds are independent of $n$ and $N$.
Innovations are bounded martingale differences with
conditional covariance $\boldsymbol\Sigma_{p,s}$,
and regressors are uniformly bounded.
For all stock pairs, the parameter changes satisfy
\begin{equation}
\begin{aligned}
\|\mathbf B_{p,s}-\mathbf B_{p,t}\|_2
&\leq L_B|t-s|,\\
\|\boldsymbol\Sigma_{p,s}
-\boldsymbol\Sigma_{p,t}\|_2
&\leq L_\Sigma|t-s|.
\end{aligned}
\label{eq:tv_parameter_drift}
\end{equation}
Here, $L_B$ and $L_\Sigma$ measure the rate at which the VAR coefficients and innovation covariances change over time. We further assume that there exists a constant
$g>0$, independent of $n$ and $N$, such that
for every admissible window length $n$,
\begin{equation}
\mathbb P\left(
\min_p\lambda_{\min}(\mathbf G_{p,t}(n))
\geq g
\right)
\geq1-\eta.
\end{equation}
Let $\widehat{\mathbf B}_{p,t}(n)$ denote the rolling OLS estimator based on $n$ effective regression observations ending at date $t$. Since one lag is used in the VAR, $n$ counts the usable response--lag pairs rather than the total number of return observations $W$.

The difference between the rolling estimate and the current population coefficients can be decomposed as
\begin{equation}
\begin{aligned}
\widehat{\mathbf B}_{p,t}(n)-\mathbf B_{p,t}
={}&
\left[
\frac{1}{n}
\sum_{s\in\mathcal I_{t,n}}
\mathbf u_s^{(p)}
\mathbf z_{s-1}^{(p)\top}
\right]
\mathbf G_{p,t}^{-1}\\
&+
\left[
\frac{1}{n}
\sum_{s\in\mathcal I_{t,n}}
(\mathbf B_{p,s}-\mathbf B_{p,t})
\mathbf z_{s-1}^{(p)}
\mathbf z_{s-1}^{(p)\top}
\right]
\mathbf G_{p,t}^{-1},
\end{aligned}
\label{eq:tv_ols_decomposition}
\end{equation}
where
\begin{equation}
\mathbf G_{p,t}
=
\frac{1}{n}
\sum_{s\in\mathcal I_{t,n}}
\mathbf z_{s-1}^{(p)}
\mathbf z_{s-1}^{(p)\top}.
\label{eq:tv_regressor_matrix}
\end{equation}
The first component represents sampling uncertainty, while the second reflects the difference between historical coefficients and their current values. Increasing the estimation window provides more observations and reduces the sampling component, but it may also increase the effect of parameter variation when market relationships change over time. A similar decomposition applies to the residual covariance estimator. Since GFEVD depends on both the VAR coefficients and the innovation covariance matrix, these two sources of estimation error also affect the estimated network.

\noindent\textbf{Proposition 5 (Rolling-window GFEVD estimation error).}
For a fixed horizon $H$, let $\varepsilon_0>0$ be
independent of $n$ and $N$, and sufficiently small
that the closed $\varepsilon_0$-neighborhood of
$\mathcal K$ remains in the stable and positive-definite
parameter region. We consider window lengths satisfying
\[
r_B(n,\alpha)+r_\Sigma(n,\alpha)\leq\varepsilon_0,
\]
with the parameter-error bounds defined in
Appendix~\ref{app:tv_gfevd_proof}.
Let $M=N(N-1)/2$ denote the number of distinct stock pairs.
There exist finite positive constants $C_1$, $C_2$, and $C_3$,
independent of $n$ and $N$, such that, with probability
at least $1-\alpha-\eta$,
\begin{equation}
\|\widehat{\mathbf D}_t(n)-\mathbf D_t\|_{\max}
\leq\delta_n,
\label{eq:tv_gfevd_error_bound}
\end{equation}
where
\begin{equation}
\begin{aligned}
\delta_n
={}&
C_1
\sqrt{
\frac{\log(24M/\alpha)}{n}
}\\
&+
C_2(L_B+L_\Sigma)(n-1)\\
&+
C_3
\left(
\sqrt{
\frac{\log(24M/\alpha)}{n}
}
+
L_B(n-1)
\right)^2.
\end{aligned}
\label{eq:tv_network_error_components}
\end{equation}

The result is obtained by controlling the sampling and parameter-drift components of the rolling VAR estimates, extending these bounds to the residual covariance estimates, and applying the smoothness of the finite-horizon GFEVD mapping. A detailed proof is provided in Appendix~\ref{app:theory_proofs}. Proposition 5 shows that the accuracy of the estimated network depends on both the amount of historical information and the speed at which the underlying relationships change. The first component of the bound decreases with the estimation window, while the second increases with parameter variation. The final component captures higher-order estimation effects. Therefore, a longer window does not necessarily provide a more accurate estimate of the current dependency network.

We can relate this estimation error to the MST stability conditions developed in Section~\ref{subsec:mst_stability}. Let $T_t^*$ denote the unique MST of the population cost matrix $\mathbf D_t$, and let $\gamma_t=\gamma(T_t^*,\mathbf D_t)$ be its minimum replacement margin. Combining Proposition 5 with Proposition 3 gives
\begin{equation}
2\delta_n<\gamma_t
\quad\Longrightarrow\quad
\widehat T_t(n)=T_t^*.
\label{eq:tv_mst_recovery_condition}
\end{equation}
For every window length satisfying
$r_B(n,\alpha)+r_\Sigma(n,\alpha)\leq\varepsilon_0$,
this condition recovers the population MST with
probability at least $1-\alpha-\eta$.
The degree ranking and the selected top-$k$ stock
set are then unchanged under the same tie-breaking rule. However, exact recovery of the entire tree is not always necessary for the selected portfolio to remain stable. To examine this possibility, define
\begin{equation}
q_t(\delta)
=
\#\left\{
e\notin E(T_t^*):
\gamma_e(T_t^*,\mathbf D_t)
\leq2\delta
\right\}.
\label{eq:tv_ambiguous_edges}
\end{equation}
This quantity is the population counterpart of
$q(T,\mathbf D;\delta)$ defined in Section~\ref{subsec:mst_stability}.
On the event that the maximum cost error is bounded by $\delta$,
the number of edge replacements relative to $T_t^*$ is at most
$q_t(\delta)$. Taking $\delta=\delta_n$ and applying Proposition 4 gives
\begin{equation}
\Delta_k(T_t^*)
>
2q_t(\delta_n)
\quad\Longrightarrow\quad
\widehat S_{k,t}(n)=S_{k,t}^*.
\label{eq:tv_topk_recovery_condition}
\end{equation}
Under the window-length restriction in Proposition 5, if the
stated degree-gap inequality holds, the selected stock set is
recovered with probability at least $1-\alpha-\eta$. If the
population MST is recovered, the selected stock set is also
recovered under the same tie-breaking rule, regardless of
whether the degree-gap inequality holds.
Thus, the estimated MST may differ from the population tree while still producing exactly the same selected stock set. This distinction is relevant for our portfolio construction because the final portfolio depends on the stocks selected from the degree ranking rather than on every edge in the network. Both recovery conditions are sufficient rather than necessary, so their failure does not imply that the network or portfolio must change.

The estimation-error bound also provides guidance for the choice of rolling-window length. To illustrate the trade-off, we consider the leading terms in Proposition 5 and temporarily ignore the higher-order component:
\begin{equation}
\delta_n
\approx
C_1
\sqrt{
\frac{\log(24M/\alpha)}{n}
}
+
C_2(L_B+L_\Sigma)n.
\label{eq:tv_leading_window_error}
\end{equation}
When $L_B+L_\Sigma>0$, minimizing this leading expression over a continuous window length gives
\begin{equation}
n_{\mathrm{bound}}^*
=
\left(
\frac{
C_1^2\log(24M/\alpha)
}{
4C_2^2(L_B+L_\Sigma)^2
}
\right)^{1/3}.
\label{eq:tv_optimal_window_benchmark}
\end{equation}
This expression illustrates how the preferred estimation window depends on the number of stock pairs and the rate of parameter variation. A larger number of stock pairs increases the information required for uniform estimation, while faster changes in the underlying parameters favor shorter windows. The expression is a benchmark for the leading error bound rather than an exact solution for the full estimation problem or the portfolio's expected return.

Suppose the VAR parameters evolve smoothly over
the rescaled time interval $t/T_{\mathrm{obs}}$,
with $L_B+L_\Sigma\asymp T_{\mathrm{obs}}^{-1}$.
Under $\log M=o(n)$, $n=o(T_{\mathrm{obs}})$,
and $\eta\to0$, the parameter-error bounds
converge to zero, so the compact-neighborhood
condition of Proposition 5 holds for sufficiently
large $T_{\mathrm{obs}}$.
The resulting estimation-error rate is
\begin{equation}
\delta_{n,T_{\mathrm{obs}}}
=
O_p\left(
\sqrt{\frac{\log M}{n}}
+
\frac{n}{T_{\mathrm{obs}}}
\right).
\label{eq:tv_window_rate}
\end{equation}
Under $\log M=o(T_{\mathrm{obs}})$,
balancing the two components gives the
window-length rate
\begin{equation}
n_{\mathrm{bound}}^*
\asymp
(T_{\mathrm{obs}}^2\log M)^{1/3}.
\label{eq:tv_optimal_window_rate}
\end{equation}
Although this result provides theoretical guidance for the estimation window, it does not identify a numerical optimal value for our empirical sample. The parameter-drift rates and constants in the bound are unknown, and the objective of the theoretical calculation is network estimation accuracy rather than portfolio performance. Therefore, the expression does not imply that the shortest or longest available window must produce the best investment results.

The number of selected stocks $k$ introduces a different consideration. While the estimation window affects the accuracy of the network, $k$ determines how many stocks are retained after the degree ranking. Section~\ref{subsec:mst_stability} shows that the stability of a given Top-$k$ selection depends partly on the degree gap at the cutoff. At the same time, increasing the number of holdings may provide diversification benefits, although the magnitude of these benefits depends on the covariance structure of the selected stocks and the portfolio weights. For a long-only portfolio with $w_i\geq0$ and $\sum_{i\in S_k(T)}w_i=1$, the portfolio variance is
\begin{equation}
\operatorname{Var}(R_p)
=
\sum_iw_i^2\operatorname{Var}(R_i)
+
\sum_{i\neq j}
w_iw_j\operatorname{Cov}(R_i,R_j).
\label{eq:portfolio_variance_theory}
\end{equation}
Following the diversification principle introduced by \citet{Markowitz1952}, we establish a simple upper bound on this variance.

\noindent\textbf{Proposition 6 (Diversification upper bound).}
Suppose that the variance of every selected stock is at most $\bar v$ and the covariance between distinct selected stocks is at most $\bar c$, where $0\leq\bar c\leq\bar v$. Then any long-only portfolio with weights summing to one satisfies
\begin{equation}
\operatorname{Var}(R_p)
\leq
\bar c
+
(\bar v-\bar c)
\sum_iw_i^2.
\label{eq:weighted_diversification_bound}
\end{equation}
For equal weights, this reduces to
\begin{equation}
\operatorname{Var}(R_p)
\leq
\bar c
+
\frac{\bar v-\bar c}{k}.
\label{eq:equal_weight_diversification_bound}
\end{equation}

For common bounds $\bar v$ and $\bar c$ across the
portfolios being compared, the equal-weight upper bound does
not increase with $k$, and it strictly decreases when
$\bar c<\bar v$.
However, the actual variance of a selected
portfolio need not decrease when another
stock is added. The result describes an upper bound under the stated restrictions on individual variances and covariances. Moreover, for unequal portfolio weights, the nominal number of holdings may not accurately represent the degree of diversification. We therefore define the effective number of holdings as
\begin{equation}
k_{\mathrm{eff}}
=
\frac{1}{
\displaystyle\sum_{i\in S_k(T)}w_i^2
},
\qquad
1\leq k_{\mathrm{eff}}\leq k.
\label{eq:effective_cardinality}
\end{equation}
Under equal weighting, $k_{\mathrm{eff}}=k$, while more concentrated allocations reduce the effective number of holdings. This distinction is relevant for our empirical analysis because the same selected stock set can produce different portfolio allocations under equal weighting and inverse historical VaR weighting.

The choice of $k$ therefore involves both selection stability and diversification. A larger value of $k$ may improve the diversification bound, but it can also move the selection cutoff into a group of stocks with similar degrees, making the final stock set more sensitive to changes in the MST. Neither consideration establishes that a particular portfolio size maximizes expected returns. For this reason, we evaluate alternative values of $W$ and $k$ as sensitivity comparisons rather than treating the historically best-performing combination as theoretically optimal. This distinction is particularly important when multiple parameter configurations are evaluated using the same historical sample \citep{White2000}. Overall, the theoretical framework explains when GFEVD may capture information beyond contemporaneous correlation and how estimation uncertainty can affect both network construction and stock selection. It does not imply that stronger measured dependence, greater network stability, or a particular parameter combination must generate higher investment returns.

%% file: sections/04_data.tex
\section{Data}
\label{sec:data}

In this study, we use daily adjusted closing prices for U.S. equities obtained
from Yahoo Finance. The price records cover the period from January 2020 to
October 28, 2026. Adjusted closing prices account for the vendor's historical
adjustments for stock splits and dividend distributions
\citep{yahoo_adjusted_close}. We use these prices because they provide a
consistent series for studying stock-return relationships over time. At the
same time, the adjusted price is a historical data series rather than a
verified transaction price, a distinction that is important when interpreting
the portfolio simulations later in the paper.

The initial list contains 486 equity tickers. After applying the data
reconciliation and eligibility checks used in the study, 451 stocks remain.
These checks use information available over the full historical sample.
The same 451 stocks, in the same predetermined order, are used in every
network and portfolio comparison. In particular, we do not remove a stock
because one network method produces an unfavorable result or a pairwise VAR
fit fails. This gives the three network methods a common set of potential
investments. The ticker-level reconciliation is retained as part of the
reproducibility records.

For each stock $i$, we calculate the daily simple return from its adjusted
closing price $P_{i,t}^{\mathrm{adj}}$ as

\begin{equation}
r_{i,t}
=\frac{P_{i,t}^{\mathrm{adj}}-P_{i,t-1}^{\mathrm{adj}}}
{P_{i,t-1}^{\mathrm{adj}}}.
\label{eq:data_daily_return}
\end{equation}

Here, $t-1$ denotes the previous trading session on the common market
calendar, not necessarily the previous calendar day. A return is calculated
only when the required prices are available. We retain missing observations
as missing and do not replace them with zero returns or carry the previous
price forward. The resulting return panel is the common input for the
pairwise VAR--GFEVD network and the two Pearson-correlation networks.

Although the price history begins in 2020, the portfolio evaluation does not
begin immediately. The primary estimation window contains 250 trading-day
returns, and March 17, 2021 is the first date on which all 451 stocks have
the required finite history. We construct trading signals at the end of a
completed trading week and execute the corresponding portfolios at the next
available session close. The first eligible completed-week signal is dated
March 19, 2021, and its execution date is March 22, 2021. The last completed
week in the available data ends on October 23, 2026, with execution on
October 26, 2026. The final observations on October 27 and 28 remain in the
portfolio return series, but they do not create an additional weekly signal.
This produces 293 weekly signals and a common evaluation period of 1,409
trading sessions, from March 22, 2021 through October 28, 2026.

We divide the evaluation period into two parts. The earlier period, from
March 22, 2021 through December 29, 2023, contains 700 sessions and is used
for model development and diagnostic analysis. The later period, from
January 2, 2024 through October 28, 2026, contains 709 sessions and is the
reserved evaluation period for the original primary specification. We
retain this division in the revised analysis to make the results comparable
with the original evaluation. However, since the portfolio outcomes over
the reserved period have already been examined, the additional analyses
conducted during the revision are retrospective rather than independent
out-of-sample tests. Table \ref{tab:data_sample_calendar} summarizes the
sample and the principal dates. All network methods and portfolio
weightings in the primary specification are evaluated on the same dates.
The shorter 60- and 120-day windows used in the subsequent sensitivity
analysis are not allowed to start trading earlier than the primary
specification.

\begin{table}[htbp]
\centering
\caption{Equity Sample and Evaluation Calendar}
\label{tab:data_sample_calendar}
\begin{tabular}{ll}
\toprule
Item & Sample information \\
\midrule
Initial candidate stocks & 486 \\
Stocks retained after reconciliation & 451 \\
Price-data endpoint & October 28, 2026 \\
Completed-week trading signals & 293 \\
Common evaluation period & March 22, 2021--October 28, 2026 \\
Development and diagnostic sessions & 700 \\
Reserved evaluation sessions & 709 \\
\bottomrule
\end{tabular}
\end{table}

For the excess-return calculations, we also use the daily three-month U.S.
Treasury bill secondary-market rate from the Federal Reserve Economic Data
(FRED) database, series DTB3 \citep{fred_dtb3}. This series is quoted on a
discount basis rather than as a realized daily investment return. Its use in
the risk-adjusted performance measures, together with the trading-cost and
execution conventions, is described in Section~\ref{sec:methodology}.

For an additional comparison with broad market performance, we use the
vendor-adjusted closing price series of SPY to construct a buy-and-hold
benchmark. We also construct an equal-weight portfolio containing all
451 stocks in the common universe. The former provides a market-related
reference, while the latter allows us to compare the network-based stock
selections with a more broadly diversified portfolio drawn from the same
historical universe. The SPY series is obtained from vendor data rather
than an independently verified official total-return index.

One limitation of the dataset should be emphasized. The final 451-stock
universe is reconstructed after observing the available sample and is not a
point-in-time record of historical S\&P 500 membership. It therefore may
exclude companies that disappeared, were removed from an index, or otherwise
failed the data requirements. We also do not have complete delisting-return
records for the excluded stocks. This introduces survivorship and selection
biases, which can affect historical performance comparisons
\citep{brown1992survivorship}. Although using the same stock universe
provides a common basis for comparing the three network methods, it does
not eliminate these biases.

Moreover, the portfolio paths are calculated from vendor-adjusted closing
prices and synthetic next-close executions. The holdings are represented
in adjusted-price units, which remain fixed between rebalancing dates.
These units provide a way to account for changes in portfolio value and
weights, but they do not represent verified historical share quantities
or transaction prices. The simulations also do not account for all market
frictions. Consequently, our empirical results describe the relative
behavior of the specified network-selection rules on this fixed historical
sample, rather than the performance of a directly executable investment
strategy.

%% file: sections/05_methodology.tex
\section{Empirical Methodology}
\label{sec:methodology}

In this section, we apply the network framework developed in Section~\ref{sec:theory} to construct and evaluate equity portfolios. Our main objective is to examine whether the dependence information captured by pairwise GFEVD provides additional value compared with conventional correlation-based networks. Although GFEVD incorporates lagged relationships that are not directly represented by Pearson correlation, this does not necessarily mean that it will produce better investment performance. To make the comparison consistent, we apply the three network methods to the same stock universe and use identical portfolio selection, weighting, and trading rules. For a given weighting method and specification, the only difference in the construction of the target portfolios is the dependence measure used to construct the network. This allows us to examine whether differences in network construction affect stock selection and, ultimately, portfolio performance.

\subsection{Network Construction and Stock Selection}
\label{subsec:empirical_network_construction}

To construct the dependency networks, we use the daily stock returns described in Section~\ref{sec:data}. For each completed trading week, we take the most recent $W$ daily return observations and estimate the relationships between stocks using a rolling-window approach. This allows the network structure to change over time as new market information becomes available. For the GFEVD network, we estimate a bivariate VAR(1) model for each distinct pair of stocks. With 451 stocks in the sample, there are
\begin{equation}
\binom{451}{2}=101{,}475
\label{eq:empirical_number_of_pairs}
\end{equation}
stock pairs at each estimation date. We use a forecast horizon of ten trading days and calculate the generalized forecast error variance decomposition following \citet{PesaranShin1998}. The decomposition measures the contribution associated with innovations in one stock to the forecast error variance of another stock. Following the connectedness framework of \citet{DieboldYilmaz2012}, we normalize the GFEVD shares within each bivariate system. Because the relationship between two stocks may differ across directions, we transform the directional GFEVD values into a symmetric measure before constructing an undirected network. In this study, we use the minimum-cost symmetrization rule:
\begin{equation}
d_{ij,t}^{G}
=
1-\max\left\{
\omega_{j\rightarrow i,t},
\omega_{i\rightarrow j,t}
\right\},
\label{eq:empirical_gfevd_cost}
\end{equation}
where $\omega_{j\rightarrow i,t}$ and $\omega_{i\rightarrow j,t}$ are the normalized directional GFEVD shares between stocks $i$ and $j$. Under this definition, a smaller cost indicates stronger dependence in at least one direction. The rule emphasizes the stronger directional relationship, although it does not require both directions to have the same degree of dependence.

For comparison, we construct two additional networks using Pearson correlation coefficients calculated from the same rolling return windows. The first uses absolute correlation, while the second retains the sign of the correlation. Let $\rho_{ij,t}$ denote the Pearson correlation between stocks $i$ and $j$ at time $t$. The corresponding edge orderings are based on $|\rho_{ij,t}|$ and $\rho_{ij,t}$, respectively. These orderings are equivalent to using the conventional correlation distances
\begin{equation}
d_{ij,t}^{A}
=
\sqrt{2\left(1-\left|\rho_{ij,t}\right|\right)},
\qquad
d_{ij,t}^{S}
=
\sqrt{2\left(1-\rho_{ij,t}\right)}.
\label{eq:empirical_correlation_distances}
\end{equation}
In the absolute-correlation network, both strong positive and strong negative correlations represent strong relationships. In the signed-correlation network, more positively correlated stock pairs receive greater priority. These two approaches provide benchmarks for examining whether the additional lagged information captured by GFEVD affects the estimated network structure. It is important to note that the empirical correlation networks are constructed from observed stock returns, whereas the theoretical equivalence result in Section~\ref{subsec:gfevd_correlation} concerns the correlation between VAR innovations.

After obtaining the cost matrix for each method, we construct a Minimum Spanning Tree (MST) using Kruskal's algorithm. The MST connects all stocks while minimizing the total edge cost and avoiding cycles. With 451 stocks, each tree contains 450 edges. This approach follows the application of minimum spanning trees in financial market network analysis \citep{Mantegna1999}. Once the MST has been constructed, we use degree centrality to identify the stocks included in the portfolio. Degree centrality is one of the simplest measures of a node's importance within a network, as it counts the number of direct connections associated with each stock. For stock $i$, we define
\begin{equation}
g_{i,t}
=
\sum_{j\neq i}
\mathbf{1}
\left\{
(i,j)\in E(T_t)
\right\},
\label{eq:empirical_degree}
\end{equation}
where $T_t$ denotes the MST at time $t$ and $E(T_t)$ is its edge set. We rank stocks in descending order of degree and select the top $k$ stocks as portfolio candidates. A stock with a higher degree is connected to more stocks in the MST, suggesting a more central position within the estimated network. However, higher degree does not necessarily imply a higher expected return. In this study, degree centrality is used as a predetermined stock selection rule rather than as a measure of portfolio optimality. Since degree values are integers, multiple stocks may have the same degree. To ensure that the selection can be reproduced, we break these ties using the fixed ordering of stocks in the sample. The same ordering is also used to resolve exact cost ties when constructing the MST.

For the primary specification, we reconstruct the costs of all candidate edges at each estimation date using the original estimation procedure. We compare the resulting trees, degree rankings, selected stocks, and target weights with the saved checkpoints before using the reconstructed costs in the stability analysis. The original checkpoints retain the selected tree edges but not the complete set of rejected edge costs.

\subsection{Portfolio Weighting}
\label{subsec:empirical_weighting}

After identifying the top $k$ stocks from each network, we assign portfolio weights using two different approaches. The first is equal weighting, which distributes capital evenly among the selected stocks. The second is inverse historical Value at Risk (VaR) weighting, which adjusts the allocation according to the estimated downside risk of each stock. We include both approaches because the network determines which stocks enter the portfolio, while the weighting method determines how capital is distributed among them. Using the same weighting methods across the three networks allows us to examine whether differences in performance remain when the allocation rule is held constant. Under equal weighting, each selected stock receives the same proportion of total portfolio wealth:
\begin{equation}
w_{i,t}^{EW}
=
\frac{1}{k},
\qquad i\in S_{k,t},
\label{eq:empirical_equal_weight}
\end{equation}
where $S_{k,t}$ represents the set of $k$ stocks selected at time $t$. Equal weighting provides a simple benchmark because it does not require an additional estimate of individual stock risk. Therefore, differences between equally weighted portfolios constructed from different networks arise from their selected stocks and the transactions required to maintain those portfolios.

For the second approach, we use historical VaR to measure the downside risk of each selected stock. VaR estimates a loss threshold associated with a given confidence level. In this study, we use a one-day horizon and a 95\% confidence level. Specifically, we calculate the fifth percentile of each stock's historical daily returns over the same rolling window used for network estimation:
\begin{equation}
\mathrm{VaR}_{i,t}
=
-q_{0.05}
\left(
r_{i,t-W+1},\ldots,r_{i,t}
\right),
\label{eq:empirical_historical_var}
\end{equation}
where $q_{0.05}$ denotes the empirical fifth percentile. The resulting value represents the estimated magnitude of a lower-tail daily loss based on the historical observations in the window. We then use the inverse of each stock's VaR to determine its portfolio weight:
\begin{equation}
w_{i,t}^{IV}
=
\frac{1/\mathrm{VaR}_{i,t}}
{\displaystyle
\sum_{j\in S_{k,t}}
1/\mathrm{VaR}_{j,t}},
\qquad i\in S_{k,t}.
\label{eq:empirical_inverse_var_weight}
\end{equation}
This means that a stock with a larger estimated downside loss receives a smaller portfolio weight, while a stock with a smaller estimated loss receives a larger weight. The final weights are normalized to sum to one, and the strategy does not take short positions.

To avoid unstable calculations, we require at least 60 finite return observations and use the type-7 empirical quantile implemented in R to estimate historical VaR. A stock receives zero inverse-VaR weight when its historical return sample is insufficient, its VaR estimate is not finite, or its estimated VaR is not strictly positive. The remaining eligible weights are normalized to sum to one, with any unallocated capital remaining in cash. In the six primary portfolio strategies, the target weights sum to one at every rebalancing date. If no selected stock has a valid positive VaR estimate, the target portfolio is held in cash. The VaR calculation uses the same estimation window as the network construction, so changing $W$ may affect both the selected stocks and their risk-based allocations. The inverse-VaR approach is used only to distribute capital among the selected stocks and does not impose a separate portfolio-level VaR constraint.

\subsection{Trading Simulation}
\label{subsec:empirical_trading}

We evaluate the selected portfolios through a weekly trading simulation. Instead of reconstructing the portfolio every trading day, we update the network after each completed trading week. The resulting stock selections and target weights form the trading signal for the next available trading session. To avoid using information before it becomes available, we separate the signal date from the execution date. For example, when a signal is formed after the close of a Friday trading session, the corresponding portfolio is rebalanced at the close of the next available trading session, typically Monday. The existing portfolio continues to earn returns until that execution close, and the new portfolio begins earning returns only afterward. This timing convention ensures that the return earned before execution is not assigned to a portfolio constructed using information that was unavailable during that holding period. The simulation contains 293 completed weekly signals, beginning on March 19, 2021 and ending on October 23, 2026. The first corresponding execution takes place on March 22, 2021, and the final execution occurs on October 26, 2026. We continue evaluating portfolio wealth through October 28, 2026, but do not introduce an additional trade based on an incomplete final trading week.

Each strategy begins with an initial portfolio wealth of one, held in cash until the first execution. We incorporate transaction costs whenever the portfolio is rebalanced. Between execution dates, the quantities held in adjusted-price units remain fixed, while the dollar values and portfolio weights change with asset returns. Let $h_{i,t-1}$ denote the dollar value of the position in stock $i$ after the previous session's close, and let $b_{t-1}$ denote the cash balance. Before any trading at the current close, the positions and total wealth are given by
\[
h_{i,t}^{-}=h_{i,t-1}(1+r_{i,t}),
\qquad
V_t^{-}=\sum_i h_{i,t}^{-}+b_{t-1}.
\]
Here, cash earns zero. On a day without rebalancing, we set $h_{i,t}=h_{i,t}^{-}$, $b_t=b_{t-1}$, and $V_t=V_t^{-}$. The transaction-cost rate is fixed at ten basis points of gross traded value. Let $Q_t$ denote the total value of purchases and sales required to move from the existing holdings to the new target portfolio. The transaction cost is
\begin{equation}
C_t
=
0.001 Q_t.
\label{eq:empirical_transaction_cost}
\end{equation}

The gross traded value is determined jointly with the transaction fee. Let $V_t^{-}$ denote portfolio wealth immediately before rebalancing, $h_{i,t}^{-}$ denote the dollar value of the existing position in stock $i$, and $w_{i,t}^{*}$ denote its new target weight. We calculate gross traded value as
\begin{equation}
Q_t
=
\sum_i
\left|
h_{i,t}^{-}
-
w_{i,t}^{*}(V_t^{-}-C_t)
\right|.
\label{eq:empirical_gross_traded_value}
\end{equation}
Since the target holdings depend on wealth after transaction costs, we solve this expression together with Equation~\eqref{eq:empirical_transaction_cost}. After the fee is deducted, the wealth, stock positions, and cash balance at the execution close are
\[
\begin{aligned}
V_t &= V_t^{-}-C_t,\\
h_{i,t} &= w_{i,t}^{*}V_t,\\
b_t &= V_t-\sum_i h_{i,t}.
\end{aligned}
\]
Turnover is calculated as $Q_t/V_t^{-}$. On each execution date, the existing portfolio earns the return ending at that day's close before rebalancing takes place, while the new portfolio begins earning returns in the following session. The same calculation applies to the initial purchase from cash. Even if the new target weights are identical to the previous target weights, a trade may still be required because the actual portfolio weights have changed between execution dates.

The original portfolio simulation held target weights constant between weekly execution dates and charged transaction costs only for changes between successive target weight vectors. Maintaining constant weights when individual stock returns differ would require additional trades, which were not charged in that calculation. In the revised analysis, we retain the original network estimates and target weights but use the holding-value recursion described above (Track B). We retain the original constant-weight calculation (Track A) and a further calculation that charges the cost of restoring the target weights at every trading-session close (Track C) for comparison. These alternatives allow us to examine the effect of the accounting convention without changing the underlying stock selection rule.

Because pairwise GFEVD requires estimating a large number of VAR models, some estimation windows may produce unsuccessful fits. When this occurs, the corresponding GFEVD strategy does not construct a new portfolio at that signal date. Instead, it continues holding the existing positions until the next successful signal, without trading to restore the previous target weights. We record these failures rather than removing individual stocks from the sample or modifying the network estimation procedure. The correlation-based strategies continue to follow their original trading rules whenever their own networks can be constructed. Therefore, a failed GFEVD estimation can result in different rebalance activity across methods, which is particularly relevant for the shorter-window experiments. We retain these observations in the performance analysis and examine their influence when interpreting the results. In the additional equal-weight failure-date comparison, the Pearson strategies also omit trades on the reserved-period dates corresponding to unsuccessful GFEVD estimates. The simulation is based on adjusted daily prices and the execution convention described above. Although it provides a consistent environment for comparing portfolio strategies, the resulting wealth paths are simulated outcomes rather than records of actual transactions. In particular, the analysis does not separately model market impact, intraday liquidity, or all other frictions associated with live trading.

\subsection{Experimental Design}
\label{subsec:empirical_design}

Our primary experiment uses a portfolio size of $k=5$ and an estimation window of $W=250$ daily return observations. The pairwise VAR lag is fixed at one, and the GFEVD forecast horizon is ten trading days. All primary portfolios follow the weekly execution schedule and are evaluated using the same ten-basis-point transaction-cost assumption. We use one fixed primary specification so that the comparison between GFEVD and correlation networks is not determined by selecting whichever combination happens to produce the highest historical return. The theoretical analysis in Section~\ref{subsec:parameter_foundations} explains how portfolio size and estimation window length may affect network estimation and stock selection, but it does not establish that one combination must maximize investment returns. The sample is divided into the diagnostic and reserved periods described in Section~\ref{sec:data}. We use the diagnostic period to examine the behavior of the network construction procedure, while the reserved period provides the main evaluation of the frozen portfolio specification. For the primary comparison, we construct portfolios using three network methods: pairwise GFEVD, absolute Pearson correlation, and signed Pearson correlation. Each network is evaluated using both equal weighting and inverse historical VaR weighting, producing six primary portfolio strategies.

In addition to the primary experiment, we examine whether the results are sensitive to the number of selected stocks and the estimation window length. We consider
\begin{equation}
k\in\{3,5,10\},
\qquad
W\in\{60,120,250\}.
\label{eq:empirical_parameter_grid}
\end{equation}
The original parameter grid contains 54 portfolio configurations, covering the three networks and both weighting methods for each combination of $k$ and $W$. In the revised analysis, we reconstruct the equal-weight targets from the saved degree rankings and evaluate the resulting 27 configurations using the revised accounting procedure. The six primary portfolios remain available under both weighting methods, but the inverse-VaR sensitivity results outside the primary specification are not reported under the revised accounting because the corresponding target weights have not been reconstructed. All evaluated configurations use the same trading calendar. In particular, a strategy using a shorter estimation window does not receive an earlier trading start simply because it requires fewer historical observations. The parameter grid allows us to investigate how network and portfolio behavior changes under different specifications. However, these comparisons are treated as exploratory sensitivity analyses, and the primary model remains $k=5$ and $W=250$ regardless of whether another combination produces a higher return. This distinction is important because the sensitivity analysis was conducted after the reserved-period performance of the primary strategy had already been examined. The additional configurations therefore cannot be interpreted as independent out-of-sample confirmations of a newly selected optimal model. We also examine the saved GFEVD alternatives using a one-step horizon and different rules for combining the two directional shares.

For additional descriptive comparisons, we consider portfolios formed from the lowest-degree stocks using the same weighting and accounting rules, an equal-weight portfolio containing all 451 stocks that follows the weekly rebalancing schedule, and a buy-and-hold portfolio based on the vendor-adjusted SPY price series with one initial purchase. The broader portfolios are not subject to the five-stock restriction and are included as economic reference points rather than as tests that isolate the effect of the dependence measure.

\subsection{Performance Evaluation and Statistical Inference}
\label{subsec:empirical_evaluation}

To evaluate the portfolio strategies, we use their daily after-cost wealth paths. Let $V_t$ denote portfolio wealth at the end of trading day $t$. The corresponding daily portfolio return is
\begin{equation}
R_{p,t}
=
\frac{V_t}{V_{t-1}}-1.
\label{eq:empirical_net_return}
\end{equation}
We use these daily returns to calculate cumulative return, annualized geometric return, annualized volatility, the Sharpe ratio, the Sortino ratio, and maximum drawdown. We also report portfolio turnover and total transaction costs to examine whether differences in performance are associated with different trading activity. Annualized return and volatility are calculated using 252 trading sessions per year. For risk-adjusted performance, we calculate the Sharpe and Sortino ratios using daily excess returns relative to the three-month Treasury bill rate obtained from the Federal Reserve Bank of St. Louis (FRED). Since the DTB3 series reports a discount yield rather than a directly investable return, we convert the quoted rate into the corresponding Treasury bill price. Let $q_{u(s)}$ denote the latest available DTB3 quote dated strictly before the beginning of the equity return interval at session $s$. For a 91-day Treasury bill and a 360-day discount-rate convention, the implied price is
\begin{equation}
P_s
=
1-\frac{q_{u(s)}}{100}\frac{91}{360}.
\label{eq:empirical_tbill_price}
\end{equation}
We then calculate the corresponding risk-free investment return between consecutive equity sessions $s$ and $t$ as
\begin{equation}
R_{f,s,t}
=
P_s^{-\Delta_{s,t}/91}-1,
\label{eq:empirical_risk_free_return}
\end{equation}
where $\Delta_{s,t}$ represents the number of calendar days between the two sessions. This calculation accounts for weekends and holidays when determining the return over the corresponding holding period. We subtract the resulting risk-free return from the portfolio's after-cost daily return to obtain the excess return. The Sharpe ratio is calculated as the arithmetic mean of daily excess returns divided by their sample standard deviation, multiplied by $\sqrt{252}$. The Sortino ratio uses the same mean excess return and annualization factor, but divides by the square root of the full-sample mean of squared negative excess returns, with nonnegative excess returns contributing zero. Maximum drawdown measures the largest decline in portfolio wealth relative to its previous peak and is calculated as
\begin{equation}
MDD
=
\min_t
\left\{
\frac{V_t}
{\displaystyle\max_{s\leq t}V_s}
-1
\right\}.
\label{eq:empirical_max_drawdown}
\end{equation}
A more negative value indicates a larger loss relative to the portfolio's previous highest value.

Although these measures provide different views of portfolio performance, our primary criterion for incremental network value is reserved-period final wealth. Specifically, the GFEVD portfolio must achieve higher final wealth than both correlation-based portfolios under both weighting methods. Let $\Delta_m$ denote the difference between the final wealth of a GFEVD portfolio and the corresponding correlation benchmark, where $m=1,\ldots,4$ represents the four comparisons across the two correlation networks and the two weighting methods. The primary criterion is
\begin{equation}
\Delta_m>0
\qquad
\text{for all }m=1,\ldots,4.
\label{eq:empirical_wealth_criterion}
\end{equation}
Final wealth is measured over the same reserved period, with the wealth paths rebased to one at the December 29, 2023 close. This normalization does not reset the holdings or initiate a new trade at the beginning of the reserved period. This criterion requires a consistent advantage across the four comparisons. A higher Sharpe ratio or a positive wealth difference against only one benchmark is not sufficient to satisfy the primary condition. We use this rule because the main question is whether GFEVD provides incremental investment value beyond both correlation networks, rather than whether it performs well under one particular comparison.

To examine the uncertainty associated with the observed performance differences, we use paired strategy returns and time-series inference methods. The original empirical analysis includes 80\% circular block-bootstrap intervals and Newey--West statistics \citep{PolitisRomano1992,NeweyWest1987}, followed by 95\% inference using 5,000 circular block-bootstrap replications. In the revised analysis, we apply the latter procedure to the daily after-cost returns obtained under the revised accounting convention. The main bootstrap uses a block length of ten trading sessions, while block lengths of 5, 21, and 63 sessions are examined to assess sensitivity to the choice of block length. For each replication, the same resampled dates are used across the compared portfolio strategies, preserving the paired structure of the experiment. We distinguish the uncertainty in daily return differences from the primary final-wealth criterion. The 95\% confidence intervals describe the mean paired daily return differences, while the tests of incremental network value are based on the four reserved-period wealth differences. Because the primary criterion requires all four wealth differences to be positive, we apply a one-sided intersection--union test. Let $p_m$ be the one-sided p-value for testing a non-positive wealth advantage in comparison $m$. The combined p-value is
\begin{equation}
p_{\mathrm{IUT}}
=
\max_{m=1,\ldots,4}p_m.
\label{eq:empirical_iut}
\end{equation}
Under this approach, evidence for the joint claim requires each of the four individual comparisons to support a positive wealth advantage. If one comparison does not provide sufficient evidence, the joint test cannot reject the null hypothesis.

The block-bootstrap procedure relies on the approximation
that the joint strategy return series is sufficiently
stationary and weakly dependent over the evaluation period,
with suitable moment conditions. The sensitivity analysis
across different block lengths does not establish that
these assumptions hold. In addition, final wealth is a
nonlinear function of the sequence of portfolio returns.
We therefore interpret the bootstrap results as an
assessment of the historical wealth comparisons under the
adopted resampling procedure, rather than as direct
inference on expected future terminal wealth.

For each comparison, the one-sided p-value is calculated from the recentered circular-block bootstrap distribution of the final-wealth difference. If $\widehat{\Delta}_m$ denotes the observed wealth difference and $\Delta_{m,b}^{*}$ denotes its value in bootstrap replication $b$, we calculate
\begin{equation}
p_m
=
\frac{
1+\displaystyle\sum_{b=1}^{B}
\mathbf{1}
\left\{
\Delta_{m,b}^{*}
\geq 2\widehat{\Delta}_m
\right\}
}{
B+1
},
\label{eq:empirical_bootstrap_pvalue}
\end{equation}
where $B=5000$. The reported 95\% confidence intervals are obtained from the type-7 empirical 2.5th and 97.5th percentiles of the bootstrap distribution of mean daily return differences. 
We apply the Holm adjustment to the four one-sided
p-values \citep{Holm1979}, while the Romano--Wolf
procedure uses standardized final-wealth differences
with Newey--West standard errors and a lag equal to
the bootstrap block length \citep{RomanoWolf2005}.
The standard errors use a linearization of terminal
wealth in terms of log returns, with the corresponding wealth levels used to scale the paired log-return contributions. These standard errors are calculated from the original sample and held fixed across bootstrap replications rather than re-estimated for each resampled path. These adjustments provide additional information about the individual comparisons, but they do not replace the joint final-wealth criterion. Furthermore, the 95\% bootstrap analysis was introduced after the primary reserved-period results had already been observed. We therefore interpret it as an additional assessment of statistical uncertainty rather than as a new experiment conducted on an untouched test sample. The same consideration applies to the revised portfolio accounting and the exploratory parameter grid. The statistical results from the original constant-weight calculation are kept separate from those obtained using the revised accounting procedure.

\subsection{Network Stability and Selection Analysis}
\label{subsec:empirical_network_stability}

In addition to portfolio returns, we examine how the estimated networks differ across methods and over time. This is important because two dependence measures may produce different edge costs without necessarily selecting different stocks for the portfolio. To compare network structures, we calculate the Jaccard index between two MST edge sets:
\begin{equation}
J(E_1,E_2)
=
\frac{|E_1\cap E_2|}
{|E_1\cup E_2|},
\label{eq:empirical_edge_jaccard}
\end{equation}
where $E_1$ and $E_2$ are the sets of edges in the two trees. A value of one indicates that the two MSTs contain exactly the same edges, while a smaller value indicates less overlap. We use this measure both to compare different network methods on the same date and to examine changes in a single network from one completed trading week to the next. However, differences in MST edges do not necessarily lead to differences in selected stocks. Since the portfolios are constructed from the highest-degree nodes, we also compare the sets of stocks selected by each network. For two portfolios containing $k$ stocks, selection overlap is defined as
\begin{equation}
O(S_1,S_2)
=
\frac{|S_1\cap S_2|}{k},
\label{eq:empirical_selection_overlap}
\end{equation}
where $S_1$ and $S_2$ are the two selected stock sets. This measure indicates the proportion of selected stocks shared by the two portfolios. We also calculate the proportion of stocks retained from one week to the next to examine the persistence of portfolio selection over time.

Finally, we investigate the degree values near the selection cutoff. When the degree of the last selected stock is the same as that of the first excluded stock, the selection depends on the predetermined tie-breaking rule. We record the frequency of these ties and the difference between the two degree values to examine how clearly the selected stocks are separated from the remaining universe. We also record whether the two selected stock sets are exactly the same. To examine the role of cutoff ties, we compare the selections obtained under the original ordering with those obtained under the reverse ordering when resolving degree ties, holding the tree itself fixed.

The network stability analysis uses the saved MST edges and degree rankings, while unsuccessful network fits remain identified in the corresponding records. For the primary estimation window of 250 daily returns, we also use the reconstructed cost matrices to calculate the replacement gaps defined in Equation~\eqref{eq:global_mst_margin}. Each tree has 101,025 non-tree edges, and we calculate their replacement gaps and the minimum replacement margin for all 293 formation dates under each network method. These margins refer to the estimated cost matrices, not to the unknown population costs. To distinguish strictly positive margins from numerical ties, we use a tolerance of $10^{-12}\max\{1,\max_{i<j}|d_{ij,t}|\}$ for each method and date.

For a given estimated tree $T_t$ and its cost matrix $\widehat{\mathbf D}_t$, we examine the number of non-tree edges $q(T_t,\widehat{\mathbf D}_t;\delta)$ defined in Section~\ref{subsec:mst_stability}. We consider perturbation radii of the form $\delta=a\,\operatorname{IQR}(\widehat{\mathbf D}_t)$, where the interquartile range is calculated across all undirected candidate edge costs using the type-7 empirical quantiles and
\[
a\in\{0,10^{-10},10^{-8},10^{-6},10^{-4},10^{-3},10^{-2}\}.
\]
These radii are used to describe local sensitivity and are not estimates of sampling uncertainty. Scaling by the interquartile range accounts for differences in the scale of edge costs, although it does not make the dependence measures equivalent.

For consecutive weekly networks constructed using the same method, we calculate the observed cost change
\[
\varepsilon_t=\|\widehat{\mathbf D}_{t+1}-\widehat{\mathbf D}_t\|_{\max}.
\]
We compare $2\varepsilon_t$ with the replacement margin of the earlier tree, count the observed edge replacements $s(T_t,T_{t+1})$, and examine whether this count is bounded by $q(T_t,\widehat{\mathbf D}_t;\varepsilon_t)$. We then compare the degree gap at the selection cutoff with twice the observed number of edge replacements and, separately, with twice the corresponding bound. We record the frequency with which these sufficient conditions hold and compare them with the observed changes in the trees and selected stocks. We also examine the degree-gap condition for pairs of network methods on the same date, using the first tree in each comparison as the reference tree. A failure to satisfy a sufficient condition does not imply that the corresponding tree or selected stock set changes.

Together, the portfolio performance and network stability analyses allow us to examine both the economic and structural consequences of using pairwise GFEVD. The first determines whether the resulting portfolios perform differently from their correlation benchmarks, while the second helps explain how differences in the estimated networks translate into changes in stock selection. Neither the observed cost changes nor the hypothetical perturbation radii provide a probability of recovering the population MST.

%% file: sections/06_results.tex
\section{Empirical Results}
\label{sec:results}

In this section, we present the empirical results of the three network construction methods and examine whether pairwise GFEVD provides additional value compared with conventional correlation-based networks. We focus first on the frozen primary specification, using five selected stocks and an estimation window of 250 trading days. We then examine the uncertainty associated with the observed portfolio performance. In addition to investment performance, we compare the structures of the estimated networks and the stability of their stock selections. The sensitivity of these findings to alternative parameter choices and implementation assumptions is examined in Section~\ref{sec:robustness}. The main economic results use the revised weekly share-drift accounting (Track B), with portfolio returns calculated after transaction costs.

\subsection{Primary Portfolio Performance}
\label{subsec:results_primary}

To evaluate the investment performance of the three network methods, we first consider the frozen primary network and selection specification with $k=5$ and $W=250$. For each network, we select the five stocks with the highest MST degree centrality and construct portfolios using both equal weighting and inverse historical VaR weighting. All strategies follow the same weekly trading calendar and transaction-cost assumptions described in Section~\ref{sec:methodology}. Table~\ref{tab:results_primary_performance} reports the results over the reserved evaluation period from January 2, 2024 to October 28, 2026, with portfolio wealth normalized to one at the December 29, 2023 close, immediately before the reserved evaluation period. The holdings at that date are inherited from the earlier simulation rather than reinitialized.

\begin{table}[htbp]
\centering
\caption{Primary Portfolio Performance During the Reserved Period}
\label{tab:results_primary_performance}
\small
\begin{tabular}{llrrrr}
\toprule
Network & Weighting & Final wealth & Ann. return (\%) &
Sharpe & Max DD (\%) \\
\midrule
GFEVD & Equal & 1.2972 & 9.69 & 0.337 & -27.10 \\
Absolute Pearson & Equal & 1.2370 & 7.85 & 0.257 & -26.86 \\
Signed Pearson & Equal & 1.3040 & 9.89 & 0.351 & -26.86 \\
\midrule
GFEVD & Inverse VaR & 1.2715 & 8.91 & 0.310 & -25.86 \\
Absolute Pearson & Inverse VaR & 1.2649 & 8.71 & 0.304 & -26.21 \\
Signed Pearson & Inverse VaR & 1.3169 & 10.28 & 0.379 & -26.21 \\
\bottomrule
\end{tabular}
\end{table}

As shown in Table~\ref{tab:results_primary_performance}, pairwise GFEVD does not consistently outperform the correlation-based networks. Under equal weighting, the GFEVD portfolio achieves a final wealth of 1.2972, representing an advantage of 0.0602 over the absolute Pearson benchmark but a disadvantage of 0.0068 relative to the signed Pearson benchmark. A similar result is observed under inverse VaR weighting, with wealth differences of 0.0066 and $-0.0454$, respectively. Since our primary criterion requires GFEVD to outperform both correlation benchmarks under both weighting methods, the observed results do not satisfy the predetermined condition for incremental network value.

\begin{figure}[htbp]
\centering
\includegraphics[width=\linewidth]
{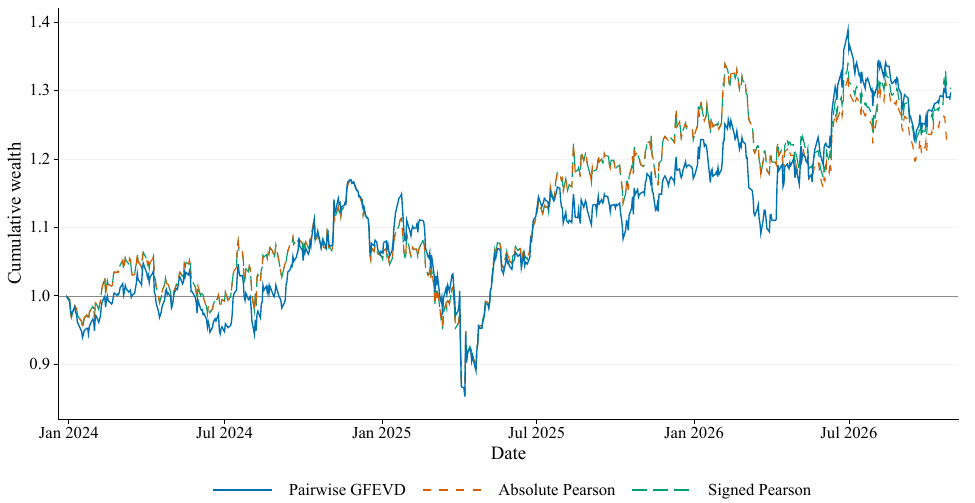}
\caption{Cumulative portfolio wealth under equal
weighting during the reserved evaluation period
from January 2, 2024, through October 28, 2026.
All wealth paths are net of transaction costs
and normalized to one at the December 29, 2023
close.}
\label{fig:results_cumulative_wealth}
\end{figure}

\begin{figure}[htbp]
\centering
\includegraphics[width=\linewidth]
{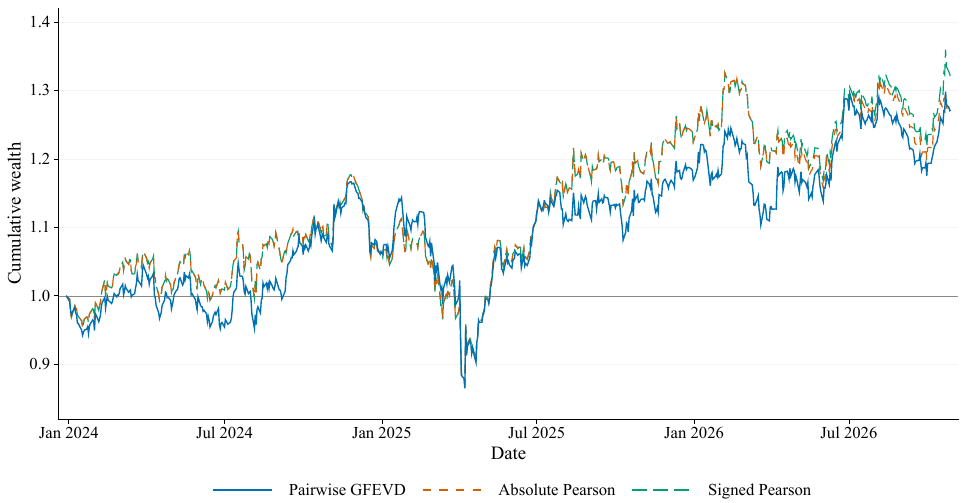}
\caption{Cumulative portfolio wealth under inverse
historical VaR weighting during the reserved
evaluation period from January 2, 2024, through
October 28, 2026. All wealth paths are net of
transaction costs and normalized to one at the
December 29, 2023 close.}
\label{fig:results_cumulative_wealth_iv}
\end{figure}

Figures~\ref{fig:results_cumulative_wealth}
and~\ref{fig:results_cumulative_wealth_iv}
illustrate how the relative performance of
the three network methods changes over time.
The differences are not constant throughout
the reserved evaluation period, and GFEVD
does not maintain a consistent wealth
advantage over both Pearson benchmarks.

The corresponding wealth differences are
reported in Appendix~\ref{app:relative_wealth}.
In particular, the final wealth advantage
over absolute Pearson under inverse VaR
weighting is only 0.0066, while GFEVD finishes
below signed Pearson under both weighting
methods. The two Pearson wealth paths also
coincide for most of the reserved period
and begin to diverge in April 2026.
These pathwise comparisons are descriptive;
their statistical interpretation is examined
in the following subsection.

Trading activity provides an additional comparison across the three network methods. During the reserved period, cumulative turnover is 65.84, 66.24, and 67.04 for GFEVD, absolute Pearson, and signed Pearson under equal weighting, respectively. Under inverse VaR weighting, the corresponding values are 69.92, 69.16, and 69.99. Total transaction costs, expressed in units of the initial full-sample wealth, are 0.1029, 0.1050, and 0.1068 under equal weighting, and 0.1104, 0.1139, and 0.1158 under inverse VaR weighting, respectively. GFEVD does not uniformly generate lower turnover, although its total transaction costs are lower than those of both Pearson benchmarks under both weighting methods. Since turnover is measured relative to pre-trade wealth, a lower sum of transaction costs does not necessarily imply a lower cumulative turnover.

The risk-adjusted performance measures provide a similar comparison. As shown in Table~\ref{tab:results_primary_performance}, GFEVD achieves a higher Sharpe ratio than absolute Pearson under both weighting methods, while signed Pearson produces the highest Sharpe ratio. Maximum drawdowns are relatively similar across the networks. Under inverse VaR weighting, GFEVD has a slightly smaller maximum drawdown than either Pearson portfolio, but this improvement is not accompanied by the highest final wealth or Sharpe ratio. These findings suggest that differences in estimated dependence do not automatically translate into better portfolio performance.

An interesting observation concerns the two correlation-based networks. Under the primary specification, the absolute and signed Pearson methods select identical five-stock portfolios on 290 of the 293 weekly signals. Their selections differ on only three signals in 2026, with one stock differing on each of those occasions. Nevertheless, the resulting portfolios have different final wealth values. This illustrates how relatively small differences in portfolio selection can affect realized performance through subsequent returns and changes in portfolio weights. It also shows that similarity in network structure or stock selection does not necessarily imply identical investment outcomes. However, the observed return differences alone do not establish which network provides a more accurate representation of the underlying dependence structure.

The simpler portfolio benchmarks provide additional context for these results. During the same reserved period, the vendor-adjusted SPY buy-and-hold portfolio and the broad equal-weight portfolio containing all 451 stocks achieve final wealth of 1.6780 and 1.5090, respectively. All six primary network portfolios finish below these benchmarks, while they finish above the corresponding low-degree selection controls, whose final wealth ranges from 0.7869 to 0.8008. The broad portfolios are not subject to the same five-stock selection restriction. These comparisons are therefore descriptive and do not isolate the contribution of the dependence measure. They also remain subject to the data limitations discussed in Section~\ref{sec:data}.

The relative ordering of the three network methods in the reserved period is unchanged from the original constant-mix calculation (Track A). However, the final wealth values are lower under the weekly share-drift accounting used here. We examine this difference separately in Section~\ref{sec:robustness}.

\subsection{Statistical Comparison of Portfolio Returns}
\label{subsec:results_inference}

Although the primary results show differences in realized portfolio wealth, these differences do not necessarily indicate that one network method provides a persistent investment advantage. To examine the uncertainty associated with the observed performance, we compare the GFEVD portfolio with each Pearson benchmark using paired daily strategy returns. The comparisons are conducted separately for equal weighting and inverse VaR weighting, giving four primary comparisons in total. Table~\ref{tab:results_primary_inference} reports the final-wealth differences, mean daily after-cost return differences, and 95\% bootstrap confidence intervals. The one-sided p-values refer to the final-wealth comparisons rather than to the mean daily return differences.

\begin{table}[htbp]
\centering
\caption{Paired Comparisons Under the Primary Specification}
\label{tab:results_primary_inference}
\small
\setlength{\tabcolsep}{4pt}
\begin{tabular}{llrrlr}
\toprule
Weighting & Benchmark & Wealth gap &
Mean daily gap & 95\% interval & One-sided $p$ \\
\midrule
Equal & Absolute & 0.0602 & 0.000073 &
$[-0.000243,\ 0.000398]$ & 0.3413 \\
Equal & Signed & -0.0068 & -0.000002 &
$[-0.000312,\ 0.000311]$ & 0.5181 \\
Inverse VaR & Absolute & 0.0066 & 0.000012 &
$[-0.000246,\ 0.000274]$ & 0.4729 \\
Inverse VaR & Signed & -0.0454 & -0.000046 &
$[-0.000304,\ 0.000208]$ & 0.6649 \\
\bottomrule
\end{tabular}
\end{table}

A positive difference in Table~\ref{tab:results_primary_inference} indicates that the GFEVD portfolio performs better than the corresponding correlation benchmark. The two positive final-wealth differences are associated with relatively small mean daily return differences. More importantly, all four 95\% bootstrap confidence intervals for the mean daily return differences contain zero. These intervals do not imply that the networks have equivalent performance, but they indicate that the observed differences in average daily returns cannot be reliably distinguished from zero under the reported bootstrap procedure.

Because the primary criterion requires GFEVD to achieve higher final wealth in all four comparisons, we further examine the joint hypothesis using the intersection--union test introduced in Section~\ref{subsec:empirical_evaluation}. The resulting p-value is
\begin{equation}
p_{\mathrm{IUT}}=0.6649.
\label{eq:results_primary_iut}
\end{equation}
The test does not reject the joint null hypothesis at the 5\% significance level. The sensitivity of this conclusion to the bootstrap block length is examined in Section~\ref{subsec:robustness_bootstrap}. The individual Holm-adjusted p-values are all equal to one, and the Romano--Wolf adjustments also do not identify a significant positive advantage for GFEVD. Therefore, both the realized final-wealth comparison and the additional statistical inference fail to establish consistent incremental investment value for pairwise GFEVD under the frozen primary specification. The corresponding intersection--union p-value under the original constant-mix accounting is 0.6717, which does not apply to the revised wealth paths. Since the revised inference is conducted on a period whose portfolio outcomes have already been examined, we interpret these tests as a supplementary assessment of statistical uncertainty rather than as an independent confirmation using an untouched evaluation sample.

\subsection{Network Structure and Portfolio Selection}
\label{subsec:results_network_structure}

We further examine whether the differences in portfolio performance are accompanied by differences in the estimated network structures and selected stocks. We compare the edge sets of GFEVD and Pearson networks constructed on the same dates and investigate how the networks and selected portfolios change from one completed trading week to the next. The comparison uses the Jaccard index and Top-$k$ selection overlap measures introduced in Section~\ref{subsec:empirical_network_stability}. Table~\ref{tab:results_network_stability} summarizes the main structural results for five-stock portfolios. The reported averages use available network estimates, while unsuccessful fits remain recorded and are examined in Section~\ref{subsec:results_failed_fits}.

\begin{table}[htbp]
\centering
\caption{MST and Portfolio-Selection Stability}
\label{tab:results_network_stability}
\small
\setlength{\tabcolsep}{5pt}
\begin{tabular}{rrrrrr}
\toprule
$W$ &
Cross-network edge &
Top-5 overlap &
Weekly edge &
Top-5 retention &
Cutoff ties \\
&
Jaccard &
GFEVD--Absolute &
Jaccard &
GFEVD &
GFEVD \\
\midrule
60 & 0.486 & 0.459 & 0.505 & 0.489 & 0.628 \\
120 & 0.676 & 0.648 & 0.685 & 0.618 & 0.669 \\
250 & 0.813 & 0.799 & 0.815 & 0.779 & 0.625 \\
\bottomrule
\end{tabular}
\end{table}

As shown in Table~\ref{tab:results_network_stability}, the GFEVD and absolute Pearson networks become more similar when longer estimation windows are used. At $W=60$, the average edge Jaccard index between the two MSTs is 0.486, indicating that the networks share only part of their estimated tree structures. The index increases to 0.676 at $W=120$ and reaches 0.813 at $W=250$. A similar pattern appears in the selected portfolios. For $k=5$, the average overlap between GFEVD and absolute Pearson selections increases from 0.459 under the 60-day window to 0.648 under the 120-day window and 0.799 under the 250-day window. These results suggest that GFEVD and Pearson correlation identify more similar network structures and portfolio candidates when longer historical windows are used. However, this is an empirical relationship observed within the sample, rather than evidence that the two methods converge to the same population network. As discussed in Section~\ref{subsec:gfevd_correlation}, their underlying dependence measures are generally different when lagged relationships are present.

The estimation window also affects how frequently the networks and portfolios change over time. For GFEVD, the average week-to-week MST edge Jaccard increases from 0.505 at $W=60$ to 0.815 at $W=250$, indicating that a larger proportion of the estimated tree remains unchanged between consecutive completed weeks. The selected portfolios exhibit a similar pattern. When five stocks are selected, mean weekly retention increases from 0.489 to 0.779 over the same range of windows. This means that the 60-day GFEVD portfolio replaces approximately 2.55 stocks per successful weekly transition, compared with approximately 1.11 stocks under the 250-day window. Although longer windows produce more persistent networks and stock selections, greater stability does not necessarily translate into higher investment returns. As reported in Section~\ref{subsec:results_parameter_sensitivity}, the shorter-window GFEVD strategies achieve favorable relative wealth differences in several exploratory equal-weight comparisons despite their less persistent selections.

Another important finding concerns the degree ranking used for stock selection. Under the frozen $k=5$, $W=250$ specification, the final selected stock and the first excluded stock have the same MST degree on 62.5\% of the GFEVD weekly signals and 63.1\% of the signals under each Pearson method. The average difference between their degree values is only 0.42. Therefore, the predetermined tie-breaking rule frequently plays a role in determining which stocks enter the portfolio. This is particularly relevant because two networks may share most of their edges but still produce different Top-$k$ selections when the degree ranking near the cutoff changes. For example, at $W=250$, the average edge Jaccard between GFEVD and absolute Pearson is 0.813, but their five-stock sets are exactly identical on only 23.9\% of the weekly signals. By comparison, the two Pearson networks are much more similar to each other. Their average edge Jaccard reaches 0.999, and their five-stock portfolios are identical on 99.0\% of the weekly signals.

\begin{figure}[htbp]
\centering
\includegraphics[width=\linewidth]
{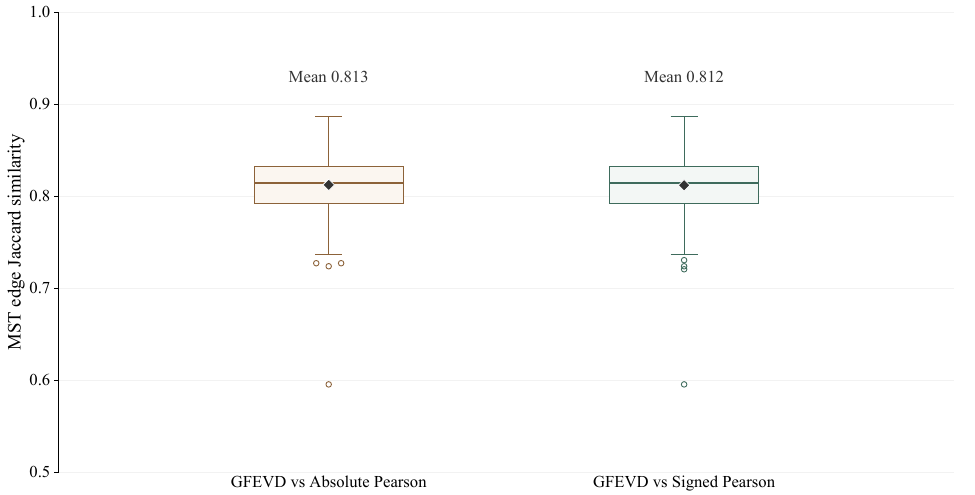}
\caption{Distribution of MST edge Jaccard
similarity between GFEVD and the two Pearson
benchmarks under the primary specification
across 293 weekly signals.}
\label{fig:results_network_selection}
\end{figure}

Figure~\ref{fig:results_network_selection}
shows the distribution of edge Jaccard
similarity between GFEVD and the two
Pearson benchmarks. The mean edge Jaccard
indices are 0.813 and 0.812 for absolute
and signed Pearson, respectively.
Although the estimated MSTs share a
substantial proportion of their edges,
this similarity does not necessarily
extend to the selected portfolios.

\begin{figure}[htbp]
\centering
\includegraphics[width=\linewidth]
{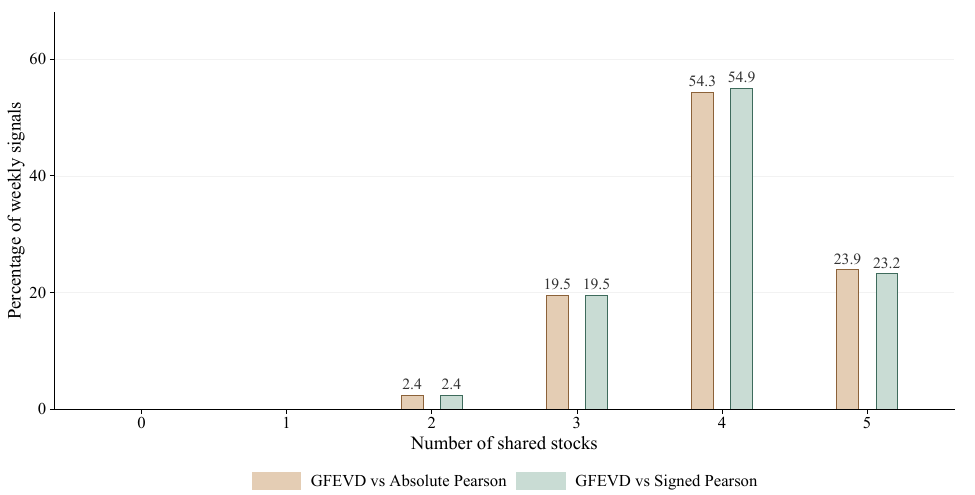}
\caption{Distribution of the number of shared
stocks between GFEVD and each Pearson benchmark
under the primary specification across 293
weekly signals. Stock overlap is calculated
using unordered selections, independently
of degree ranking.}
\label{fig:results_top5_overlap}
\end{figure}

As shown in Figure~\ref{fig:results_top5_overlap},
the same five stocks are selected in only
23.9\% and 23.2\% of the weekly signals
when GFEVD is compared with absolute and
signed Pearson, respectively. The most
common outcome is four shared stocks,
occurring in 54.3\% and 54.9\% of the
weeks, respectively. Thus, although the
networks frequently generate different
five-stock portfolios, their selections
generally remain substantially overlapping.
This distinction is consistent with the
theoretical discussion in
Section~\ref{subsec:mst_stability}:
changes in the MST may affect the degree
ranking near the selection cutoff without
producing an entirely different portfolio.

We also examine whether changes in the MST are associated with changes in the selected portfolio. For GFEVD with $k=5$, the Spearman correlations between week-to-week edge Jaccard and Top-5 retention are 0.332, 0.241, and 0.258 for estimation windows of 60, 120, and 250 trading days, respectively. These positive but moderate correlations indicate that weeks with more persistent network edges tend to have more persistent stock selections, although the relationship is not deterministic. This finding is consistent with the distinction developed in Section~\ref{subsec:mst_stability}: a change in an MST does not necessarily produce a change in the selected stock set, and changes in selected stocks can occur even when the overall network structure remains relatively similar. These results describe the observed association between tree changes and stock selection, rather than providing a sufficient condition for stability.

To examine the conditions in Section~\ref{subsec:mst_stability} more directly, we reconstruct the complete candidate edge-cost matrices under the primary specification. The reconstruction reproduces the original MSTs and target weights. Since the rejected candidate costs were not saved in the original experiment, the replacement margins are calculated from the reconstructed matrices. For each of the 879 method-date networks, we calculate the gaps for all 101,025 non-tree edges, giving 88,800,975 replacement gaps in total. All 879 minimum replacement margins are strictly positive. Table~\ref{tab:results_replacement_stability} summarizes these margins and the observed changes between consecutive weekly networks.

\begin{table}[htbp]
\centering
\caption{Replacement Margins and Adjacent-Week Stability Under the Primary Specification}
\label{tab:results_replacement_stability}
\small
\setlength{\tabcolsep}{5pt}
\begin{tabular}{lrrrr}
\toprule
Network & Median $\gamma$ & Median $\varepsilon$ & Median $s$ & Same Top-5 \\
\midrule
GFEVD & $2.68\times10^{-5}$ & 0.0905 & 43 & 58/292 \\
Absolute Pearson & $5.15\times10^{-5}$ & 0.1573 & 41 & 69/292 \\
Signed Pearson & $5.14\times10^{-5}$ & 0.1598 & 41 & 67/292 \\
\bottomrule
\end{tabular}
\par\smallskip
\begin{minipage}{0.98\linewidth}
\footnotesize Notes: Each network has 293 sample margins and 292 adjacent-week comparisons. $\varepsilon$ is the observed maximum absolute change in candidate edge costs and $s$ is the number of replaced tree edges. The conditions $2\varepsilon<\gamma$ and $\Delta_5>2s$ hold in none of the adjacent-week comparisons. Cost magnitudes are specific to each dependence measure.
\end{minipage}
\end{table}

Although all estimated MSTs have positive replacement margins, the sufficient condition in Proposition 3 is not satisfied in any of the 876 adjacent-week comparisons. In fact, the tree edge set changes in every comparison. The median number of replaced edges is 43 for GFEVD and 41 for each Pearson network. The bound $s\leq q(\varepsilon)$ holds in all comparisons, but the difference between the bound and the observed number of replacements is at least 1,110. Thus, while the bound correctly limits the number of possible edge replacements, it is conservative for the cost changes observed between consecutive weeks. Since the different dependence measures use different cost transformations, the magnitudes of their replacement margins alone do not establish which network is more stable.

The corresponding results for stock selection are somewhat different. The sufficient condition in Proposition 4 is not satisfied in any of the 876 adjacent-week comparisons, yet the selected five-stock set remains unchanged in 194 cases. In the same-date comparisons across network methods, the edge set remains unchanged in 232 of 879 comparisons; the tree and degrees change but the Top-5 set remains unchanged in another 196 comparisons; and the Top-5 set changes in the remaining 451. The condition $\Delta_5>2s$ holds in 87 of these cross-method comparisons, all between absolute and signed Pearson using absolute Pearson as the reference tree, and none of the certified selections changes. These results illustrate the difference between a sufficient guarantee of stability and the observed persistence of stock selection.

The replacement margins and perturbation magnitudes in these comparisons are calculated from estimated sample cost matrices. In particular, the observed difference between two consecutive matrices is not an estimate of the error relative to an unknown population cost matrix. The results therefore do not establish exact recovery of the underlying population network. Rather, they show how changes in estimated dependence can affect the tree and the stocks selected from it, and why these two forms of stability need to be examined separately.

Additional results for the hypothetical perturbation
radii and the reverse tie-breaking rule are reported
in Appendix~\ref{app:additional_stability}.
The sufficient conditions provide stronger stability
guarantees for small hypothetical perturbations,
while the reverse-order comparison shows that the
selected stock sets are sensitive to the ordering
rule when ties occur at the degree cutoff.
These results further distinguish the stability
of the estimated tree from the stability of the
portfolio selection rule.

%% file: sections/07_robustness.tex
\section{Robustness Checks}
\label{sec:robustness}

We examine the robustness of the empirical findings to alternative portfolio sizes, estimation windows, GFEVD forecast horizons, symmetrization rules, bootstrap block lengths, and the treatment of numerical estimation failures. These checks assess whether the primary conclusion
depends on particular implementation choices and whether
the favorable results observed under some alternative
specifications remain reliable. The primary specification
of $k=5$ and $W=250$ remains unchanged throughout the
analysis. Since the alternative configurations were
examined after the primary reserved-period results
became available, their performance is treated as
exploratory rather than as independent out-of-sample
confirmation. The portfolio results reported in this section
use the revised weekly share-drift accounting (Track B),
except where the original constant-mix calculation (Track A)
or the daily-rebalancing sensitivity (Track C) is explicitly
identified.

\subsection{Sensitivity to Portfolio Size and Estimation Window}
\label{subsec:results_parameter_sensitivity}

We consider portfolio sizes
$k\in\{3,5,10\}$ and estimation windows
$W\in\{60,120,250\}$. Each combination is evaluated
using the same three network construction methods
and reserved evaluation period. The revised parameter
grid uses equal weighting, since the inverse historical
VaR target weights outside the primary specification
were not stored and have not been reconstructed.
For each parameter combination, we report the two
GFEVD-minus-Pearson final-wealth differences under
equal weighting. The $W=60$ comparisons also omit
the three reserved-period GFEVD failure signals from
all three methods, as described in
Section~\ref{subsec:results_failed_fits}. A positive
difference against both benchmarks does not by itself
satisfy the four-comparison criterion, which also
requires positive differences under inverse VaR weighting.

\begin{table}[htbp]
\centering
\caption{Portfolio-Size and Estimation-Window Sensitivity under Equal Weighting}
\label{tab:results_parameter_sensitivity}
\small
\begin{tabular}{rrrrc}
\toprule
Window $W$ & Portfolio size $k$ &
\multicolumn{2}{c}{GFEVD minus Pearson wealth} & Both positive \\
\cmidrule(lr){3-4}
 & & Absolute & Signed & \\
\midrule
60 & 3 & 0.2976 & 0.2076 & Yes \\
60 & 5 & 0.6851 & 0.6392 & Yes \\
60 & 10 & 0.1503 & 0.0907 & Yes \\
\midrule
120 & 3 & -0.0630 & -0.0414 & No \\
120 & 5 & 0.1019 & -0.0437 & No \\
120 & 10 & -0.0326 & -0.0272 & No \\
\midrule
250 & 3 & -0.0478 & 0.0289 & No \\
250 & 5 & 0.0602 & -0.0068 & No \\
250 & 10 & 0.1661 & 0.1533 & Yes \\
\bottomrule
\end{tabular}
\par\smallskip
\begin{minipage}{0.95\linewidth}
\footnotesize Notes: Results use Track B and equal weighting. For $W=60$,
all three methods omit the three reserved-period GFEVD failure signals.
The inverse-VaR sensitivity grid has not been reconstructed.
\end{minipage}
\end{table}

Table~\ref{tab:results_parameter_sensitivity}
shows substantial sensitivity to the estimation window
and portfolio size. Under equal weighting, GFEVD achieves
positive wealth differences against both Pearson benchmarks
for all three portfolio sizes when $W=60$. None of the
three portfolio sizes has positive differences against
both benchmarks when $W=120$, while only $k=10$ does
so when $W=250$. These results concern the equal-weight
comparisons only. We cannot restate the four-comparison
criterion for the full grid under the revised accounting.

The differences are particularly large for
$k=5$ and $W=60$. Under equal weighting, GFEVD
achieves reserved-period final wealth of 1.8216,
compared with 1.1365 for absolute Pearson and
1.1825 for signed Pearson. These results indicate that
the relative economic performance of GFEVD can
change substantially with the estimation window,
although the $W=60$ comparison also uses the
failure-signal treatment described below. The positive
differences against both benchmarks at $W=60$ are not
maintained at $k=5$ under the frozen $W=250$ specification.

The variation across estimation windows is
consistent with the theoretical trade-off discussed
in Section~\ref{subsec:time_varying_recovery}.
A shorter window uses more recent observations
but provides less information for estimating
the pairwise VAR models. A longer window uses
more observations but may incorporate dependence
relationships that have changed over time.
The empirical wealth differences do not directly
measure the estimation error of the underlying
population networks and cannot establish that
a particular window is theoretically optimal.
The equal-weight comparison avoids changes in the
allocation arising from the estimation of individual
VaR. However, changes in the selected stocks and
the resulting trading costs may both contribute to
the observed wealth differences.

\subsection{Bootstrap Block-Length Sensitivity}
\label{subsec:robustness_bootstrap}

The primary statistical comparison uses
5,000 paired circular block-bootstrap replications
with a reported block length of ten trading
sessions. We additionally examine block lengths
of 5, 21, and 63 sessions to assess whether
the inference depends on the assumed persistence
of daily return differences. The same resampled
date indices are applied to the compared
portfolios within each replication.

\begin{table}[htbp]
\centering
\caption{Bootstrap Block-Length Sensitivity Under the Primary Specification}
\label{tab:robustness_block_length}
\begin{tabular}{rr}
\toprule
Block length & IUT $p$-value \\
\midrule
5 & 0.6557 \\
10 & 0.6649 \\
21 & 0.6555 \\
63 & 0.6461 \\
\bottomrule
\end{tabular}
\end{table}

Table~\ref{tab:robustness_block_length}
shows that the intersection--union p-value
remains between 0.6461 and 0.6649 across the
four block lengths. None of these specifications
rejects the joint null hypothesis at the
five percent significance level. The primary
finding therefore does not depend on the
reported block length of ten sessions.
Together with the individual paired comparisons
in Table~\ref{tab:results_primary_inference},
these results provide no statistically
supported evidence that GFEVD consistently
outperforms both Pearson benchmarks under
both weighting methods.

The exploratory parameter grid requires
a separate qualification. The original Track A analysis
at $k=5$ and $W=60$ produced an intersection--union
$p$-value of 0.0850 with a block length of ten sessions
and 0.0384 with a block length of 63 sessions. None
of the nine original parameter combinations rejected
at the five percent level with a block length of ten.
These results belong to the original accounting and
are not tests of the revised parameter grid. Since the
inverse-VaR sensitivity grid has not been reconstructed,
we do not report a new four-comparison test for these
alternative parameter combinations.

\subsection{Estimation Failures and Additional Checks}
\label{subsec:results_failed_fits}

Shorter estimation windows introduce
additional numerical difficulties for
the pairwise VAR procedure. Across the
293 weekly signals, the $W=60$ GFEVD
method records five method-level fitting
failures, including two in the diagnostic
period and three in the reserved period.
No corresponding failures are recorded
for the $W=120$ or $W=250$ GFEVD
networks. Under the predetermined
execution rule, an unsuccessful GFEVD
estimation does not generate a new
portfolio. The previously held positions
remain in place until the next successful signal,
and their values and portfolio weights continue
to change with asset returns. Failed estimates
are retained in the experiment and
are not replaced by networks constructed
from a reduced stock universe.

The failures also affect the structural
statistics reported in
Section~\ref{subsec:results_network_structure}.
Weekly comparisons requiring an
available GFEVD tree exclude the
corresponding failed observations.
The five isolated failures affect
ten adjacent transitions in the
$W=60$ weekly persistence calculation.
These exclusions apply only to
structural statistics that cannot
be calculated without an estimated
network. The associated trading
dates remain in the portfolio
evaluation.

The treatment of failed signals
could influence relative investment
performance if the Pearson strategies
continue to rebalance when the GFEVD
method fails. In the revised equal-weight
$W=60$ comparisons, the Pearson portfolios
also retain their existing positions on the
execution dates corresponding to the three
reserved-period GFEVD failure signals.
The GFEVD portfolio omits five signals over
the full sample, while each Pearson portfolio
omits the three reserved-period signals.
No rebalancing to the previous target takes
place on the omitted execution dates.
The calculation replays previously saved
stock selections and return records without
re-estimating any networks.

Under this treatment, the GFEVD portfolio
finishes ahead of both Pearson portfolios
under equal weighting for all three values
of $k$ at $W=60$. Thus, the observed
reserved-period wealth differences persist
when the three corresponding rebalancing
opportunities are omitted for all methods.
However, the ordinary $W=60$ replay with
Pearson trading on those dates and the
inverse-VaR failure comparison have not been
reconstructed under Track B. This diagnostic
does not establish statistical significance
or isolate the economic contribution of
dynamic dependence, and it does not alter
the conclusion from the frozen primary
specification.

We also examine the sensitivity of the primary
wealth comparisons to the accounting convention.
The original constant-mix calculation (Track A)
maintains target weights between weekly execution
dates without charging for the trades implied by
these restorations. Track B instead holds positions
between execution dates, while Track C maintains
target weights daily and charges for the implied
trades. Under both weighting methods, all six
reserved-period final wealth values are lower in
Track B and Track C than in Track A. However, the
relative ordering of the three network methods
remains the same under all three conventions.
The revised accounting therefore changes the
levels of final wealth but not the descriptive
conclusion of the primary comparison.

\subsection{Dynamic Dependence Ablation}
\label{subsec:robustness_ablation}

To examine the contribution of multi-step dynamic dependence, we compare the original GFEVD network with an alternative constructed using a one-step forecast horizon. Both networks use the same unrestricted pairwise VAR(1) estimates and innovation covariance matrices, while the original specification uses $H=10$ and the alternative uses $H=1$. At the one-step horizon, the decomposition depends only on contemporaneous innovation covariance and does not include the propagation of shocks through subsequent periods. The comparison therefore isolates the contribution of multi-step propagation to the estimated network. The one-step network is not equivalent to the Pearson networks constructed from stock returns in our primary experiment. Before conducting the comparison, we also reconstructed the original $H=10$ minimum-cost networks and verified that their MSTs and selected portfolios matched the saved checkpoints across all 293 weekly signals.

\begin{table}[htbp]
\centering
\caption{Network Sensitivity to Forecast Horizon and Symmetrization}
\label{tab:robustness_gfevd_variants_overlap}
\small
\setlength{\tabcolsep}{4pt}
\begin{tabular}{lrrrr}
\toprule
Variant &
Edge Jaccard &
Top-5 overlap &
Same Top-5 (\%) &
Weekly retention \\
\midrule
$H=1$, minimum cost & 0.849 & 0.810 & 23.5 & 0.777 \\
$H=10$, mean cost & 0.910 & 0.875 & 44.7 & 0.762 \\
$H=10$, maximum cost & 0.848 & 0.818 & 28.3 & 0.777 \\
\bottomrule
\end{tabular}
\end{table}

Table~\ref{tab:robustness_gfevd_variants_overlap} reports the structural comparisons relative to the original $H=10$ minimum-cost network. The one-step alternative has an average MST edge Jaccard of 0.849 and an average Top-5 selection overlap of 0.810. Although the two networks share a substantial proportion of their structures, none of the 293 weekly MSTs is exactly identical. The same five stocks are selected on 69 weekly signals, representing 23.5\% of the sample, while the complete degree ranking of these five stocks is identical on only 17 signals. These results indicate that multi-step propagation affects the network construction and can change the stocks selected through MST degree centrality, even when the same underlying VAR estimates are used.

The one-step network nevertheless exhibits similar week-to-week selection stability. Its average Top-5 retention is 0.777, compared with 0.779 for the original specification, corresponding to approximately 1.12 and 1.11 stocks replaced per weekly transition, respectively. The difference in forecast horizon therefore changes the estimated networks and selected portfolios without producing a substantial difference in their temporal persistence. The exploratory investment results also depend on the weighting method. During the reserved period, the one-step portfolio achieves lower final wealth under equal weighting but higher final wealth under inverse VaR weighting. These findings show that multi-step propagation affects portfolio construction, but its contribution to realized investment performance is not uniformly positive across the two allocation methods.

\subsection{Alternative Symmetrization Rules}
\label{subsec:robustness_symmetrization}

The original GFEVD network converts two directed dependence measures into an undirected edge cost by taking the minimum of the two directed costs. This is equivalent to selecting the larger of the two normalized directional GFEVD shares. To assess the effect of this choice, we consider two alternative symmetrization rules that use the mean and maximum of the directed costs. Both alternatives retain the original $H=10$ forecast horizon and the same pairwise VAR estimates. The comparison changes only the construction of undirected edge costs and does not introduce a different underlying dependence model.

The results in Table~\ref{tab:robustness_gfevd_variants_overlap} show that the mean-cost network is more similar to the original minimum-cost network than the maximum-cost alternative. The average MST edge Jaccard is 0.910 under mean symmetrization and 0.848 under maximum symmetrization. The corresponding average Top-5 overlaps are 0.875 and 0.818. The mean-cost rule selects exactly the same five stocks as the original specification on 131 of the 293 weekly signals, compared with 83 signals for the maximum-cost rule. Neither alternative produces an MST identical to the original on any weekly signal. The week-to-week Top-5 retention also changes from 0.779 under the original specification to 0.762 under mean symmetrization and 0.777 under maximum symmetrization. These differences indicate that the choice of symmetrization rule affects both the estimated tree structure and the stocks selected from it.

\begin{table}[htbp]
\centering
\caption{Exploratory Portfolio Performance Under Alternative GFEVD Specifications}
\label{tab:robustness_gfevd_variant_performance}
\small
\setlength{\tabcolsep}{4pt}
\begin{tabular}{llrrr}
\toprule
Variant & Weighting &
Final wealth & Sharpe & Costs \\
\midrule
$H=10$, minimum & Equal & 1.2972 & 0.337 & 0.1029 \\
$H=10$, minimum & Inverse VaR & 1.2715 & 0.310 & 0.1104 \\
\midrule
$H=1$, minimum & Equal & 1.2471 & 0.271 & 0.1207 \\
$H=1$, minimum & Inverse VaR & 1.3068 & 0.365 & 0.1303 \\
\midrule
$H=10$, mean & Equal & 1.2328 & 0.250 & 0.1132 \\
$H=10$, mean & Inverse VaR & 1.2382 & 0.262 & 0.1213 \\
\midrule
$H=10$, maximum & Equal & 1.2679 & 0.300 & 0.1137 \\
$H=10$, maximum & Inverse VaR & 1.3105 & 0.370 & 0.1190 \\
\bottomrule
\end{tabular}
\par\smallskip
\begin{minipage}{0.95\linewidth}
\footnotesize Notes: Reserved-period Track B results. Costs are expressed
in units of initial full-sample wealth.
\end{minipage}
\end{table}

Table~\ref{tab:robustness_gfevd_variant_performance} reports the exploratory portfolio results for the reserved period under the same trading calendar, weighting methods, and revised portfolio accounting as the primary experiment. Under equal weighting, all three alternative specifications achieve lower final wealth than the original $H=10$ minimum-cost network. Under inverse VaR weighting, the one-step and maximum-cost alternatives achieve higher final wealth, while the mean-cost alternative remains below the original specification. All three alternatives also produce higher total transaction costs during the reserved period. The differences in final wealth therefore vary with both the network construction rule and the portfolio weighting method.

The full-sample results provide an additional qualification. The one-step alternative achieves higher final wealth than the original specification under both weighting methods over the entire registered sample. However, its equal-weight advantage is concentrated in the diagnostic period and reverses during the reserved period. The mean-cost rule produces lower full-sample final wealth under both weighting methods, and the maximum-cost rule also produces lower full-sample final wealth under both weighting methods. These comparisons demonstrate that the original network construction is sensitive to the forecast horizon and the treatment of directional dependence. They do not establish that an alternative specification provides a more reliable investment strategy. All three variants were examined after the primary results became available, and their realized performance cannot serve as independent out-of-sample evidence or replace the frozen primary specification.

%% file: sections/08_discussion_and_conclustion.tex
\section{Discussion and Conclusion}
\label{sec:discussion_conclusion}

This study examines whether pairwise Generalized Forecast Error Variance
Decomposition (GFEVD) provides additional information for equity portfolio
selection beyond contemporaneous Pearson correlation. We construct dependency
networks using pairwise VAR models and compare their Minimum Spanning Trees
(MSTs) with those obtained from absolute and signed Pearson correlations.
Our theoretical analysis shows that the relationship between GFEVD and
correlation depends on the underlying VAR dynamics. In the absence of
cross-lag relationships, normalized GFEVD is a monotonic function of the
absolute correlation between innovations, and its MST is equivalent to
the spanning tree obtained by maximizing absolute innovation correlations
under the conditions established in Section~\ref{subsec:gfevd_correlation}.
When lagged relationships are present, GFEVD can capture predictive
dependence that contemporaneous correlation does not directly measure.
However, differences in dependence measures do not necessarily produce
different spanning trees or selected portfolios. The stability conditions
developed in Section~\ref{subsec:mst_stability} show that the effect of
changes in edge costs depends on the replacement margins of the tree,
while the effect of edge replacements on stock selection depends on the
degree ranking near the selection cutoff. Therefore, the additional
dependence information captured by GFEVD does not by itself imply an
improvement in subsequent portfolio returns.

The empirical results do not establish a consistent investment advantage
for pairwise GFEVD under the primary specification. With five selected
stocks and a 250-day estimation window, GFEVD achieves higher reserved-period
final wealth than the absolute Pearson benchmark under both equal weighting
and inverse historical VaR weighting. However, it finishes below the signed
Pearson benchmark under both weighting methods. The four wealth comparisons
therefore fail to satisfy our predetermined criterion for incremental
network value. The reported bootstrap results provide
no additional evidence of a consistent investment advantage.
All four 95 percent bootstrap intervals for mean daily return differences
contain zero, and the intersection--union test based on the revised
portfolio accounting produces a $p$-value of 0.6649. We consequently
cannot reject the joint null hypothesis that GFEVD lacks a positive
advantage in at least one of the four primary comparisons. The primary
network portfolios also finish below the broad equal-weight portfolio
and the buy-and-hold benchmark based on vendor-adjusted SPY prices,
although these benchmarks do not impose the same portfolio-size restriction.
Additional exploratory comparisons with alternative portfolio construction
methods likewise do not establish a general advantage for central
GFEVD selection. These findings do not establish that GFEVD and
correlation are equivalent, nor do they rule out an advantage in other
settings. They indicate that the additional dynamic dependence measured
by GFEVD does not translate into a statistically supported improvement
under our primary experimental design.

The sensitivity analysis shows that the relative performance of the
network methods depends on the estimation window and the number of
selected stocks. Under equal weighting, GFEVD achieves higher final
wealth than both Pearson benchmarks for all three portfolio sizes
with a 60-day estimation window and for the ten-stock portfolio with
a 250-day window. These results do not extend to all parameter
combinations. The corresponding inverse historical VaR comparisons
outside the primary specification have not been reconstructed under
the revised accounting procedure, so the four-comparison criterion
cannot be evaluated over the complete parameter grid. The favorable
60-day results are also accompanied by less persistent stock selections
and five GFEVD estimation failures across the full signal calendar.
In the revised equal-weight comparison, all three network methods
omit trading on the three reserved-period GFEVD failure dates.
The positive wealth differences remain under this treatment, although
the results are exploratory because the parameter grid was examined
after the primary evaluation. The estimation-window trade-off developed
in Section~\ref{subsec:time_varying_recovery} provides a theoretical
explanation for why shorter and longer windows may produce different
networks, but the empirical results do not identify an optimal window
or establish that the observed return differences arise from changes
in population network estimation error.

The network comparisons further illustrate the distinction between
measuring dependence and obtaining investment value from that information.
GFEVD and absolute Pearson produce increasingly similar MSTs as the
estimation window becomes longer, while the GFEVD selections also become
more persistent across consecutive weeks. Under the primary specification,
the average MST edge Jaccard between GFEVD and absolute Pearson is 0.813,
but their five-stock selections are exactly identical on only 23.9
percent of the weekly signals. Moreover, the degree gap at the selection
cutoff is frequently zero, so the predetermined tie-breaking rule can
affect which stocks enter the portfolio.

The replacement-margin analysis provides further evidence of the
distinction between network stability and stock selection stability.
Across the three network methods, the sufficient condition for
preserving the MST is not satisfied in any of the 876 comparisons
between consecutive weekly networks. Although the tree changes in
every comparison, the selected five-stock set remains unchanged in
194 cases. These results illustrate that exact stability of the
network is not necessary for stability of the selected stocks.
They also show that the theoretical sufficient conditions can be
conservative when applied to observed changes in financial networks.
The replacement margins and observed cost perturbations describe
the behavior of the estimated networks rather than the recovery
of unknown population networks.

The economic value of a dependency network therefore depends not
only on the information represented by its edges but also on how
that information is converted into portfolio decisions. Greater
network stability does not necessarily correspond to higher portfolio
wealth, and changes in selected stocks do not necessarily improve
returns. The additional horizon and symmetrization comparisons
support this distinction. Both multi-step propagation and the
conversion of directed dependence measures into undirected edge
costs affect the estimated MSTs and selected portfolios. However,
their effects on realized portfolio wealth differ across weighting
methods and do not identify a consistently superior network
specification.

Several limitations should be considered when interpreting these
results. The experiment uses a fixed stock universe whose composition
is determined from the full historical sample rather than from a
point-in-time constituent list. The absence of complete historical
membership and delisting-return records introduces survivorship
and selection concerns that are not removed by using the same
stock universe across network methods. The pairwise GFEVD approach
also estimates each stock pair separately rather than constructing
a joint system for the entire market, so its directional relationships
describe statistical forecast-error dependence rather than identified
causal spillovers. In addition, the portfolio results depend on the
selected VAR specification, forecast horizon, MST degree-ranking rule,
weighting methods, and assumed trading costs. Although the revised
accounting allows portfolio weights to change between weekly
rebalancing dates and calculates transaction costs from the value
of the assets traded, the simulation is still based on vendor-adjusted
closing prices and does not capture every friction associated with
actual transactions. In particular, the proportional transaction-cost
assumption does not explicitly model asset-specific bid--ask spreads,
liquidity constraints, or the market impact of large trades.
 The revised statistical analysis and additional
sensitivity comparisons use a historical period whose portfolio
outcomes have already been examined. They therefore provide
retrospective evidence rather than independent out-of-sample
confirmation.

Future research could examine point-in-time stock universes,
alternative dynamic network measures, different portfolio-selection
rules, and execution assumptions that account more fully for liquidity
and market impact. It would also be useful to investigate whether
particular market conditions favor dynamic dependence measures
and whether changes in network structure can be linked more
directly to subsequent portfolio performance.

Overall, our findings show that pairwise GFEVD can provide a
different representation of stock dependence and generate stock
selections that differ from correlation-based alternatives.
However, differences in dependence information do not necessarily
change the filtered network, changes in the network do not always
affect stock selection, and changes in stock selection do not
guarantee better investment performance. The theoretical and
empirical results together suggest that the value of a dependence
measure should be evaluated not only by the information it captures,
but also by how that information affects the resulting investment
decisions. Under the primary specification examined in this study,
we do not find sufficient evidence that pairwise GFEVD provides
a consistent investment advantage over the correlation-based
alternatives.

%% file: appendix/A_proofs.tex
\section{Proofs for the Theoretical Framework}
\label{app:theory_proofs}

\subsection{Proof of Proposition 1: Correlation Equivalence}

\[
\mathbf A=\operatorname{diag}(a_i,a_j),
\qquad
\boldsymbol\Phi(h)=\operatorname{diag}(a_i^h,a_j^h),
\qquad
\boldsymbol\Sigma=
\begin{pmatrix}
\sigma_{11}&\sigma_{12}\\
\sigma_{12}&\sigma_{22}
\end{pmatrix}.
\]

\begin{equation}
\begin{aligned}
\theta_{1\leftarrow2}(H)
&=
\frac{
\sigma_{22}^{-1}
\displaystyle\sum_{h=0}^{H-1}
(a_i^h\sigma_{12})^2
}{
\displaystyle\sum_{h=0}^{H-1}
a_i^{2h}\sigma_{11}
}
=
\frac{\sigma_{12}^2}
{\sigma_{11}\sigma_{22}}
=
\rho_{\varepsilon,ij}^2,
\\[6pt]
\theta_{1\leftarrow1}(H)
&=
\frac{
\sigma_{11}^{-1}
\displaystyle\sum_{h=0}^{H-1}
(a_i^h\sigma_{11})^2
}{
\displaystyle\sum_{h=0}^{H-1}
a_i^{2h}\sigma_{11}
}
=1.
\end{aligned}
\end{equation}

\begin{equation}
\begin{aligned}
\widetilde\theta_{1\leftarrow2}(H)
&=
\frac{\theta_{1\leftarrow2}(H)}
{\theta_{1\leftarrow1}(H)+\theta_{1\leftarrow2}(H)}
=
\frac{\rho_{\varepsilon,ij}^2}
{1+\rho_{\varepsilon,ij}^2},
\\[4pt]
\widetilde\theta_{2\leftarrow1}(H)
&=
\frac{\rho_{\varepsilon,ij}^2}
{1+\rho_{\varepsilon,ij}^2}.
\end{aligned}
\end{equation}

For the spanning-tree equivalence,
\begin{equation}
\begin{aligned}
d_{ij}
&=
1-
\max\left\{
\widetilde\theta_{1\leftarrow2}(H),
\widetilde\theta_{2\leftarrow1}(H)
\right\}\\
&=
\frac{1}{1+\rho_{\varepsilon,ij}^2}.
\end{aligned}
\end{equation}

\[
|\rho_{\varepsilon,ij}^{(1)}|
>
|\rho_{\varepsilon,ij}^{(2)}|
\quad\Longleftrightarrow\quad
d_{ij}^{(1)}<d_{ij}^{(2)}.
\]

The two edge orderings are therefore equivalent, including their ties.
Kruskal's algorithm can use the same ordering and tie-breaking rule
under either measure. Since every optimal spanning tree can be obtained
by an appropriate ordering of tied edges, the two measures also give
the same set of optimal trees.

For the observed return correlation, we write the return equations
after subtracting their stationary means:
\[
r_{i,t}=a_i r_{i,t-1}+\varepsilon_{i,t},
\qquad
r_{j,t}=a_j r_{j,t-1}+\varepsilon_{j,t}.
\]

\begin{equation}
\begin{aligned}
\operatorname{Var}(r_i)
&=\frac{\sigma_{11}}{1-a_i^2},
&
\operatorname{Var}(r_j)
&=\frac{\sigma_{22}}{1-a_j^2},
\\[4pt]
\operatorname{Cov}(r_i,r_j)
&=\frac{\sigma_{12}}{1-a_i a_j}.
\end{aligned}
\end{equation}

\begin{equation}
\begin{aligned}
\rho_{r,ij}
&=
\frac{
\operatorname{Cov}(r_i,r_j)
}{
\sqrt{
\operatorname{Var}(r_i)
\operatorname{Var}(r_j)
}
}\\
&=
\rho_{\varepsilon,ij}
\frac{
\sqrt{(1-a_i^2)(1-a_j^2)}
}{
1-a_i a_j
}.
\end{aligned}
\end{equation}

At $H=1$,
\[
\boldsymbol\Phi(0)=\mathbf I_2,
\qquad
\widetilde\theta_{1\leftarrow2}(1)
=
\widetilde\theta_{2\leftarrow1}(1)
=
\frac{\rho_{\varepsilon,ij}^2}
{1+\rho_{\varepsilon,ij}^2}.
\]

\subsection{Proof of Proposition 2: Dynamic Dependence}

\[
\boldsymbol\Sigma
=
\operatorname{diag}(\sigma_{11},\sigma_{22}),
\qquad
\phi_{ab}(h)
=
\mathbf e_a^\top\mathbf A^h\mathbf e_b.
\]

\begin{equation}
\begin{aligned}
\mathbf e_1^\top
\boldsymbol\Phi(h)
\boldsymbol\Sigma
\mathbf e_2
&=
\sigma_{22}\phi_{12}(h),
\\[4pt]
V_1(H)
&=
\sigma_{11}
\sum_{h=0}^{H-1}\phi_{11}(h)^2
+
\sigma_{22}
\sum_{h=0}^{H-1}\phi_{12}(h)^2.
\end{aligned}
\end{equation}

Since $\boldsymbol\Sigma$ is diagonal, the sum of the two raw
generalized contributions for variable 1 is one: both numerators
sum to $V_1(H)$. Row normalization therefore leaves the
contributions unchanged, and
\begin{equation}
\widetilde\theta_{1\leftarrow2}(H)
=
\frac{
\sigma_{22}
\displaystyle\sum_{h=0}^{H-1}
\phi_{12}(h)^2
}{
V_1(H)
}.
\end{equation}

\[
\phi_{12}(0)=\mathbf e_1^\top\mathbf e_2=0.
\]

\[
\begin{aligned}
\widetilde\theta_{1\leftarrow2}(H)>0
&\iff
\sum_{h=0}^{H-1}\phi_{12}(h)^2>0\\
&\iff
\exists h\in\{1,\ldots,H-1\}:
\phi_{12}(h)\neq0.
\end{aligned}
\]

For Equation~\eqref{eq:one_way_dynamic_example},
\[
\mathbf A^2=\mathbf0,
\qquad
\boldsymbol\Sigma=\mathbf I_2.
\]

\begin{equation}
\begin{aligned}
\widetilde\theta_{1\leftarrow2}(1)&=0,
\\
\widetilde\theta_{1\leftarrow2}(H)
&=
\frac{\beta^2}{1+\beta^2},
\qquad H\geq2,
\\
\widetilde\theta_{2\leftarrow1}(H)&=0.
\end{aligned}
\end{equation}

\begin{equation}
\begin{aligned}
\operatorname{Cov}(r_{1,t},r_{2,t})
&=
\operatorname{Cov}
\left(
\beta\varepsilon_{2,t-1}
+
\varepsilon_{1,t},
\varepsilon_{2,t}
\right)
=0,
\\[4pt]
\operatorname{Cov}(r_{1,t},r_{2,t-1})
&=\beta.
\end{aligned}
\end{equation}

\subsection{Proof of Proposition 3: MST Stability}

For the unique MST $T$,
\[
d_e>
\max_{f\in P_T(e)}d_f,
\qquad e\notin E(T).
\]

\[
\|\widehat{\mathbf D}-\mathbf D\|_{\max}
\leq\varepsilon,
\]

\begin{equation}
\begin{aligned}
\widehat d_e
-
\max_{f\in P_T(e)}
\widehat d_f
&\geq
d_e-\varepsilon
-
\left(
\max_{f\in P_T(e)}d_f+\varepsilon
\right)
\\
&=
\gamma_e(T,\mathbf D)-2\varepsilon
\\
&\geq
\gamma(T,\mathbf D)-2\varepsilon.
\end{aligned}
\end{equation}

\[
2\varepsilon<\gamma(T,\mathbf D)
\quad\Longrightarrow\quad
\widehat d_e>
\max_{f\in P_T(e)}
\widehat d_f
\quad
\forall e\notin E(T).
\]

By the cycle property, every edge outside $E(T)$ is excluded
from every MST of $\widehat{\mathbf D}$. Since an MST must
contain $N-1$ edges, $T$ remains the unique MST:
\[
\widehat T=T.
\]

For the GFEVD cost transformation,
\begin{equation}
\left|
\max\{x_1,x_2\}
-
\max\{y_1,y_2\}
\right|
\leq
\max\{
|x_1-y_1|,
|x_2-y_2|
\}.
\end{equation}

\begin{equation}
\begin{aligned}
|\widehat d_{ij}-d_{ij}|
\leq
\max\bigl\{
&|\widehat\omega_{j\to i}
-\omega_{j\to i}|,
\\
&|\widehat\omega_{i\to j}
-\omega_{i\to j}|
\bigr\}.
\end{aligned}
\end{equation}

\subsection{Proof of Proposition 4: Top-$k$ Stability}

Let $a_i$ and $b_i$ denote the numbers of added and removed edges incident to stock $i$.

\[
0\leq a_i,b_i\leq s(T,T').
\]

\begin{equation}
\begin{aligned}
g_i(T')-g_i(T)&=a_i-b_i,
\\
|g_i(T')-g_i(T)|
&\leq s(T,T').
\end{aligned}
\end{equation}

For $i\in S_k(T)$ and $j\notin S_k(T)$,
\begin{equation}
\begin{aligned}
g_i(T')-g_j(T')
&\geq
g_i(T)-g_j(T)-2s(T,T')
\\
&\geq
g_{(k)}(T)-g_{(k+1)}(T)
-2s(T,T')
\\
&=
\Delta_k(T)-2s(T,T').
\end{aligned}
\end{equation}

\[
\Delta_k(T)>2s(T,T')
\quad\Longrightarrow\quad
g_i(T')>g_j(T').
\]

\[
S_k(T')=S_k(T).
\]

For the perturbation bound in Section~\ref{subsec:mst_stability},
let $T'$ be an MST of $\mathbf D'$ with
$\|\mathbf D'-\mathbf D\|_{\max}\leq\delta$.
For every $e\notin E(T)$ with $\gamma_e(T,\mathbf D)>2\delta$,
the same argument as in the proof of Proposition 3 gives
\[
d'_e-\max_{f\in P_T(e)}d'_f
\geq\gamma_e(T,\mathbf D)-2\delta>0.
\]
By the cycle property, such an edge cannot enter $T'$.
It follows that
\[
E(T')\setminus E(T)
\subseteq
\{e\notin E(T):\gamma_e(T,\mathbf D)\leq2\delta\},
\]
and hence
\[
s(T,T')=|E(T')\setminus E(T)|
\leq q(T,\mathbf D;\delta).
\]
Applying Proposition 4 gives the sufficient condition
$\Delta_k(T)>2q(T,\mathbf D;\delta)$.
If $q(T,\mathbf D;\delta)=0$, then $T'=T$ and the selected
stock sets are identical under the same tie-breaking rule,
regardless of the degree gap.

\subsection{Proof of Proposition 5: Rolling-Window GFEVD Estimation Error}
\label{app:tv_gfevd_proof}

\begin{proof}

Let
\[
\mathcal I_{t,n}=\{t-n+1,\ldots,t\}.
\]

For every pair $p$ and $s\in\mathcal I_{t,n}$, assume
\begin{equation}
\|\mathbf z_{s-1}^{(p)}\|_\infty\leq K,
\qquad
\|\mathbf u_s^{(p)}\|_\infty\leq U,
\label{eq:app_uniform_bounds}
\end{equation}
and
\[
\lambda I_2
\preceq
\boldsymbol\Sigma_{p,s}
\preceq
\Lambda I_2,
\qquad
0<\lambda\leq\Lambda<\infty.
\]

All sums below are over $s\in\mathcal I_{t,n}$.

\begin{equation}
\widehat{\mathbf B}_{p,t}(n)
=
\left(
\frac{1}{n}
\sum_s
\mathbf x_s^{(p)}
\mathbf z_{s-1}^{(p)\top}
\right)
\mathbf G_{p,t}^{-1},
\label{eq:app_rolling_ols}
\end{equation}

\[
\mathbf G_{p,t}
=
\frac{1}{n}
\sum_s
\mathbf z_{s-1}^{(p)}
\mathbf z_{s-1}^{(p)\top}.
\]

\begin{equation}
\begin{aligned}
\widehat{\mathbf B}_{p,t}(n)-\mathbf B_{p,t}
={}&
\left[
\frac{1}{n}\sum_s
\mathbf u_s^{(p)}
\mathbf z_{s-1}^{(p)\top}
\right]\mathbf G_{p,t}^{-1}\\
&+
\left[
\frac{1}{n}\sum_s
(\mathbf B_{p,s}-\mathbf B_{p,t})
\mathbf z_{s-1}^{(p)}
\mathbf z_{s-1}^{(p)\top}
\right]\mathbf G_{p,t}^{-1}.
\end{aligned}
\label{eq:app_tv_ols_error}
\end{equation}

Since
\[
\mathbb E[
\mathbf u_s^{(p)}
\mid\mathcal F_{s-1}
]=0,
\qquad
\mathbf z_{s-1}^{(p)}
\in\mathcal F_{s-1},
\]
the Azuma--Hoeffding inequality and a union bound over
the $6M$ coefficient entries give
\begin{equation}
\begin{aligned}
\mathbb P\Bigg(
\max_p
\left\|
\frac{1}{n}\sum_s
\mathbf u_s^{(p)}
\mathbf z_{s-1}^{(p)\top}
\right\|_2
>
\sqrt{6}UK
\sqrt{\frac{2\log(24M/\alpha)}{n}}
\Bigg)
\leq\frac{\alpha}{2}.
\end{aligned}
\label{eq:app_tv_sampling_bound}
\end{equation}

The drift condition implies
\begin{equation}
\begin{aligned}
&
\left\|
\frac{1}{n}\sum_s
(\mathbf B_{p,s}-\mathbf B_{p,t})
\mathbf z_{s-1}^{(p)}
\mathbf z_{s-1}^{(p)\top}
\right\|_2\\
&\qquad\leq
\frac{3K^2L_B}{n}
\sum_{h=0}^{n-1}h\\
&\qquad=
\frac{3K^2L_B(n-1)}{2}.
\end{aligned}
\label{eq:app_tv_drift_bound}
\end{equation}

On the event
\[
\min_p\lambda_{\min}(\mathbf G_{p,t})\geq g,
\]
we have
\[
\max_p\|\mathbf G_{p,t}^{-1}\|_2\leq g^{-1}.
\]

Consequently,
\begin{equation}
\max_p
\|\widehat{\mathbf B}_{p,t}(n)-\mathbf B_{p,t}\|_2
\leq r_B(n,\alpha),
\label{eq:app_tv_B_bound}
\end{equation}

\begin{equation}
\begin{aligned}
r_B(n,\alpha)
={}&
\frac{\sqrt{6}UK}{g}
\sqrt{\frac{2\log(24M/\alpha)}{n}}\\
&+
\frac{3K^2L_B(n-1)}{2g}.
\end{aligned}
\label{eq:app_tv_rB}
\end{equation}

The estimated residual satisfies
\[
\widehat{\mathbf u}_s^{(p)}
=
\mathbf u_s^{(p)}+\mathbf e_s^{(p)},
\]

\begin{equation}
\mathbf e_s^{(p)}
=
\left(
\mathbf B_{p,s}
-
\widehat{\mathbf B}_{p,t}(n)
\right)
\mathbf z_{s-1}^{(p)}.
\label{eq:app_residual_perturbation}
\end{equation}

\[
\begin{aligned}
\|\mathbf e_s^{(p)}\|_2
&\leq
\|\mathbf B_{p,s}-\mathbf B_{p,t}\|_2
\|\mathbf z_{s-1}^{(p)}\|_2\\
&\quad+
\|\mathbf B_{p,t}-\widehat{\mathbf B}_{p,t}(n)\|_2
\|\mathbf z_{s-1}^{(p)}\|_2\\
&\leq
\sqrt{3}K
\left[
r_B(n,\alpha)+L_B(n-1)
\right].
\end{aligned}
\]

Define
\begin{equation}
E_n
=
\sqrt{3}K
\left[
r_B(n,\alpha)+L_B(n-1)
\right].
\label{eq:app_tv_En}
\end{equation}

The residual covariance estimator is
\begin{equation}
\widehat{\boldsymbol\Sigma}_{p,t}
=
\frac{1}{n}
\sum_s
\widehat{\mathbf u}_s^{(p)}
\widehat{\mathbf u}_s^{(p)\top}.
\label{eq:app_tv_cov_estimator}
\end{equation}

\[
\begin{aligned}
\widehat{\boldsymbol\Sigma}_{p,t}
-\boldsymbol\Sigma_{p,t}
={}&
\frac{1}{n}\sum_s
\left(
\mathbf u_s^{(p)}
\mathbf u_s^{(p)\top}
-\boldsymbol\Sigma_{p,s}
\right)\\
&+
\frac{1}{n}\sum_s
\left(
\boldsymbol\Sigma_{p,s}
-\boldsymbol\Sigma_{p,t}
\right)\\
&+
\frac{1}{n}\sum_s
\left(
\widehat{\mathbf u}_s^{(p)}
\widehat{\mathbf u}_s^{(p)\top}
-
\mathbf u_s^{(p)}
\mathbf u_s^{(p)\top}
\right).
\end{aligned}
\]

Since
\[
\mathbb E[
\mathbf u_s^{(p)}
\mathbf u_s^{(p)\top}
\mid\mathcal F_{s-1}
]
=
\boldsymbol\Sigma_{p,s},
\]
the Azuma--Hoeffding inequality and a union bound over
the $4M$ covariance entries give
\begin{equation}
\begin{aligned}
\mathbb P\Bigg(
\max_p
\left\|
\frac{1}{n}\sum_s
\left(
\mathbf u_s^{(p)}
\mathbf u_s^{(p)\top}
-
\boldsymbol\Sigma_{p,s}
\right)
\right\|_2
>
2(U^2+\Lambda)
\sqrt{\frac{2\log(16M/\alpha)}{n}}
\Bigg)
\leq\frac{\alpha}{2}.
\end{aligned}
\label{eq:app_tv_cov_sampling}
\end{equation}

\begin{equation}
\begin{aligned}
\left\|
\frac{1}{n}\sum_s
\boldsymbol\Sigma_{p,s}
-\boldsymbol\Sigma_{p,t}
\right\|_2
&\leq
\frac{L_\Sigma}{n}
\sum_{h=0}^{n-1}h\\
&=
\frac{L_\Sigma(n-1)}{2}.
\end{aligned}
\label{eq:app_tv_cov_drift}
\end{equation}

\begin{equation}
\begin{aligned}
&
\widehat{\mathbf u}_s^{(p)}
\widehat{\mathbf u}_s^{(p)\top}
-
\mathbf u_s^{(p)}
\mathbf u_s^{(p)\top}\\
&=
\mathbf u_s^{(p)}
\mathbf e_s^{(p)\top}
+
\mathbf e_s^{(p)}
\mathbf u_s^{(p)\top}
+
\mathbf e_s^{(p)}
\mathbf e_s^{(p)\top}.
\end{aligned}
\label{eq:app_residual_covariance_expansion}
\end{equation}

\[
\begin{aligned}
&
\left\|
\widehat{\mathbf u}_s^{(p)}
\widehat{\mathbf u}_s^{(p)\top}
-
\mathbf u_s^{(p)}
\mathbf u_s^{(p)\top}
\right\|_2\\
&\qquad\leq
2\|\mathbf u_s^{(p)}\|_2
\|\mathbf e_s^{(p)}\|_2
+
\|\mathbf e_s^{(p)}\|_2^2\\
&\qquad\leq
2\sqrt{2}UE_n+E_n^2.
\end{aligned}
\]

\begin{equation}
\max_p
\|
\widehat{\boldsymbol\Sigma}_{p,t}
-
\boldsymbol\Sigma_{p,t}
\|_2
\leq r_\Sigma(n,\alpha),
\label{eq:app_tv_sigma_bound}
\end{equation}

\begin{equation}
\begin{aligned}
r_\Sigma(n,\alpha)
={}&
2(U^2+\Lambda)
\sqrt{\frac{2\log(16M/\alpha)}{n}}\\
&+
\frac{L_\Sigma(n-1)}{2}
+
2\sqrt{2}UE_n
+
E_n^2.
\end{aligned}
\label{eq:app_tv_rSigma}
\end{equation}

The two concentration events each have failure probability at
most $\alpha/2$, while the regressor-invertibility event has
failure probability at most $\eta$. By a union bound, the
coefficient and covariance bounds hold jointly with probability
at least $1-\alpha-\eta$ for the given formation date and
window length. No independence across stock pairs is required.

Let
\[
F_H(\mathbf A,\boldsymbol\Sigma)
=
\widetilde{\boldsymbol\Theta}(H).
\]

The population parameters belong to the compact set
$\mathcal K$. Define

\[
\operatorname{dist}(\vartheta,\mathcal K)
=
\inf_{(\mathbf A',\boldsymbol\Sigma')\in\mathcal K}
\left(
\|\mathbf A-\mathbf A'\|_2
+
\|\boldsymbol\Sigma-\boldsymbol\Sigma'\|_2
\right),
\qquad
\vartheta=(\mathbf A,\boldsymbol\Sigma).
\]

\[
\mathcal K_{\varepsilon_0}
=
\left\{
\vartheta:
\operatorname{dist}(\vartheta,\mathcal K)
\leq\varepsilon_0
\right\}.
\]

For every
$(\mathbf A,\boldsymbol\Sigma)\in
\mathcal K_{\varepsilon_0}$,
there exist uniform constants
$A_*<\infty$, $0<\lambda_*\leq\Lambda_*<\infty$
such that
\[
\|\mathbf A\|_2\leq A_*,
\qquad
\lambda_*I_2
\preceq\boldsymbol\Sigma
\preceq\Lambda_*I_2.
\]

For $a\in\{1,2\}$, let
\[
D_a(H)
=
\sum_{h=0}^{H-1}
\mathbf e_a^\top
\mathbf A^h\boldsymbol\Sigma
(\mathbf A^h)^\top
\mathbf e_a.
\]

\[
\begin{aligned}
0<\lambda_*
&\leq D_a(H)\\
&\leq
\Lambda_*
\sum_{h=0}^{H-1}A_*^{2h}
=:D_*<\infty.
\end{aligned}
\]

\[
\begin{aligned}
\sum_{b=1}^2
\theta_{a\leftarrow b}(H)
&\geq
\theta_{a\leftarrow a}(H)\\
&\geq
\frac{\lambda_*}{D_*}>0.
\end{aligned}
\]

For fixed $H$, $F_H$ is continuously differentiable on
$\mathcal K_{\varepsilon_0}$, with
\[
C_H
=
\sup_{\vartheta\in\mathcal K_{\varepsilon_0}}
\|DF_H(\vartheta)\|
<\infty.
\]

Under
\[
r_B(n,\alpha)+r_\Sigma(n,\alpha)
\leq\varepsilon_0,
\]
the estimated parameters and the line segments joining
them to the population parameters lie in
$\mathcal K_{\varepsilon_0}$ on the concentration and
regressor-invertibility events.

\begin{equation}
\begin{aligned}
&
\left\|
F_H(
\widehat{\mathbf A}_{p,t},
\widehat{\boldsymbol\Sigma}_{p,t}
)
-
F_H(
\mathbf A_{p,t},
\boldsymbol\Sigma_{p,t}
)
\right\|_2\\
&\qquad\leq
C_H
\left[
r_B(n,\alpha)+r_\Sigma(n,\alpha)
\right].
\end{aligned}
\label{eq:app_tv_gfevd_lipschitz}
\end{equation}

For the symmetric edge cost,
\[
d_{ij,t}
=
1-\max\{
\omega_{j\to i,t},
\omega_{i\to j,t}
\},
\]
and
\[
|\max\{a,b\}-\max\{c,d\}|
\leq
\max\{|a-c|,|b-d|\}.
\]

\begin{equation}
\|\widehat{\mathbf D}_t(n)-\mathbf D_t\|_{\max}
\leq
C_H
\left[
r_B(n,\alpha)+r_\Sigma(n,\alpha)
\right].
\label{eq:app_tv_cost_bound}
\end{equation}

Let
\[
S_n
=
\sqrt{\frac{\log(24M/\alpha)}{n}}.
\]

\begin{equation}
\begin{aligned}
C_H
\left[
r_B(n,\alpha)+r_\Sigma(n,\alpha)
\right]
\leq{}&
C_1S_n\\
&+
C_2(L_B+L_\Sigma)(n-1)\\
&+
C_3
\left[
S_n+L_B(n-1)
\right]^2.
\end{aligned}
\label{eq:app_tv_final_bound}
\end{equation}

The constants $C_1,C_2,C_3$ depend only on the
uniform parameter bounds, $g$, and $C_H$.

\[
\mathbb P\left(
\|\widehat{\mathbf D}_t(n)-\mathbf D_t\|_{\max}
\leq\delta_n
\right)
\geq1-\alpha-\eta.
\]

\end{proof}

\subsection{Derivation of the Window-Length Benchmark}
\label{app:tv_window_derivation}

We now derive the window-length benchmark introduced in Section~\ref{subsec:time_varying_recovery}. The leading terms in the network estimation-error bound are
\begin{equation}
f(n)
=
C_1
\sqrt{
\frac{\log(24M/\alpha)}{n}
}
+
C_2(L_B+L_\Sigma)n.
\label{eq:app_tv_window_objective}
\end{equation}
For simplicity, define $L=L_B+L_\Sigma$ and $a=\log(24M/\alpha)$. When $L>0$, we treat $n$ as a continuous variable and write
\begin{equation}
f(n)
=
C_1\sqrt{a}\,n^{-1/2}
+
C_2Ln.
\end{equation}
The first derivative is
\begin{equation}
f'(n)
=
-\frac{C_1\sqrt{a}}{2n^{3/2}}
+
C_2L.
\end{equation}
We set the derivative equal to zero and obtain
\begin{equation}
n^{3/2}
=
\frac{C_1\sqrt{a}}{2C_2L}.
\end{equation}
The second derivative is
\begin{equation}
f''(n)
=
\frac{3C_1\sqrt{a}}{4n^{5/2}}
>0.
\end{equation}
Therefore, the stationary point is the unique minimizer of the leading estimation-error bound:
\begin{equation}
n_{\mathrm{bound}}^*
=
\left(
\frac{
C_1^2\log(24M/\alpha)
}{
4C_2^2(L_B+L_\Sigma)^2
}
\right)^{1/3}.
\label{eq:app_tv_optimal_window}
\end{equation}
If $L_B+L_\Sigma=0$, the leading bound decreases as $n$ increases, so its minimum over a feasible set of window lengths occurs at the largest admissible window.

The expression above minimizes only the leading terms in the estimation-error bound. The complete bound in Equation~\eqref{eq:tv_network_error_components} also contains a quadratic component, so the continuous solution does not necessarily minimize the full finite-sample expression. In empirical applications, the window length must also be an integer and must remain within the available historical sample. The formula therefore provides a theoretical benchmark rather than a directly implementable optimal investment window.

We further examine the behavior of the bound when the population parameters change smoothly over rescaled time $t/T_{\mathrm{obs}}$. Suppose the parameter-drift rates satisfy
\begin{equation}
L_B+L_\Sigma
\asymp T_{\mathrm{obs}}^{-1}.
\end{equation}
For sequences satisfying $\log M=o(n)$ and $n=o(T_{\mathrm{obs}})$, the quadratic remainder becomes asymptotically smaller than the leading terms. If the uniform regressor-matrix event also has probability approaching one, the network estimation-error rate is
\begin{equation}
\delta_{n,T_{\mathrm{obs}}}
=
O_p\left(
\sqrt{\frac{\log M}{n}}
+
\frac{n}{T_{\mathrm{obs}}}
\right).
\label{eq:app_tv_asymptotic_rate}
\end{equation}
Balancing the sampling and parameter-drift components gives
\begin{equation}
n_{\mathrm{bound}}^*
\asymp
(T_{\mathrm{obs}}^2\log M)^{1/3}.
\end{equation}
If $\log M=o(T_{\mathrm{obs}})$, this benchmark satisfies $n_{\mathrm{bound}}^*/T_{\mathrm{obs}}\to0$, and the corresponding leading estimation-error rate becomes
\begin{equation}
O_p\left(
\left(
\frac{\log M}{T_{\mathrm{obs}}}
\right)^{1/3}
\right).
\label{eq:app_tv_optimal_rate}
\end{equation}
These rates describe the behavior of the theoretical network estimation-error bound. They do not identify a finite-sample window that maximizes portfolio returns or the Sharpe ratio. Exact recovery of the MST also requires the estimation error to be sufficiently small relative to the population replacement margin. Therefore, convergence of the network estimation error alone is not sufficient to guarantee exact recovery when the replacement margin may approach zero.

\subsection{Proof of Proposition 6: Diversification Bound}

For $w_i\geq0$ and $\sum_iw_i=1$,
\begin{equation}
\begin{aligned}
\operatorname{Var}(R_p)
&=
\sum_iw_i^2
\operatorname{Var}(R_i)
+
\sum_{i\neq j}
w_iw_j
\operatorname{Cov}(R_i,R_j)
\\
&\leq
\bar v\sum_iw_i^2
+
\bar c\sum_{i\neq j}
w_iw_j
\\
&=
\bar v\sum_iw_i^2
+
\bar c
\left(
1-\sum_iw_i^2
\right)
\\
&=
\bar c+
(\bar v-\bar c)\sum_iw_i^2.
\end{aligned}
\end{equation}

For $w_i=1/k$,
\begin{equation}
\begin{aligned}
\sum_iw_i^2
&=
k\left(\frac1k\right)^2
=
\frac1k,
\\[4pt]
\operatorname{Var}(R_p)
&\leq
\bar c+
\frac{\bar v-\bar c}{k}.
\end{aligned}
\end{equation}

By Cauchy--Schwarz,
\begin{equation}
\begin{aligned}
\left(\sum_iw_i\right)^2
&\leq k\sum_iw_i^2,
\\
\sum_iw_i^2
&\geq\frac1k.
\end{aligned}
\end{equation}

Since the weights are nonnegative and sum to one,
\[
\sum_iw_i^2\leq\left(\sum_iw_i\right)^2=1.
\]

\begin{equation}
1
\leq
k_{\mathrm{eff}}
=
\frac1{\sum_iw_i^2}
\leq k.
\end{equation}

For common bounds $\bar v$ and $\bar c$ across the portfolios
being compared, the equal-weight variance bound is nonincreasing
in $k$, and strictly decreasing if $\bar c<\bar v$. This does not
imply that the actual variance decreases with the number of holdings.

%% file: appendix/B_additional_results.tex
\section{Additional Results}
\label{app:additional_results}

\subsection{Illustrative MST Network Structures}
\label{app:mst_snapshot}

To provide a visual illustration of the
structural similarities and differences
among the three network methods, we compare
their MSTs for the completed-week signal
dated January 5, 2024.

\begin{figure}[htbp]
\centering
\includegraphics[width=\linewidth]
{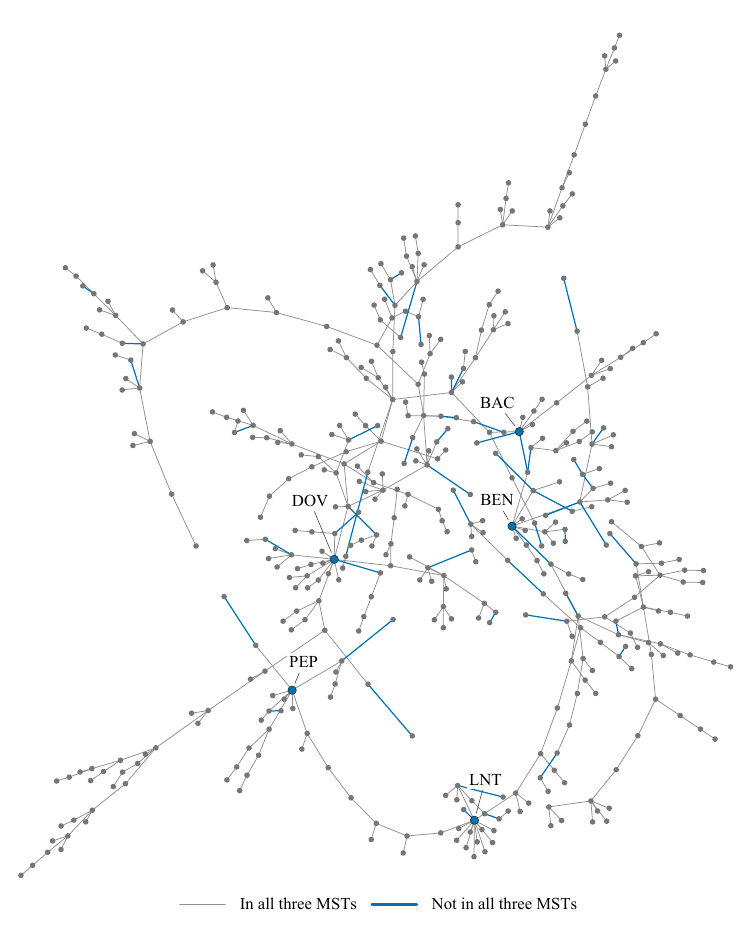}
\caption{Pairwise GFEVD MST for the
completed-week signal dated January 5, 2024.
The five selected stocks are LNT, DOV,
BEN, PEP, and BAC.}
\label{fig:appendix_mst_snapshot}
\end{figure}

\begin{figure}[htbp]
\centering
\includegraphics[width=\linewidth]
{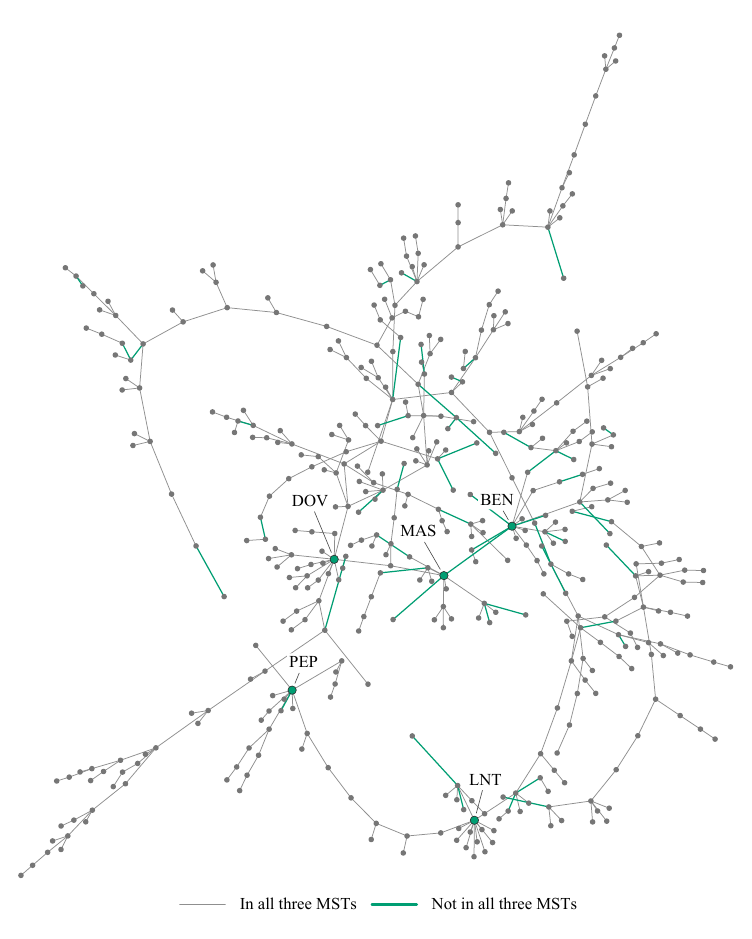}
\caption{Signed Pearson MST for the
completed-week signal dated January 5, 2024.
The five selected stocks are LNT, DOV,
BEN, PEP, and MAS.}
\label{fig:appendix_mst_snapshot_signed}
\end{figure}

Figures~\ref{fig:appendix_mst_snapshot}--%
\ref{fig:appendix_mst_snapshot_signed}
present the three networks using identical
node positions and graphical scales.
Each tree contains 451 stocks and 450 edges.
The 398 edges common to all three MSTs
are shown in gray, while the remaining
edges are highlighted in the corresponding
network color. Larger nodes indicate the
five stocks selected by the primary
degree-ranking rule. Node positions are
used only for visualization and do not
represent economic distances.

\begin{figure}[htbp]
\centering
\includegraphics[width=\linewidth]
{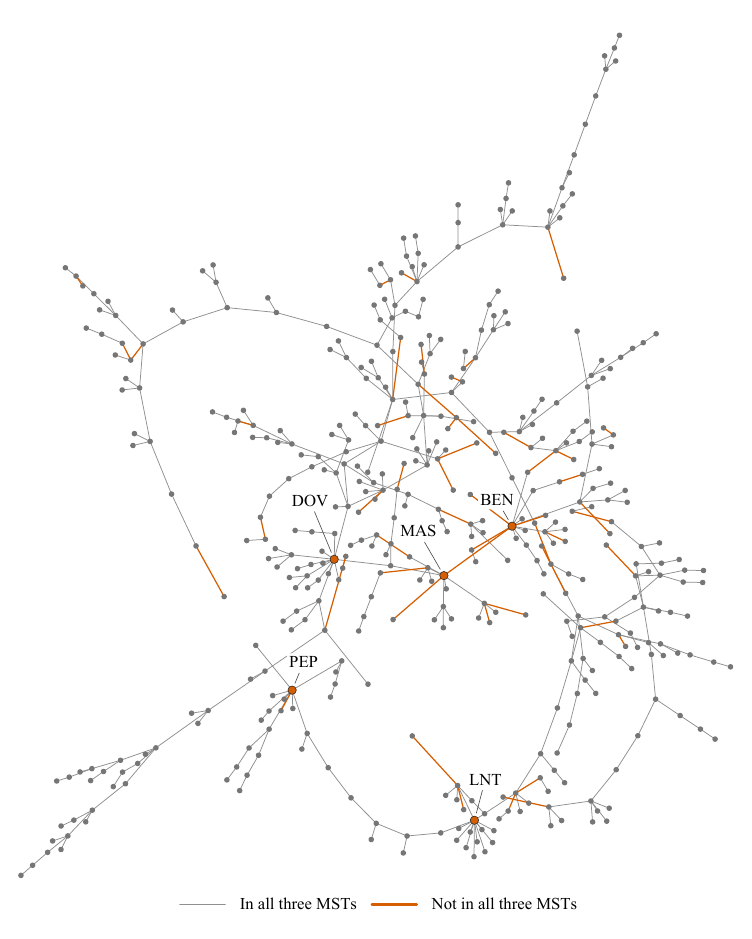}
\caption{Absolute Pearson MST for the
completed-week signal dated January 5, 2024.
The five selected stocks are LNT, DOV,
BEN, PEP, and MAS.}
\label{fig:appendix_mst_snapshot_absolute}
\end{figure}

On this signal date, the three methods
identify largely similar central stocks,
with GFEVD selecting BAC instead of MAS.
Although the three MSTs share most of
their edges, the differences in network
structure can affect the selected stocks.
These figures illustrate one realized
network configuration and should not
be interpreted as representative of
all weekly signals.

\clearpage

\subsection{Alternative Portfolio Accounting Conventions}
\label{app:accounting_comparison}

We compare the three portfolio accounting conventions discussed in
Section~\ref{subsec:empirical_trading}. Track A is the original
constant-mix calculation, which applies the target weights to daily
returns but does not charge for the implied restorations between
scheduled executions. Track B, used for the main empirical results,
allows the portfolio weights to drift between weekly executions and
calculates transaction costs from the drifted positions. Track C
restores the target weights at each daily close and charges for these
trades. On a scheduled execution date, Track C trades directly from
the drifted positions to the new target weights, without a separate
restoration to the previous target. All three calculations use the
same saved weekly target weights and ten-basis-point cost rate.

\begin{table}[htbp]
\centering
\caption{Portfolio Accounting Comparison During the Reserved Period}
\label{tab:appendix_accounting_comparison}
\small
\setlength{\tabcolsep}{4pt}
\begin{tabular}{llrrrrrr}
\toprule
& & \multicolumn{3}{c}{Final wealth} &
\multicolumn{3}{c}{Recorded costs} \\
\cmidrule(lr){3-5}\cmidrule(lr){6-8}
Network & Weighting & A & B & C & A & B & C \\
\midrule
GFEVD & Equal & 1.3035 & 1.2972 & 1.2948 & 0.0992 & 0.1029 & 0.1092 \\
Absolute Pearson & Equal & 1.2462 & 1.2370 & 1.2379 & 0.1015 & 0.1050 & 0.1114 \\
Signed Pearson & Equal & 1.3135 & 1.3040 & 1.3048 & 0.1033 & 0.1068 & 0.1132 \\
\midrule
GFEVD & Inverse VaR & 1.2740 & 1.2715 & 1.2674 & 0.1098 & 0.1104 & 0.1173 \\
Absolute Pearson & Inverse VaR & 1.2697 & 1.2649 & 1.2632 & 0.1137 & 0.1139 & 0.1214 \\
Signed Pearson & Inverse VaR & 1.3215 & 1.3169 & 1.3148 & 0.1156 & 0.1158 & 0.1233 \\
\bottomrule
\end{tabular}
\end{table}

Table~\ref{tab:appendix_accounting_comparison} reports final wealth
rebased to one at the December 29, 2023 close. Recorded costs are
summed over the reserved period and expressed in units of the original
full-sample initial wealth. The wealth values are lower under both
Track B and Track C than under Track A for all six portfolios.
However, under each accounting convention, GFEVD finishes above
absolute Pearson and below signed Pearson under both weighting methods.
The accounting choice therefore affects the level of portfolio wealth
without changing the signs of the four primary comparisons.

For completeness, Table~\ref{tab:appendix_track_a_performance}
retains the performance measures from the original constant-mix
calculation. These measures describe returns after recorded costs,
not returns after charging for every implied restoration of the
target weights. They are not the corrected performance measures
reported in Section~\ref{subsec:results_primary}.

\begin{table}[htbp]
\centering
\caption{Original Track A Portfolio Performance During the Reserved Period}
\label{tab:appendix_track_a_performance}
\small
\begin{tabular}{llrrrr}
\toprule
Network & Weighting & Final wealth & Ann. return (\%) &
Sharpe & Max DD (\%) \\
\midrule
GFEVD & Equal & 1.3035 & 9.88 & 0.346 & -27.06 \\
Absolute Pearson & Equal & 1.2462 & 8.14 & 0.271 & -26.83 \\
Signed Pearson & Equal & 1.3135 & 10.18 & 0.364 & -26.83 \\
\midrule
GFEVD & Inverse VaR & 1.2740 & 8.99 & 0.314 & -25.84 \\
Absolute Pearson & Inverse VaR & 1.2697 & 8.86 & 0.311 & -26.20 \\
Signed Pearson & Inverse VaR & 1.3215 & 10.42 & 0.386 & -26.20 \\
\bottomrule
\end{tabular}
\end{table}

\subsection{Original Parameter and Bootstrap Sensitivity Results}
\label{app:track_a_sensitivity}

We retain the original parameter grid to distinguish the historical
Track A results from the revised equal-weight comparisons in
Section~\ref{subsec:results_parameter_sensitivity}. Under Track A,
both equal weighting and inverse historical VaR weighting were
examined for each combination of $W$ and $k$. The minimum wealth gap
is the smallest of the four GFEVD-minus-Pearson final-wealth differences.
A positive value indicates that all four comparisons favor GFEVD.

\begin{table}[htbp]
\centering
\caption{Original Track A Portfolio-Size and Estimation-Window Sensitivity}
\label{tab:appendix_track_a_grid}
\small
\begin{tabular}{rrrrc}
\toprule
Window $W$ & Size $k$ &
Minimum gap & IUT $p$-value &
All four positive \\
\midrule
60 & 3 & 0.2649 & 0.2388 & Yes \\
60 & 5 & 0.6417 & 0.0850 & Yes \\
60 & 10 & 0.1074 & 0.3063 & Yes \\
\midrule
120 & 3 & -0.1394 & 0.7730 & No \\
120 & 5 & -0.0513 & 0.6255 & No \\
120 & 10 & -0.0396 & 0.6425 & No \\
\midrule
250 & 3 & -0.0556 & 0.6829 & No \\
250 & 5 & -0.0475 & 0.6717 & No \\
250 & 10 & 0.1115 & 0.1934 & Yes \\
\bottomrule
\end{tabular}
\end{table}

As shown in Table~\ref{tab:appendix_track_a_grid}, four of the nine
original parameter combinations satisfy the descriptive four-comparison
criterion. None rejects the intersection--union null at the five percent
level using a block length of ten trading sessions. The original
$W=60$, $k=5$ comparison produces a $p$-value of 0.0384 when the block
length is increased to 63 sessions, compared with 0.0850 at ten sessions.
This result depends on the block length and on an exploratory parameter
choice. It does not provide independent confirmation of incremental
investment value.

\begin{table}[htbp]
\centering
\caption{Original Track A Bootstrap Block-Length Sensitivity}
\label{tab:appendix_track_a_bootstrap}
\begin{tabular}{rr}
\toprule
Block length & IUT $p$-value \\
\midrule
5 & 0.6619 \\
10 & 0.6717 \\
21 & 0.6631 \\
63 & 0.6519 \\
\bottomrule
\end{tabular}
\end{table}

Table~\ref{tab:appendix_track_a_bootstrap} reports the original
block-length sensitivity for the primary specification. The results
in both tables are retained as historical records of Track A. They
cannot be transferred to Track B because the underlying wealth paths
and trading costs differ. In particular, the inverse historical VaR
target weights outside the primary parameter specification were not
stored and have not been reconstructed for the revised parameter
grid. The corresponding four-comparison tests under Track B therefore
remain unavailable.

\subsection{Additional Exploratory Comparisons}
\label{app:exploratory_comparisons}

The historical portfolio comparisons in
Section~\ref{subsec:results_primary} are supplemented by a small
set of alternative selection and allocation rules. During the
reserved period, the vendor-adjusted SPY buy-and-hold benchmark
and the equal-weight portfolio containing all 451 stocks achieve
final wealth of 1.6780 and 1.5090, respectively, under Track B.
The low-degree portfolios finish with wealth between 0.7869 and
0.8008. The primary high-degree portfolios finish above these
low-degree portfolios but below the two broader benchmarks.
These comparisons differ in their selection rules or portfolio
sizes and do not separately identify the contribution of the
dependence measure. The broad and low-degree portfolios use the
same ex-post stock universe, and the SPY series is based on
vendor-adjusted prices rather than an official total-return index.

We also conducted separate exploratory comparisons in simulated
markets using a risk-and-trading-cost objective. The archived
comparisons report lower values of this objective for peripheral
selection, a cardinality-constrained minimum-variance search,
hierarchical risk parity, equal risk contribution, minimum-variance
portfolios using sample and Ledoit--Wolf covariance estimates,
and equal-weight allocation than for central GFEVD selection.
Random selection and the long-only mean--variance comparison
produce higher values of the objective. The associated
intersection--union tests based on loss and certainty equivalent
do not reject in favor of GFEVD. An additional cardinality-grid
comparison is excluded from the maximum-ten-stock comparison
because it permits more than ten holdings.

These simulation results are exploratory and are not pooled with
the historical portfolio results. The simulation protocols,
outputs, and the adaptive-cardinality experiments are retained
in the reproducibility records rather than treated as additional
main-text contributions. The comparisons do not establish that
central GFEVD selection is generally preferable to alternative
portfolio construction methods.

\subsection{Relative Portfolio Wealth}
\label{app:relative_wealth}

To provide a more detailed comparison of
the relative performance of the network
methods, we report the differences between
the cumulative wealth of GFEVD and those
of the two Pearson benchmarks during the
reserved evaluation period.

\begin{figure}[htbp]
\centering
\includegraphics[width=\linewidth]
{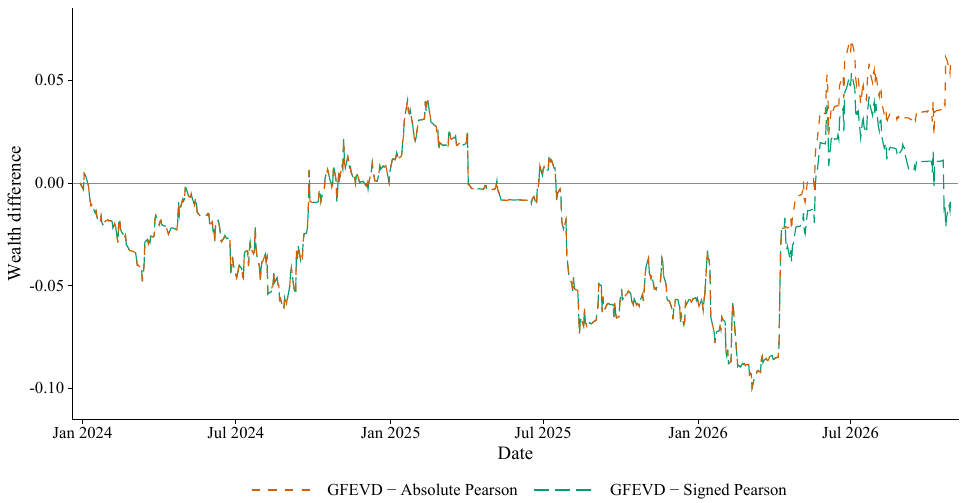}
\caption{Cumulative wealth differences between
GFEVD and the two Pearson benchmarks under
equal weighting. Positive values indicate
higher cumulative wealth for GFEVD, while
negative values indicate higher cumulative
wealth for the corresponding Pearson benchmark.
All differences are calculated from the
rebased after-cost wealth paths.}
\label{fig:appendix_wealth_diff_equal}
\end{figure}

\begin{figure}[htbp]
\centering
\includegraphics[width=\linewidth]
{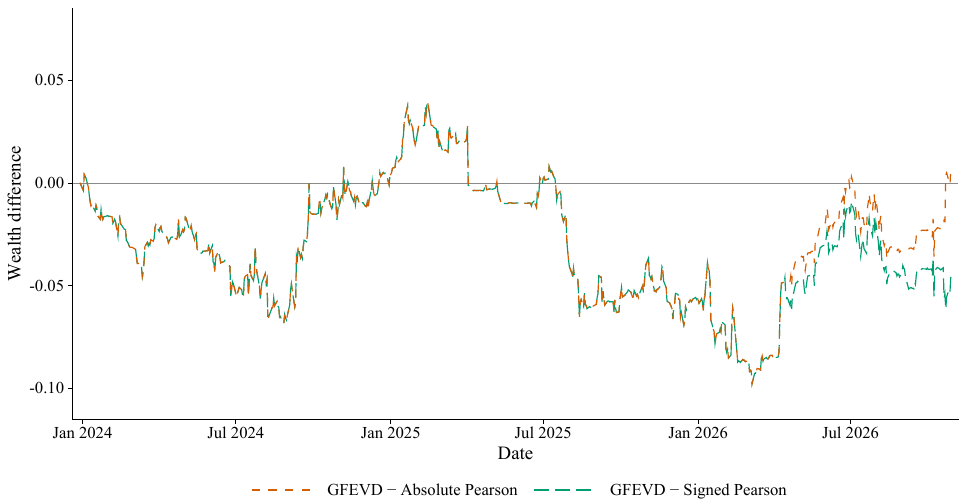}
\caption{Cumulative wealth differences between
GFEVD and the two Pearson benchmarks under
inverse historical VaR weighting. Positive
values indicate higher cumulative wealth
for GFEVD, while negative values indicate
higher cumulative wealth for the corresponding
Pearson benchmark. All differences are
calculated from the rebased after-cost
wealth paths.}
\label{fig:appendix_wealth_diff_iv}
\end{figure}

Figures~\ref{fig:appendix_wealth_diff_equal}
and~\ref{fig:appendix_wealth_diff_iv}
show that the relative wealth differences
vary over time under both weighting methods.
Although GFEVD finishes ahead of absolute
Pearson in both cases, its relative wealth
advantage is not maintained throughout the
evaluation period. These comparisons provide
a descriptive illustration of the wealth
paths rather than a separate statistical
test of investment performance.

\subsection{Additional Stability and Tie-Breaking Results}
\label{app:additional_stability}

We first examine the stability conditions under the
hypothetical perturbation radii described in
Section~\ref{subsec:empirical_network_stability}.
For each estimated cost matrix, the perturbation
radius is defined as
\[
\delta=a\,\operatorname{IQR}(\widehat{\mathbf D}_t).
\]
Table~\ref{tab:appendix_perturbation_certificates}
reports the corresponding replacement counts
$q(\delta)$ and the frequency with which the
sufficient conditions for MST and Top-5 stability
are satisfied.

\begin{table}[htbp]
\centering
\caption{Stability Conditions Under Hypothetical
Cost Perturbations}
\label{tab:appendix_perturbation_certificates}
\small
\setlength{\tabcolsep}{5pt}
\begin{tabular}{rrrrrr}
\toprule
Scale $a$ &
Median $q$ &
Maximum $q$ &
$q=0$ &
$\Delta_5>2q$ &
Certified Top-5 \\
\midrule
$0$        & 0  & 0   & 879 & 326 & 879 \\
$10^{-10}$ & 0  & 0   & 879 & 326 & 879 \\
$10^{-8}$  & 0  & 0   & 879 & 326 & 879 \\
$10^{-6}$  & 0  & 1   & 872 & 322 & 872 \\
$10^{-4}$  & 0  & 4   & 531 & 185 & 531 \\
$10^{-3}$  & 5  & 18  & 11  & 5   & 11  \\
$10^{-2}$  & 51 & 112 & 0   & 0   & 0   \\
\bottomrule
\end{tabular}
\par\smallskip
\begin{minipage}{0.98\linewidth}
\footnotesize
Notes: Each row contains 879 method-date observations
under the primary estimation specification.
The final three columns report observation counts.
Top-5 stability is certified when $q=0$ or
$\Delta_5>2q$. The former condition preserves
the MST and hence the selected stock set under
the same tie-breaking rule. The perturbation
radii are hypothetical and do not represent
estimated sampling errors.
\end{minipage}
\end{table}

For perturbation scales no greater than $10^{-8}$,
$q(\delta)$ is zero for all 879 estimated networks.
As the perturbation scale increases, more non-tree
edges fall within the corresponding replacement
threshold. At $a=10^{-4}$, the sufficient conditions
preserve the selected five-stock set in 531 cases,
while at $a=10^{-2}$ neither condition is satisfied
in any of the estimated networks.

For the perturbation radii considered here, all
observations satisfying the degree-gap condition
also satisfy $q=0$. The degree-gap condition
therefore provides no additional stability
guarantees beyond those obtained from exact
MST preservation in this comparison.
These results describe the local sensitivity
of the estimated networks and should not be
interpreted as probabilities of recovering
the underlying population networks.

We next examine the sensitivity of stock selection
to the predetermined ordering used to resolve
degree ties. Holding each estimated MST and its
degree values fixed, we compare the selected
stocks under the original universe ordering
with those obtained under the reverse ordering.

\begin{table}[htbp]
\centering
\caption{Stock Selection Sensitivity to Reverse
Tie-Breaking}
\label{tab:appendix_reverse_tiebreak}
\small
\setlength{\tabcolsep}{4pt}
\begin{tabular}{lrrrrr}
\toprule
Network & $k$ & Tied dates & Changed sets &
Mean overlap & Tied overlap \\
\midrule
GFEVD            & 3  & 133 & 133 & 0.838 & 0.644 \\
                 & 5  & 183 & 183 & 0.850 & 0.760 \\
                 & 10 & 224 & 224 & 0.869 & 0.829 \\
\midrule
Absolute Pearson & 3  & 127 & 127 & 0.838 & 0.627 \\
                 & 5  & 185 & 185 & 0.849 & 0.761 \\
                 & 10 & 236 & 236 & 0.868 & 0.836 \\
\midrule
Signed Pearson   & 3  & 125 & 125 & 0.841 & 0.627 \\
                 & 5  & 185 & 185 & 0.848 & 0.759 \\
                 & 10 & 234 & 234 & 0.869 & 0.835 \\
\bottomrule
\end{tabular}
\par\smallskip
\begin{minipage}{0.98\linewidth}
\footnotesize
Notes: Each network and portfolio-size combination
contains 293 formation dates with $W=250$.
Overlap is measured as the number of stocks shared
by the original and reverse-order selections,
divided by $k$. Mean overlap is calculated across
all formation dates, while tied overlap is
conditional on a tied degree cutoff.
\end{minipage}
\end{table}

As shown in Table~\ref{tab:appendix_reverse_tiebreak},
reversing the tie-breaking order changes the
selected stock set whenever the degree cutoff
is tied, while selections with a strictly
positive degree gap remain unchanged.
Under the primary five-stock specification,
the conditional mean overlap is approximately
0.76 for each network method.
Thus, when a cutoff tie occurs, reversing
the ordering rule replaces approximately
1.2 stocks on average.

The frequency of tied cutoffs increases with
portfolio size over the three values of $k$
considered here. However, the proportion of
shared stocks conditional on a tie also
increases. These findings indicate that the
degree-ranking rule may be sensitive to the
ordering of stocks even when the estimated
network remains unchanged. The comparisons
are restricted to stock selection and do not
establish the effect of alternative tie-breaking
rules on subsequent portfolio returns.

%% file: appendix/C_data_reproducibility.tex
\section{Data and Reproducibility}
\label{app:data_reproducibility}

This appendix provides additional information on the construction of the
return panel, the reconstruction of the estimated networks, and the
portfolio calculations used in the revised empirical analysis. We retain
the original network specification and the corresponding target weights
so that the effect of the revised portfolio accounting can be examined
separately from changes in network estimation.

\subsection{Data Construction and Sample Reconciliation}
\label{app:data_construction}

The return panel is constructed from the saved Yahoo Finance adjusted
closing prices and the ticker-level reconciliation records described in
Section~\ref{sec:data}. The original list contains 486 tickers, of which
451 satisfy the eligibility requirements over the available historical
sample. The retained stocks are ordered according to their original
universe positions. To construct the common calendar, we retain dates
with finite and positive adjusted closing prices for at least
$\lfloor 0.90N\rfloor$ of the retained stocks. Prices are then aligned
on this calendar, and daily simple returns are calculated according to
Equation~\eqref{eq:data_daily_return}. Missing observations are not
replaced by zero returns or carried forward. The daily DTB3 observations
used in the performance calculations are also retained with the data
records.

The corrected signal calendar contains 293 completed-week signals,
from March 19, 2021 through October 23, 2026. The corresponding execution
dates run from March 22, 2021 through October 26, 2026, and portfolio
wealth is evaluated through October 28, 2026. The signal associated with
the incomplete final week is not used. The same calendar and ticker
ordering are retained in the subsequent reconstruction. The checksum
of the reconstructed return panel agrees with that of the panel used
in the original experiment.

Although the estimation windows use returns available through each
signal date, the eligibility checks use information from the full
historical sample. The stock universe is therefore not a point-in-time
record of index membership. The data also do not contain a separate
series of delisting returns for the excluded stocks. These limitations
remain in the revised analysis and should be distinguished from the
timing of network estimation and portfolio execution.

\subsection{Network Estimation and Reconstruction}
\label{app:network_reconstruction}

For the primary specification, we retain the rolling window of 250
daily return observations, the pairwise VAR(1), the forecast horizon
of ten trading days, and the minimum-cost symmetrization rule. The
three network methods use the same stock universe and the same
deterministic ordering for resolving ties. Each weekly checkpoint
contains the estimated tree, its degree ranking, and the equal-weight
and inverse historical VaR target vectors for the primary five-stock
portfolio.

The original checkpoints did not retain the costs of all candidate
edges that were excluded from the MST. To calculate the replacement
margins, we reconstructed the candidate costs using the original
network estimation functions and the corrected return panel. The
reconstruction covers 293 dates and three network methods, giving
879 networks and 89,196,525 undirected candidate-edge costs. The
reconstructed MST edge sets, degree rankings, selected five-stock sets,
and target weights agree with the saved primary checkpoints. The
maximum absolute differences in the retained MST edge weights and the
two target-weight vectors are zero. The reconstruction does not,
however, establish bytewise agreement with the original non-tree
costs, since those costs were not saved in the original experiment.

The resulting cost archive is used to calculate the sample replacement
gaps, the cost perturbations between consecutive weekly estimates,
and the stability conditions described in
Section~\ref{subsec:empirical_network_stability}. These calculations
require the archived costs and trees rather than another VAR fit.
The estimated replacement margins are not population margins, and
the observed cost perturbations are not estimates of the errors
relative to an unknown population cost matrix. A failed sufficient
condition is not interpreted as evidence that the selected stocks
must change.

\subsection{Portfolio Simulation and Statistical Inference}
\label{app:portfolio_reproduction}

The revised portfolio calculations use the saved target weights and
execute the trades at the next available session close following the
signal date. The dollar values of the existing positions are updated
using the returns through that close before the new target weights
are applied. Transaction costs are calculated at ten basis points of
gross traded value using the joint fee and holdings calculation in
Section~\ref{subsec:empirical_trading}. Between executions, the
quantities held in adjusted-price units remain unchanged. This is the
share-drift accounting convention (Track B). The original constant-mix
paths (Track A) and the paths with daily restoration costs (Track C)
are retained separately for comparison. The adjusted-price units
are synthetic holdings rather than verified historical brokerage
share quantities.

The six primary strategies are replayed without changing their network
estimates or target weights. The reconstructed wealth and transaction
costs agree with the earlier share-drift calculations up to numerical
precision. For the reserved period, wealth is normalized to one at
the December 29, 2023 close; positions are carried into the period
without an additional entry trade. The paired bootstrap uses the
corresponding daily after-cost returns, 5,000 replications, and block
lengths of 5, 10, 21, and 63 sessions. The seed is determined by
$51001+1000j+\ell$, where $j=1,2,3$ denotes the full, diagnostic,
and reserved periods, respectively, and $\ell$ denotes the block
length. In particular, the reserved-period calculation with ten-session
blocks uses seed 54011. The same resampled dates are used for all six
strategies within each replication. This preserves the paired
comparison without requiring independence across strategy returns.

For the parameter sensitivity analysis, the equal-weight target
vectors are reconstructed from the saved degree rankings using the
original tie-breaking rule. For the 60-day window, the three methods
omit trading on the reserved-period GFEVD failure signals, and GFEVD
also omits its two diagnostic-period failure signals. The existing
positions continue to change in value on these dates. The missing
inverse historical VaR target vectors outside the primary parameter
combination are not reconstructed in this replay. The horizon and
symmetrization comparisons use their separately saved target weights.
All of these comparisons use previously examined historical outcomes
and do not constitute a new independent evaluation sample.

\subsection{Reproducibility Records}
\label{app:reproducibility_records}

The empirical calculations are implemented in R. The original
checkpoints, reconstructed candidate-edge costs, structural statistics,
and portfolio accounting records are retained separately. The
candidate-edge reconstruction was performed using R version 4.5.0.
The final revision-stage replay reads these saved records and writes
its results to a separate directory. From the root of the preserved
research repository, the replay is run using
\begin{verbatim}
Rscript results/revision_v2/core_finalization/run_v2r1.R
\end{verbatim}
The script requires the original source functions, cached price and
interest-rate data, and the archived checkpoints. It is not a standalone
script that downloads the data and estimates all networks from the
beginning. The original manuscript and experiment records should be
retained when running the replay because the script also checks their
stored checksums.

The output records include the six daily portfolio ledgers, the
primary performance and paired inference tables, the bootstrap and
parameter sensitivity results, and the replacement-margin and
selection-stability summaries. Input and output manifests record the
checksums of individual files; the sizes of large checkpoint
directories are recorded separately rather than treating them as
bytewise checksum verification. The test records include the return
panel and calendar checks, the agreement with the saved target weights,
and the numerical reconciliation of portfolio wealth and fees.
The final summaries incorporate corrections to the aggregation of
structural comparisons and to a source-code string check after the
original replay. These corrections did not require the bootstrap to
be rerun. The records distinguish the original constant-mix results
from the revised accounting results so that the tables in the main
text and the additional results in Appendix~\ref{app:additional_results}
can be traced to their corresponding calculations.